\documentclass[preprint]{aastex}
\usepackage{graphicx}
\usepackage{float}
\usepackage[usenames]{color}
\usepackage{natbib}

\def\l{$\lambda$}
\def\a{$\alpha$}

\def\D{$\Delta$}
\def\hb{H$\beta$\/}
\def\civ{{\sc{Civ}}$\lambda$1549\/}
\def\mgii{{Mg\sc{ii}}$\lambda$2800\/}
\def\oi{{\sc{Oi}}$\lambda$1304\/}
\def\siii{ Si{\sc ii}$\lambda$1814\/}
\def\mbh{$M_{\rm BH}$\/}
\def\feii{{Fe\sc{ii}}\/}
\def\feiii{{Fe\sc{iii}}\/}
\def\kms{km s$^{-1}$\/}
\def\<{$\textless$}
\def\>{$\textgreater$}
\def\.{$\cdot$}
\def\R{$r_\mathrm{BLR}$}
\def\M{$M_\mathrm{BH}$}
\def\n{$n_\mathrm{e}$}
\def\ciii{{\sc{Ciii]}}$\lambda$1909\/}
\def\siiii{ Si{\sc iii]}$\lambda$1892\/}
\def\aliii{Al{\sc iii}$\lambda$1860\/}
\def\siii{{Si\sc{ii}}$\lambda$1814\/}
\def\civ{{\sc{Civ}}$\lambda$1549\/}
\def\siiv{Si{\sc iv}$\lambda$1397\/}
\def\oiv{O{\sc iv]}$\lambda$1402\/}
\def\caii{Ca{\sc ii}$\lambda$8579\/}
\def\nh{$n_{\mathrm{H}}$\/}
\def\nhu{$n_{\mathrm{H}}U$\/}

\def\ltsima{$\; \buildrel < \over \sim \;$}
\def\ltsim{\lower.5ex\hbox{\ltsima}}  
\def\gtsima{$\; \buildrel > \over \sim \;$}
\def\gtsim{\lower.5ex\hbox{\gtsima}} 
\def\cmq{cm$^{-2}$\/}
\def\cm3{cm$^{-3}$\/}

\shorttitle{Broad Line Region Radius of Low-$z$ Quasars }
\shortauthors{Negrete et al.}

\begin{document}
\title{Reverberation and photoionization estimates of the Broad Line Region Radius in Low-$z$ Quasars}
 
\author{C. Alenka Negrete\altaffilmark{1}}
\affil{Instituto Nacional de Astrof\'isica, \'Optica y Electr\'onica, Mexico}
\affil{Instituto de Astronom\'ia, Universidad Nacional Aut\'onoma de
M\'exico, Mexico}
\author{Deborah Dultzin\altaffilmark{2}}
\affil{Instituto de Astronom\'ia, Universidad Nacional Aut\'onoma de
M\'exico, Mexico}
\author{Paola Marziani\altaffilmark{3}}
\affil{INAF, Astronomical Observatory of Padova, Italy}
 
\and
 
\author{Jack W. Sulentic\altaffilmark{4}}
\affil{Instituto de Astrof\'isica de Andaluc\'ia, Spain}

\altaffiltext{1}{cnegrete@inaoep.mx}
\altaffiltext{2}{deborah@astro.unam.mx}
\altaffiltext{3}{paola.marziani@oapd.inaf.it}
\altaffiltext{4}{sulentic@iaa.es}
 
\begin{abstract}
Black Hole Mass estimation in quasars, especially at high redshift, involves use of single epoch spectra with s/n and resolution that permit accurate measurement of the width of a broad line assumed to be a reliable virial estimator. Coupled with an estimate of the  radius of the broad line region this yields \mbh\.  The radius of the broad line region (BLR) may be inferred from an extrapolation of the correlation between source luminosity and reverberation derived \R\ measures (the so-called Kaspi relation involving about 60 low z sources). We are exploring a different method for estimating \R\ directly from inferred physical conditions in the BLR of each source. We report here on a comparison of \R\ estimates  that come from our method and from reverberation  mapping. Our ``photoionization'' method employs diagnostic line intensity ratios in the rest-frame range 1400-2000 \AA\ (\aliii/\siiii, \civ/\aliii) that enable derivation of the  product of  density  and ionization  parameter with the BLR distance derived  from the definition of the ionization parameter.  We find good agreement between our estimates of the density, ionization parameter and \R\ and those from reverberation  mapping.  We suggest empirical corrections to improve the agreement between individual photoionization-derived \R\ values and those obtained from reverberation mapping. The results in this paper can be exploited to estimate black  hole masses  \M\ for large samples of high-$z$\ quasars using an appropriate virial broadening estimator. We show that the width of the UV intermediate emission lines are consistent with the width of \hb, therefore providing  a reliable virial broadening estimator that can be measured in large samples of high-z quasars.
\end{abstract}


\section{Introduction}
Quasars are intriguing objects whose intense activity arises in a small volume (fraction-of-parsecÊ radius) whose widely accepted interpretation involves accretion onto a central black hole \citep[e.g.][and references therein]{donofrioetal12}. An important signature of a majority of quasars involves the presence of broad emission lines in the UV-optical-IR spectrum. A major challenge involves estimation of the distance \RÊ 
from the central continuum source of the line emitting region (the broad line region: BLR). The BLR cannot be resolved with direct imaging even in the nearest sources.Ê The most direct methodÊ for estimating \R\ (hereafter also referred to as the ``radius'' for brevity) is through reverberation mapping \citep[RM;][]{peterson98,horneetal04}. This technique measures the time delay $\tau$\ inÊ the response of the broad emission lines to changes in the ionizing continuum.Ê The rest-frame distance is then defined as:

\begin{equation}
r_\mathrm{BLR}= \frac{c \cdot \tau }{ (1+z)} = c \tau_\mathrm{rf}
\label{eq:light},
\end{equation}

whereÊ $\tau_\mathrm{rf}$\ is the time delay in the quasar rest frame. For brevity we shall use only $c\tau$ to refer to this rest frame distance. \R\ estimates from reverberation mapping require a significant observational effort and haveÊ been obtained for onlyÊ $\approx$ 50 nearby objects ($z <$ 0.4) \citep{kaspietal00,petersonetal04,kaspietal05,bentzetal09,bentzetal10,denneyetal10}. An indirectÊ method for measuring \R\ was proposed by \citet{kaspietal00,kaspietal05} whoÊ found a correlation between \R\ and the optical continuum luminosity atÊ 5100\AA,

\begin{equation}
r_\mathrm{BLR} \propto L^\alpha
\label{eq:rkasp}
\end{equation}

with \a\ $\approx 0.5 -Ê 0.7$\ and recent studies favoring a value between 0.5 \citep{bentzetal09} and 0.6 \citep{marzianietal09}.Ê Apart from uncertainty of the power-law exponent (and any possible dependence of $\alpha$ on the the luminosity range considered) the rmsÊ intrinsic scatterÊ associated with \R in the \citet{bentzetal09} data is $\approx 0.28$\ dex.Ê This method has the advantage of being straightforwardly applicable to large samples of quasars for which single-epoch spectra are available requiring only moderate resolution spectrophotometry involving a broad emission line assumed to be a valid virial estimator. Eq. \ref{eq:rkasp} hasÊ paved the way towards definition of ``scaling laws'' between central black hole mass \mbh, line width and luminosity \citep[][and references therein]{shenliu12}.Ê Estimating the black hole mass is plagued by large uncertainties becauseÊ several other variables are at play: emitting region structure, orientation effects, emitting gas dynamics, etc. They will be briefly discussed in \S \ref{sec:discussion}.Ê However, estimated uncertainties in $\log$\mbh\ associated with scaling laws are $\approx \pm$0.3 -- 0.4 dex at 1 $\sigma$\ confidence level  implying that a large fraction of the scatter is due to the assumption of Eq. \ref{eq:rkasp}. 

The aim of this paper is to reduce the uncertainty in individual estimates of \R\ and hence \mbh\. Rather than relying on the \R\ -- $L$\ correlation we propose an alternate approach for \R\ estimation based on theÊ simple assumption that gas giving rise to the UV resonance lines is photoionized by the central continuum source. \R\ is then estimated not from a correlation but on an object-by-object basis.Ê The photoionization method is explained in \S \ref{sec:method}, while the sample is described in \S \ref{sec:sample}. We compare photoionization and reverberation values in \S \ref{sec:results}.Ê Discussion of the potential advantagesÊ are given \S \ref{subsec:comparison}.Ê Computations were made considering $H_0$ =70 km s$^{-1}$Mpc$^{-1}$ and a relative energy density $\Omega_\Lambda=0.7$ and $\Omega_\mathrm{M}=0.3$.

\section{Method}
\label{sec:method}

The BLR radius \R\ is linked to physical parameters such as  hydrogen  density (\nh) and ionization parameter ($U$) by  the definition of the ionization parameter itself: 

\begin{equation}
r_{BLR} = \left[ \frac {\int_{\nu_0}^{+\infty}  \frac{L_\nu} {h\nu} d\nu} {4\pi n_\mathrm{H} U c} \right]^{1/2}\label{eq:r}
\end{equation}
\defcitealias{negreteetal12}{Paper I}

where $L_{\nu}$ is the specific luminosity per unit frequency $\nu$, $h$ is the Planck constant and $c$\ the speed of light. The integral is carried out from the Lyman limit to the largest frequency on the rest frame specific flux $f_{\nu} = L_{\nu}/(4 \pi d^{2})$. For the integral we will use an average of two Spectral Energy Distributions (SEDs) described by \citet{mathewsferland87} and \citet[][]{laoretal97b}. This assumption will be further discussed in \S \ref{subsec:comparison}.  We can estimate  \R\ if we have a reasonable estimate of the product \nhu\ \citep[][]{padovanirafanelli88,padovanietal90,negrete11,negreteetal12}. Recently,   \citet{negreteetal12}   \citepalias[hereafter][]{negreteetal12} derived \nh\ and $U$\ for two high Eddington ratio sources that at low- and moderate $L$\ show the typical spectra of narrow-line Seyfert 1s (NLSy1). These findings cannot be easily generalized; however we  show in the following that the  product \nhu\ can still be estimated from a set of diagnostic ratios.

\subsection{Emission Line Ratios}

Emission line ratios such as \ciii/ \siiii\ and \aliii/ \siiii\ are important diagnostics for ranges of density that depend on their transition probabilities \citep[e.g.][]{feldmanetal92}. Emission lines originating from forbidden or semi-forbidden transitions become collisionally quenched above the critical density and hence weaker than lines for which collisional effects are negligible. \ciii/ \siiii\ is suitable as a diagnostic when \nh\ $\ltsim 10^{11}$ \cm3. The \aliii / \siiii\ ratio is well suited over the density range 10$^{11}$ -- 10$^{13}$ \cm3\ \citep[\citetalias{negreteetal12},][]{marzianietal11}.  The ratios \siii/\siiii\ and \siiv/\siiii\ are independent of metallicity and sensitive to ionization. The ratio \siiv/\civ\ is mainly sensitive to metallicity.  Measuring \siii\ is a challenge in most spectra because of its weakness. Emission line ratios involving \siii\ are therefore subject to large uncertainty and poorly constrain physical conditions. We do not use ratios involving this line. \siiv\ is a stronger line but severe blending with \oiv\ hampers its use. In the extreme sources considered in \citetalias{negreteetal12} the high inferred BC density led us to expect an insignificant \oiv\ contribution. However this might not be true for a more general population of quasars \citep{willsnetzer79}.

We exclude the \siiv+\oiv\ blend and restrict our analysis to three diagnostic ratios involving the four remaining strongest metal  lines in the spectra: 1)   \ciii/ \siiii\ which is  important because several sources show large  \ciii\ equivalent widths (unlike sources included in \citetalias{negreteetal12}); 2) \aliii/ \siiii\ that is sensitive to \nh\ and believed to reflect the densest regions which are most optically thick to the ionizing continuum; 3) \civ/\siiii\ as a marker of ionization level. A quantitative interpretation of these diagnostic ratios requires supporting photoionization simulations. {\sc cloudy} simulations \citep{ferlandetal13} at fixed \nh\ and $U$\ values allow us to study how these parameters influence the diagnostic ratios we have adopted. Our simulations span the density range $7.00 \leq \log$ \nh$ \leq 14.00$ and $-4.50 \leq \log U \leq 00.00$ in intervals of  0.25 dex assuming plane-parallel geometry, solar metallicity,  column density 10$^{23}$ \cmq\ as well as a ``standard'' quasar continuum as parameterized by \citet{mathewsferland87}. Further details are given in \citetalias{negreteetal12}. Sources in \citetalias{negreteetal12} show weak \ciii\ emission (relative to \siiii) which simplifies interpretation of the emission-line spectrum. In those cases computation of constant value contours for the diagnostic ratios in the $U$ vs. \nh\  plane show convergence towards a low ionization plus high density range. In the case of the sources considered here we show that \aliii / \siiii\  and \civ/\siiii\, along with any other ratio not involving \ciii\, converge towards a low-ionization, high \nh\ region while ratios involving \ciii\ converge to a higher ionization, low \nh\ zone (\S \ref{sec:results}).

In the photoionization computations with {\sc cloudy} we have made several simplifying assumptions. The main one consists in deriving a single value of \R\ from fixed physical conditions. It is known that the BLR is not a shell nor a sequence of nested shells; however a gradient of ionization is indicated by the shorter reverberation time responses of lines coming from ions of higher ionization potential \citep[e.g., ][]{petersonwandel99,netzer08}. The Ê\R\ values derived here are probably biased towards the inner radius of the BLR. This is likely to be true for both our method and the RM because Êgas close to the continuum source responds more strongly to the incoming ionizing radiation. The \aliii, \siiii, \civ\ lines are all emitted in the fully ionized zone of emitting gas clouds or slabs \citepalias{negreteetal12}, so that they are sensitive to the ionizing photon flux which is exactly the product \nh $U$. ÊA more realistic approach would be to consider a model that allows gas density \nh, Ê column density $N_{\rm c}$, and Ê ionization parameter $U$, to be functions of $r$\ Ê\citep{devereux13,devereuxheaton13}. As pointed out by \citet{devereux13} the \R\ derived from reverberation mapping is an abstraction that may not have a clear structural counterpart, and the RM \R\ can indeed be lower than Êan emissivity weighted average if the geometry of the BLR is thick \citep{netzer90}. Nonetheless, RM \R\ has been considered as the best approximation available for the ``virial radius'' of the BLR and widely employed as such \citep[e.g.][and references therein]{dultzinetal06,marzianisulentic12,shen13}. ÊOur approach is probably good enough to provide estimate of the ``virial radius'' Êequivalent to the one based on RM. In addition, in \citetalias{negreteetal12} we show that our method allows to solve for \nh, $U$ and metal content in NLSy1 sources that are $\approx$ 10\%\ of all quasars.

\subsection{Extraction of the Broad Component}

Step 1 of our method involves isolation of the broad component (BC) in the selected emission lines. The BC is believed to be associated, at least in part, with the region predominantly emitting  low ionization lines (LILs) such as \mgii, \feii, part of the Balmer lines, \siii, \oi, \caii\ as well as intermediate ionization lines such as \ciii, \aliii\ and \siiii\ \citep{baldwinetal04,matsuokaetal08,marzianietal10,negreteetal12}.  A BC is present in the overwhelming majority of Seyfert 1 and Type 1 quasar spectra. In order to isolate this component in UV lines we  use \hb\ BC to define a BC profile shift and width. We take advantage of the fact that the BC is the dominant component  in all LILs if  FWHM(\hb) $\la$ 4000 \kms\ \citep[Population A, ][]{zamfiretal10}.  We will use the definition of Population A (with a FWHM(BC) $\la$ 4000 \kms\ and Lorentzian BC profile) and Population B (with a FWHM(BC) \> 4000 \kms\ and Gaussian BC profile) objects as well as the Eigenvector 1 (E1) parameter space described in \citetalias{negreteetal12} and \citet[][]{sulenticetal00a}. In Population A sources \civ\ is often dominated by  blueshifted emission probably associated with a high-ionization outflow \citep{richardsetal02,sulenticetal07}. In order to extract BC \civ\ we assume that the broad profile of \civ\ can be described as the sum  of the  BC (assumed FWHM equal or larger than FWHM \hb) + a blueshifted component   \citep{marzianietal96,leighly00,marzianietal10}. This approach is further supported  by  recent work indicating that  virial and outflow motions can coexist in quasars \citep{richardsetal11,wangetal11}. For broader (Population B) sources \hb\ can be modeled as the sum of  BC and redshifted very broad components (VBC; FWHM $\sim$ 10000 \kms; as  in the spectrum of PG0052+251 shown in Fig. \ref{fig:fits}).  For both populations, we fit \ciii\ with the same components of \hb. In the case of \siiii, \aliii\ and \siii, we fit a single BC component. The sum of the three components (broad, blue-shifted, very-broad) reproduce the line profiles  of the strongest lines in low-$z$\ quasars \citep{sulenticetal00c,marzianietal10}. The relative intensity  is different in the various line components but shifts and widths are roughly consistent for all lines.  More details are given by \citet{marzianietal10} where line component trends along the E1 sequence are shown.

In practice we apply  a multi-component profile decomposition using {\sc iraf} task {\sc specfit} in order to extract the  broad line component  \citep{marzianietal10}: 1) a relatively  unshifted, symmetric BC component representative gas whose broadening is assumed to arise in a virial velocity field. Our aim is to isolate the BC from 2) a blueshifted component associated with outflow or wind emission and 3) a VBC whose strength relative to the BC is set by an inflection observed in the \hb\ profile.

\section{The Sample of Sources with \R\ Obtained by Reverberation}
\label{sec:sample}

Reverberated sources allow the product \nhu\ to be independently estimated from \R\ and the source luminosity by inverting Eq. \ref{eq:r} or, almost equivalently, \R\ derived from photoionization consideration can be directly compared to $c\tau$. We selected 13 of 35  AGN with reverberation data \citep[][henceforth P04]{petersonetal04} showing high enough S/N in the rest-frame range 1400-2000 \AA\ to allow accurate decomposition of \ciii, \siiii, \aliii,  and \civ. We extracted UV spectra from the HST archive and carried out data reduction using  standard {\sc iraf} tasks. Optical spectra were taken from \citet{marzianietal03a}. Data  were corrected for Galactic extinction.  Table \ref{tab:obs} presents the sample with IDs given in Column 1 and other columns described below. The redshift of this sample is $z$ \< 0.24, but the luminosity range is relatively large ($\lambda L_{\lambda}(5100) \sim 10^{43} - 10^{45}$ erg s$^{-1}$) and almost uniformly covered.

\section{Results}
\label{sec:results}

\begin{figure}
\includegraphics[scale=0.35]{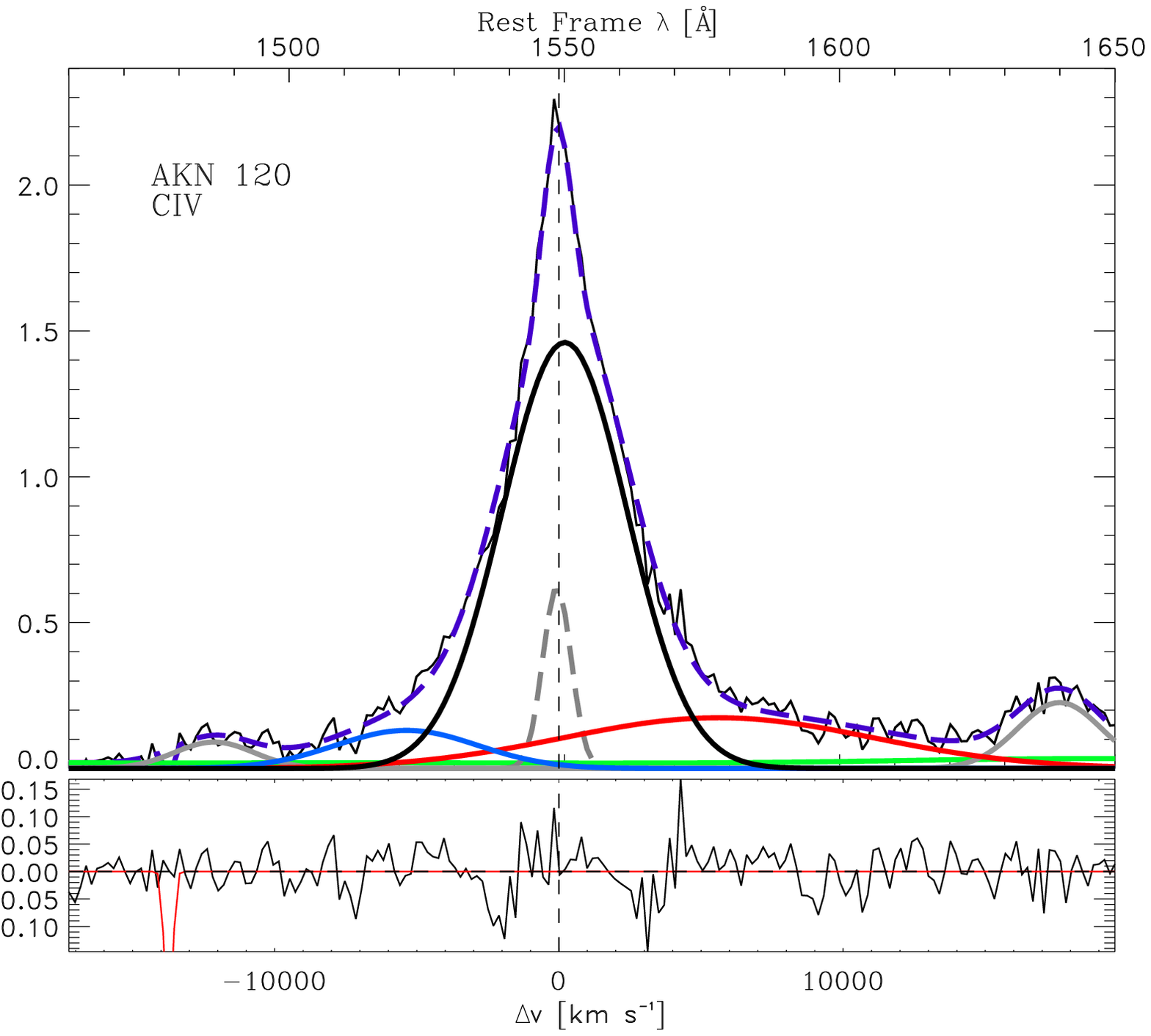}\includegraphics[scale=0.35]{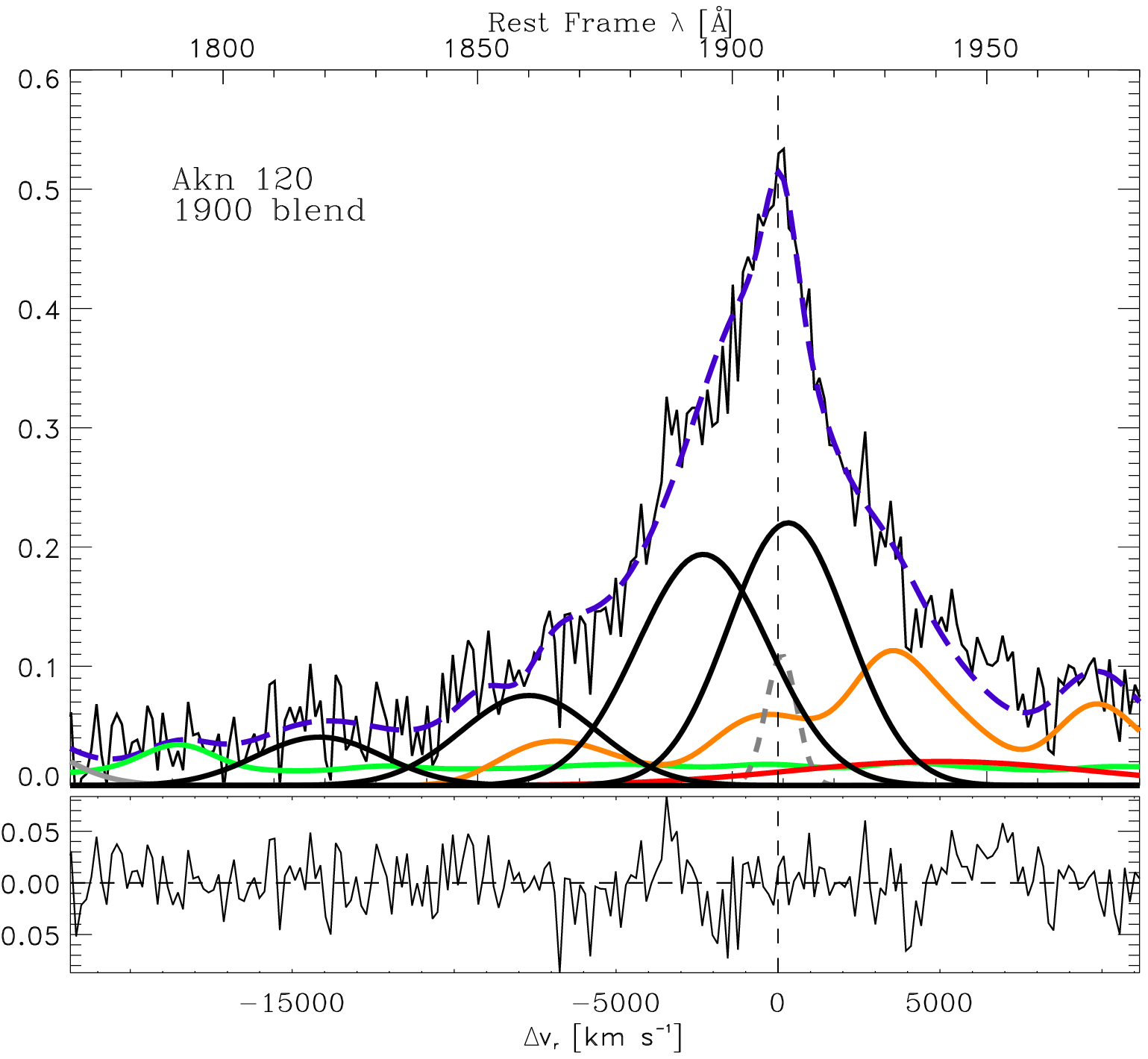}\includegraphics[scale=0.35]{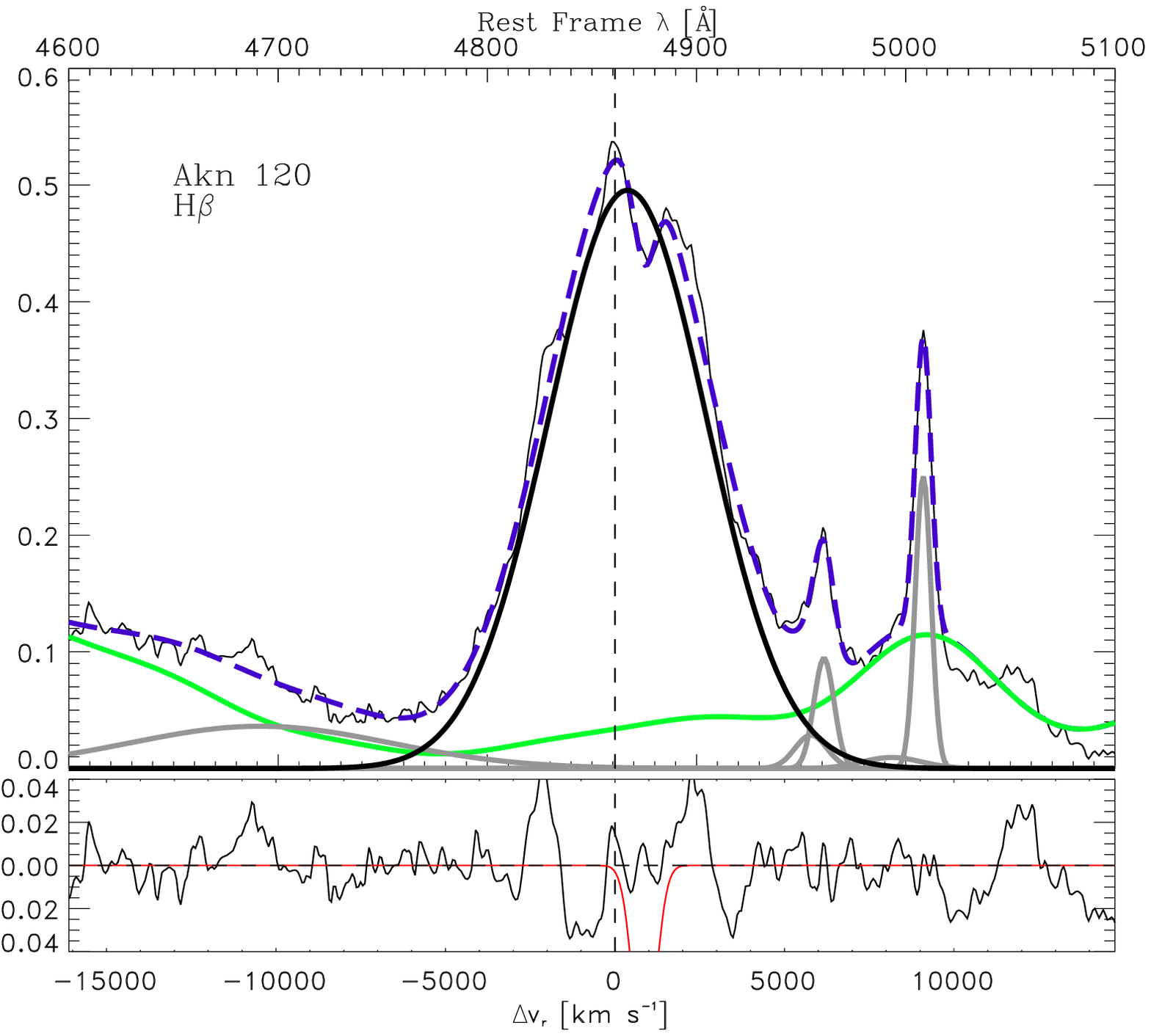}\\
\includegraphics[scale=0.35]{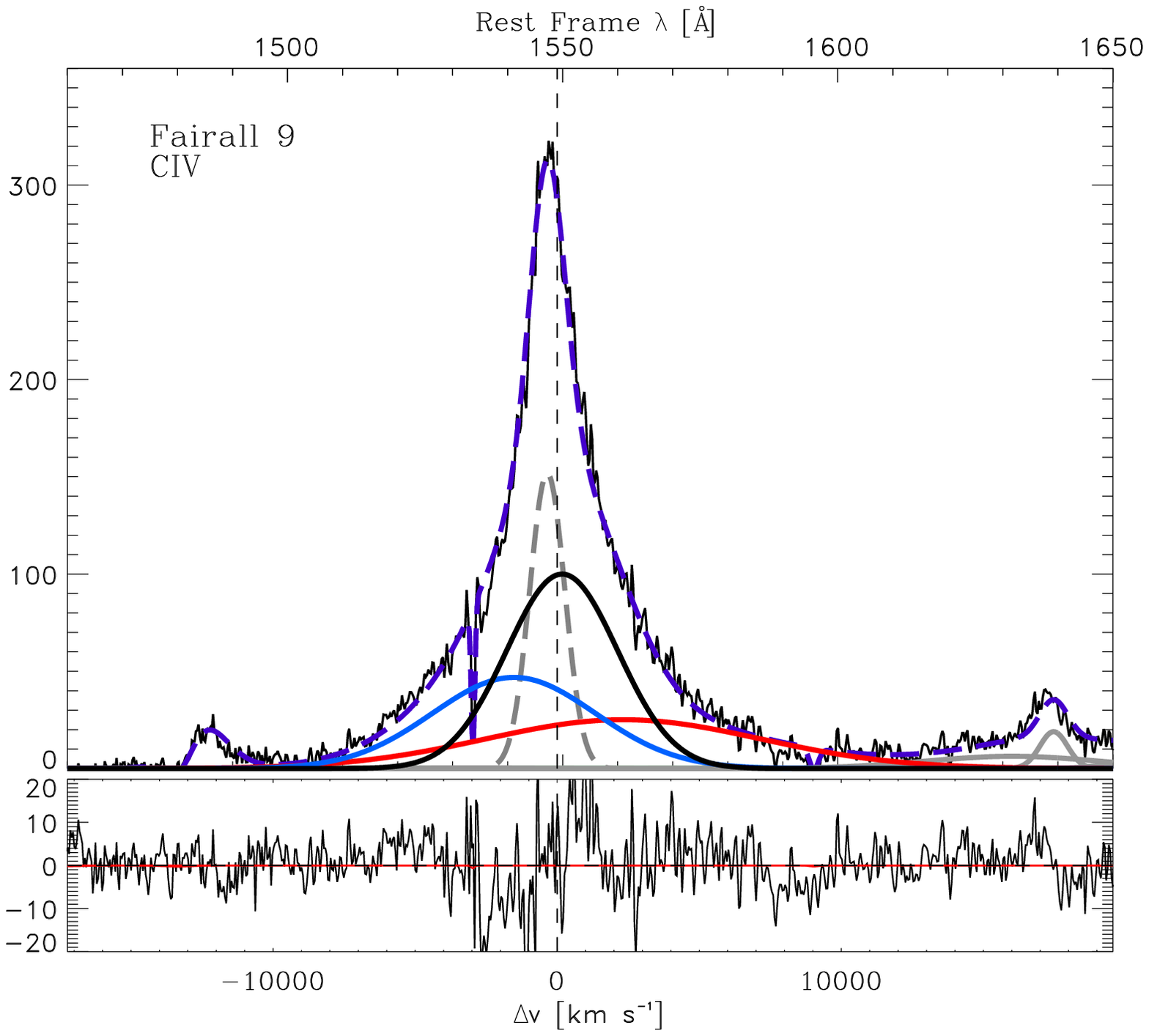}\includegraphics[scale=0.35]{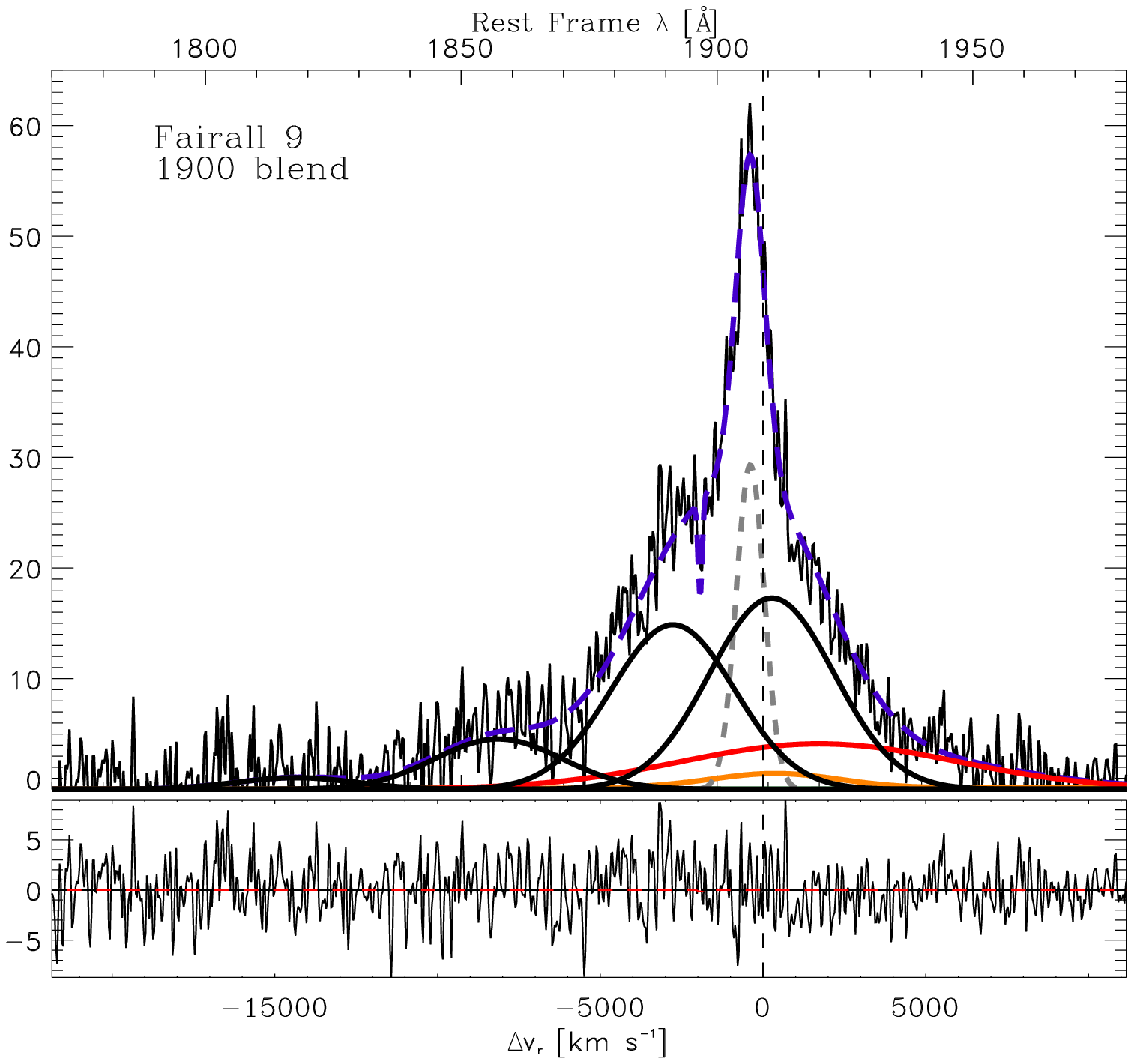}\includegraphics[scale=0.35]{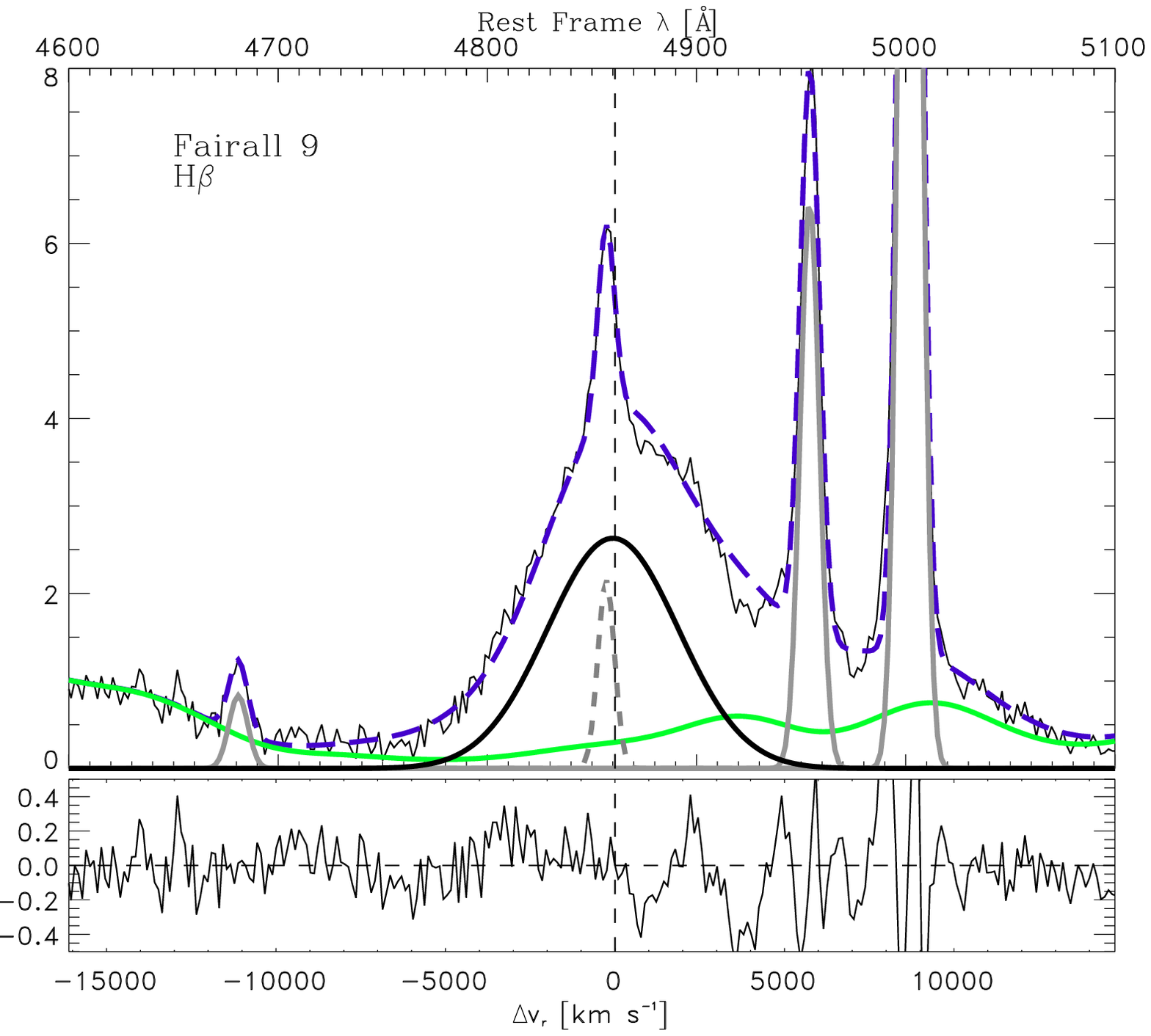}\\
\includegraphics[scale=0.35]{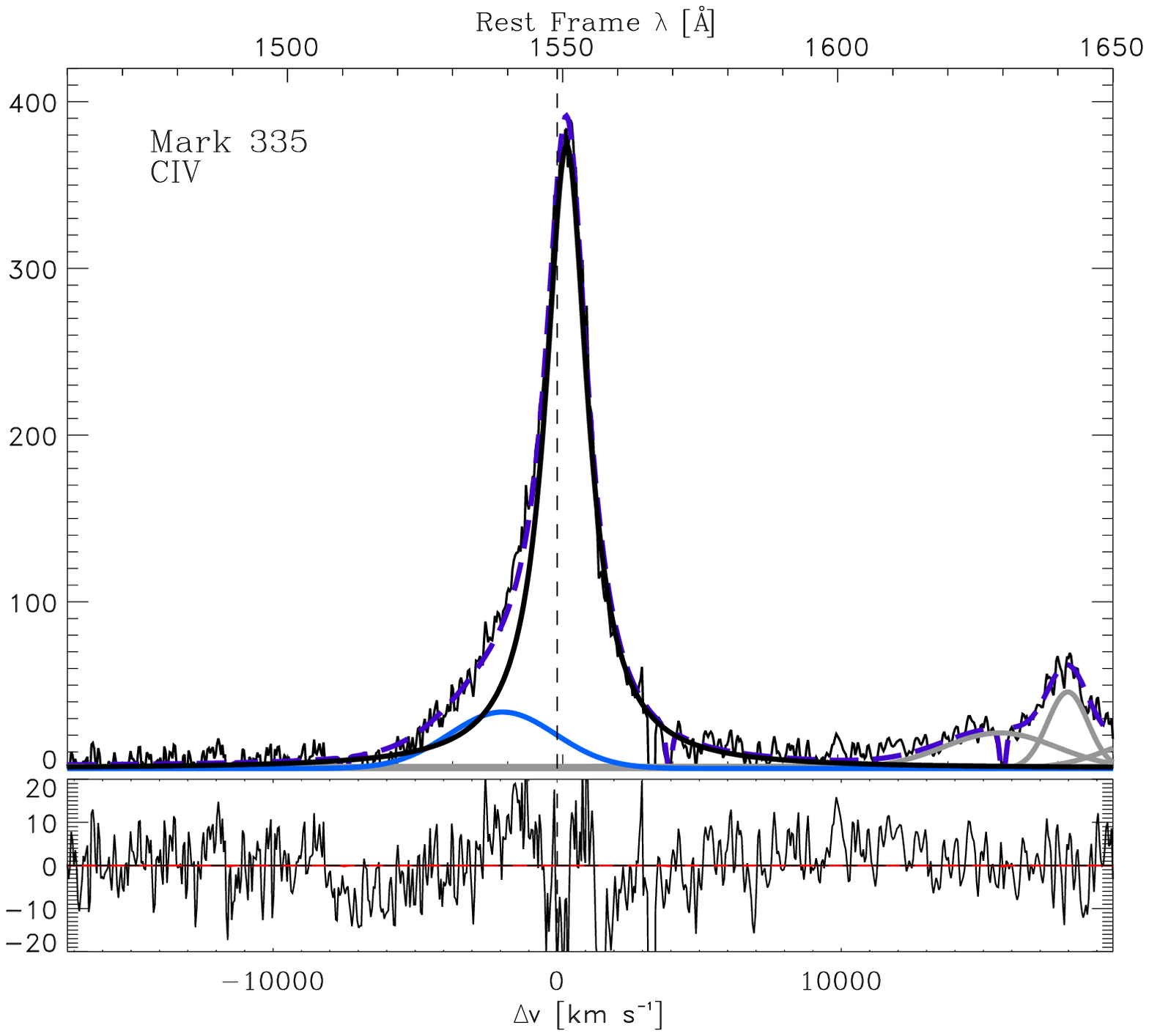}\includegraphics[scale=0.35]{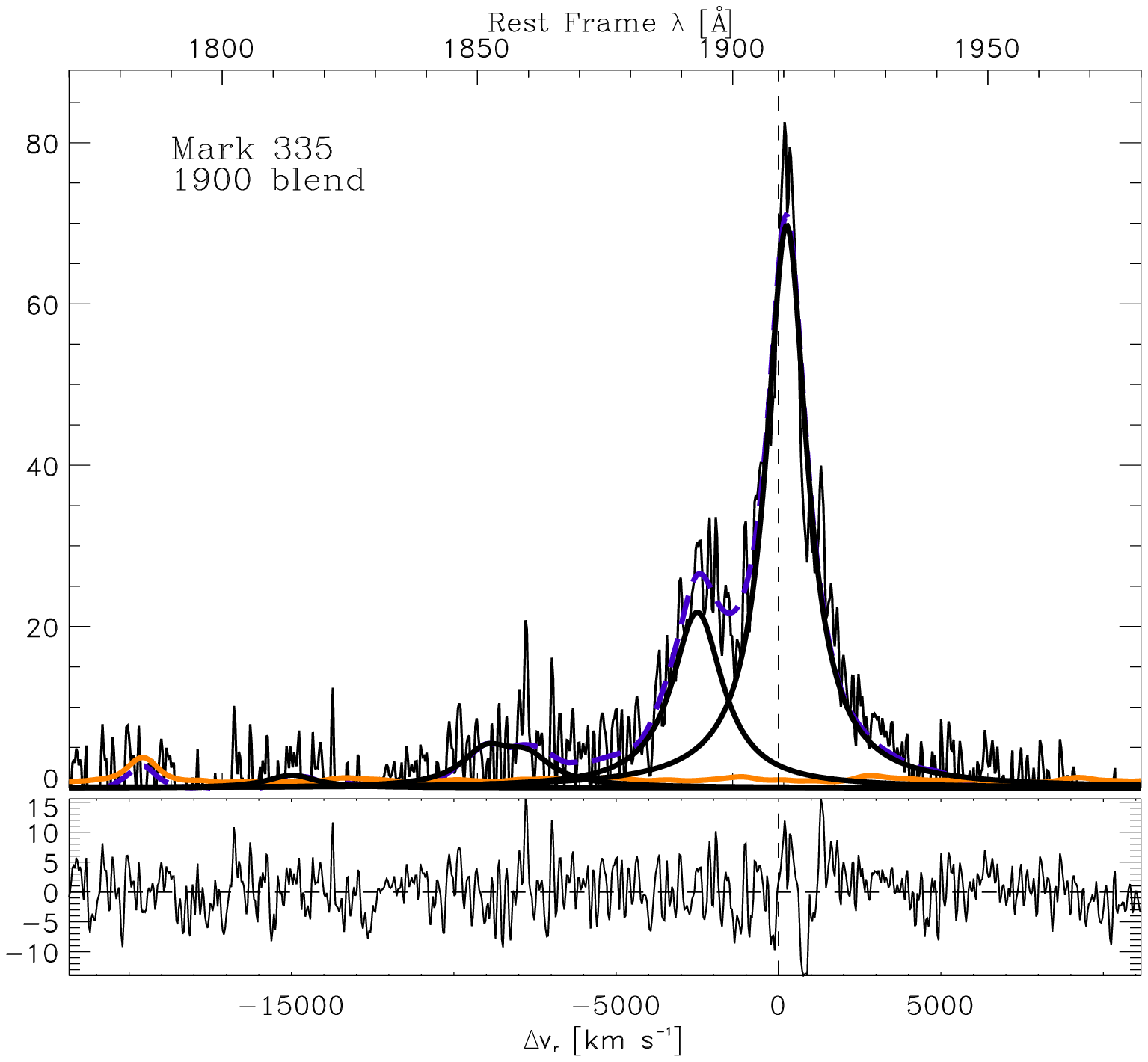}\includegraphics[scale=0.35]{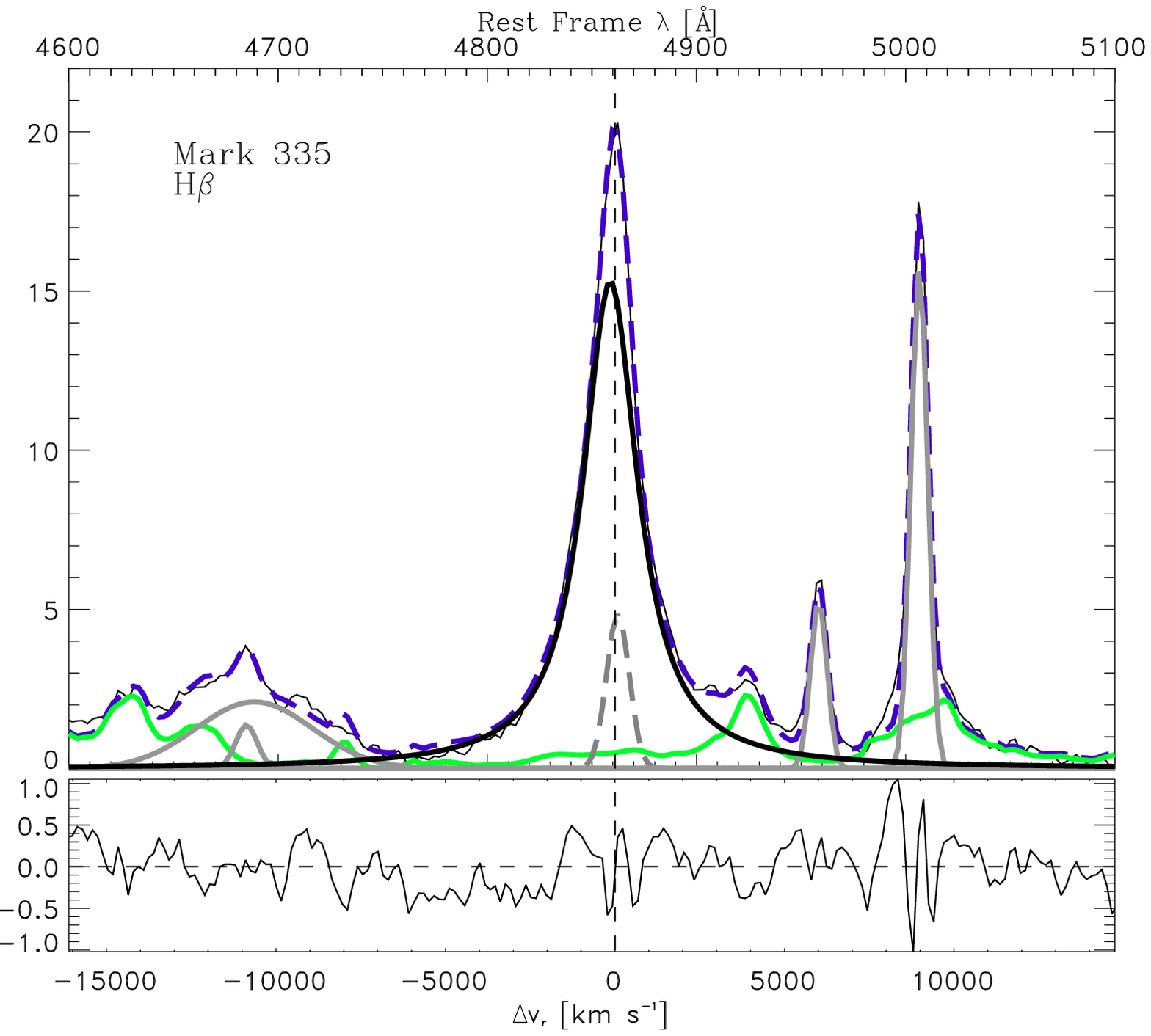}
\caption{Multicomponent fits for the 13 objects of our sample. Upper abscissa is rest frame wavelength in \AA, lower abscissa is in radial velocity units, ordinate is specific flux per unit wavelength in arbitrary units. Panels under the fits are the residuals. Vertical long dashed line is the rest frame for (left) \civ, (middle) \ciii\ and (right) \hb. Short purple dashed line is the fit to the whole spectrum. Black lines are the broad central components. Green lines represent the \feii\  template emission. The red lines are the VBC. The dashed grey lines are narrow components. The blue line in \civ\ corresponds to the BLUE component. The solid  grey lines in \hb\ and \civ\ represents the contribution of various underlying weaker emission lines. The orange line in the \l1900\AA\ blend is the \feiii\ template. In the panels of the residuals, the thin red lines are absorption lines considered in the fits.
\label{fig:fits}}
\end{figure}

Figure \ref{fig:fits} shows the spectra of Pop. A and Pop. B objects; see Col. (2) of Table \ref{tab:obs}. Fits to Pop. A sources yield line intensities of:  a) \civ\ BC + narrow component  (NC) + blue-shifted component; b) the 1900\AA\ blend that includes \ciii\ BC, \siiii\ BC, \aliii\ BC and \siii\ BC; and (c) \hb\  BC.  Narrow components are fitted whenever clearly visible. For Pop. B sources, the fits yield line intensities of a) \civ\ BC + NC + blue-shifted  + VBC; b) 1900\AA\ blend that includes \ciii\ BC + VBC, \siiii\ BC, \aliii\ BC and \siii\ BC; and c) \hb\  BC + NC + VBC. In Table \ref{tab:obs}, Col. (3) lists the rest-frame specific continuum flux at 1700\AA, Cols. (4) to (7) are the rest-frame line flux of the BC components of \civ, \aliii, \siiii\ and \ciii, Col. (8) is the FWHM of the \hb\ BC, Col. (9) is the FWHM of \siiii, \aliii\ and of the \civ\ BC. Col. (10) is the dispersion in the FWHM of the BCs. Figure \ref{fig:fits} indicates that a BC whose width is consistent with \hb\ can be straightforwardly extracted from the \aliii, and \siiii\ lines. The same approach yields also the BC component of \civ. 

\begin{figure}
\includegraphics[scale=0.35]{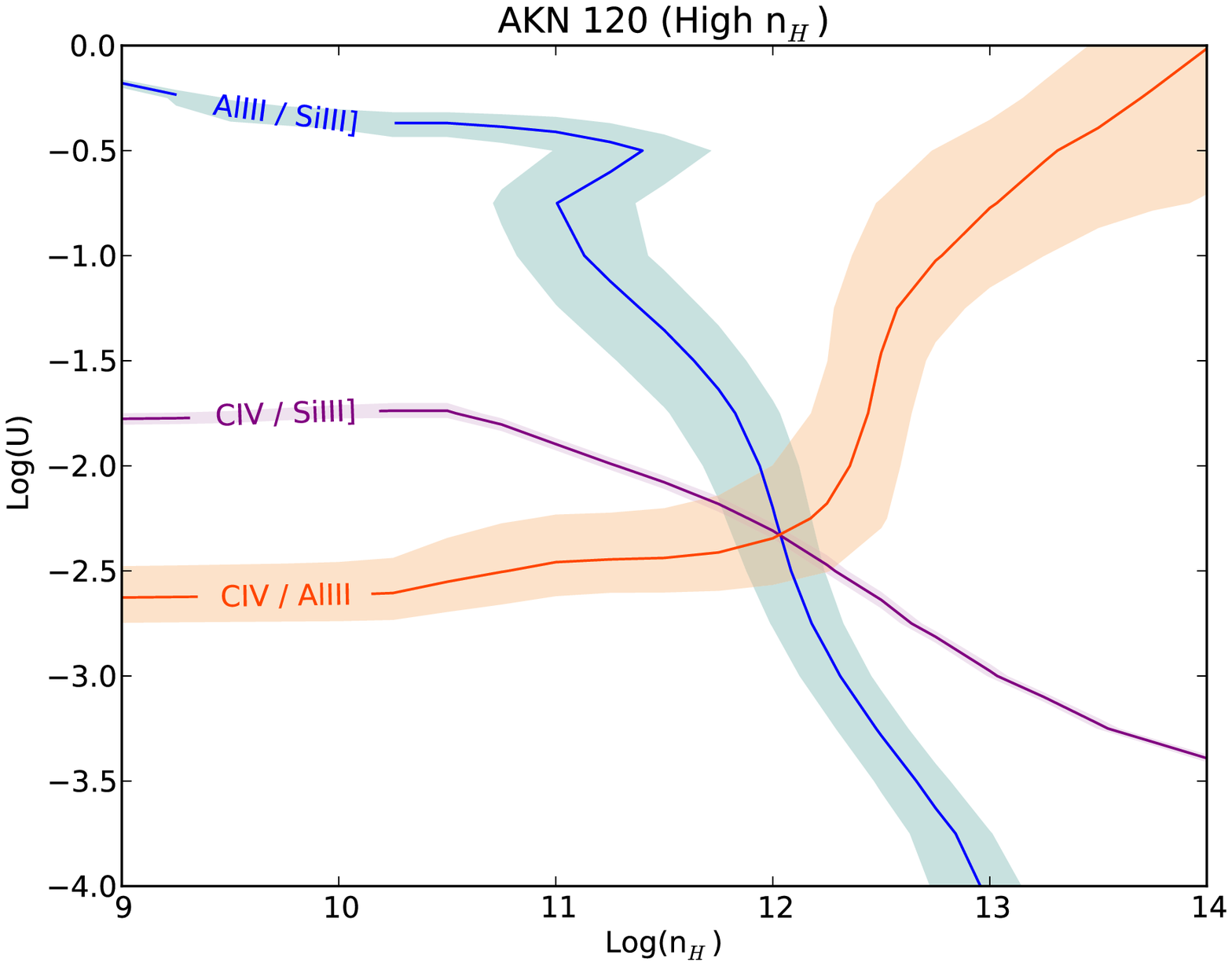}\includegraphics[scale=0.35]{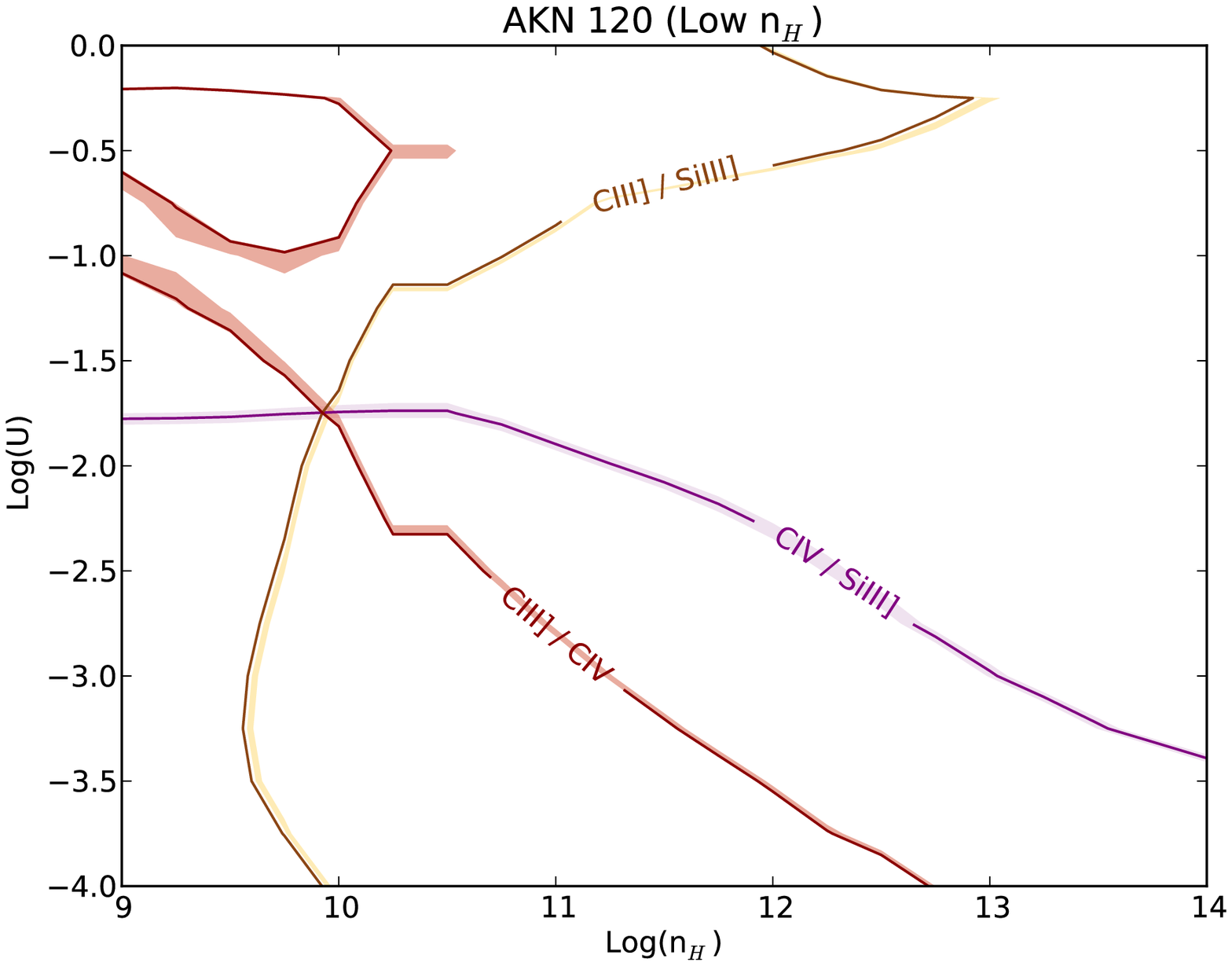}\\
\includegraphics[scale=0.35]{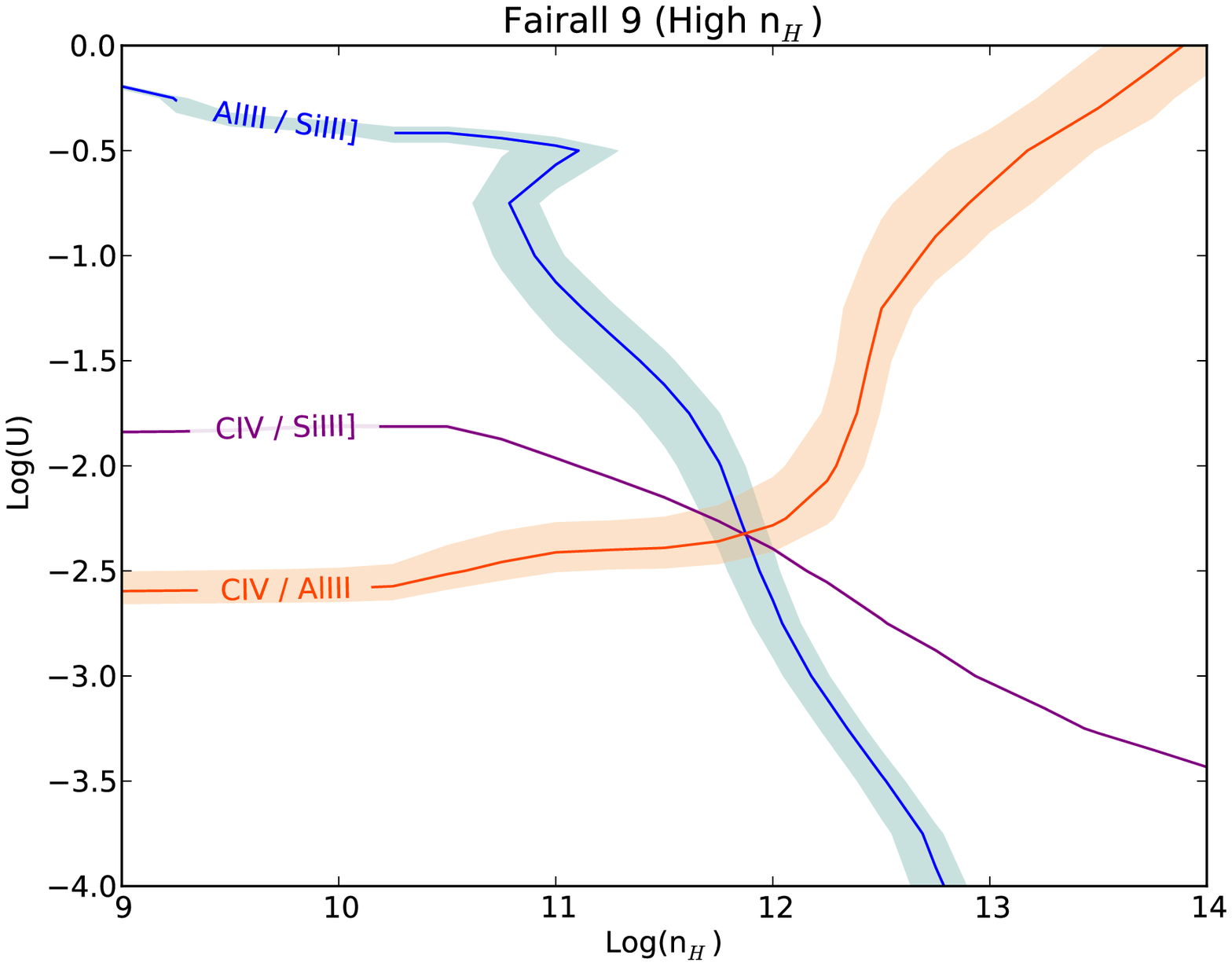}\includegraphics[scale=0.35]{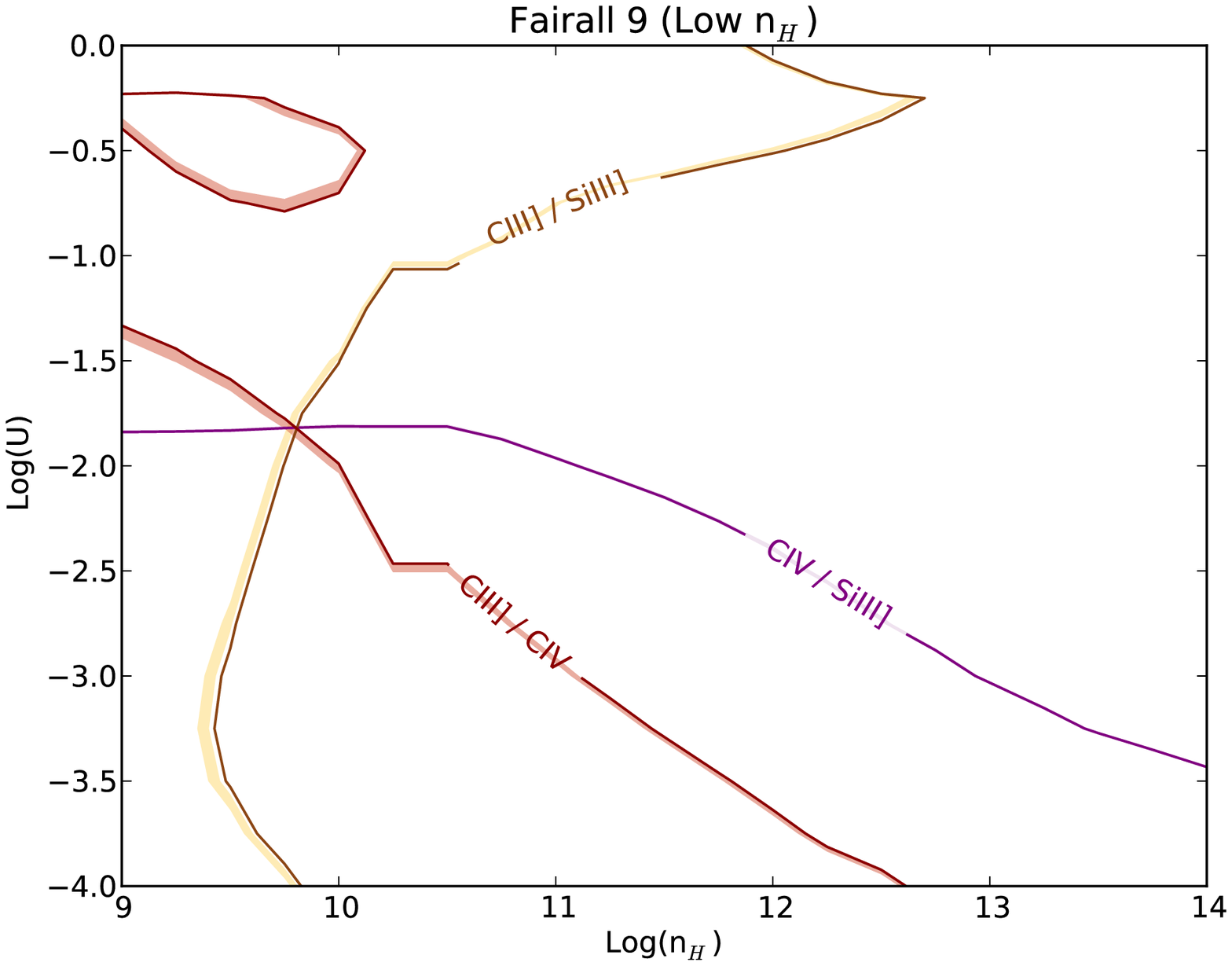}
\includegraphics[scale=0.35]{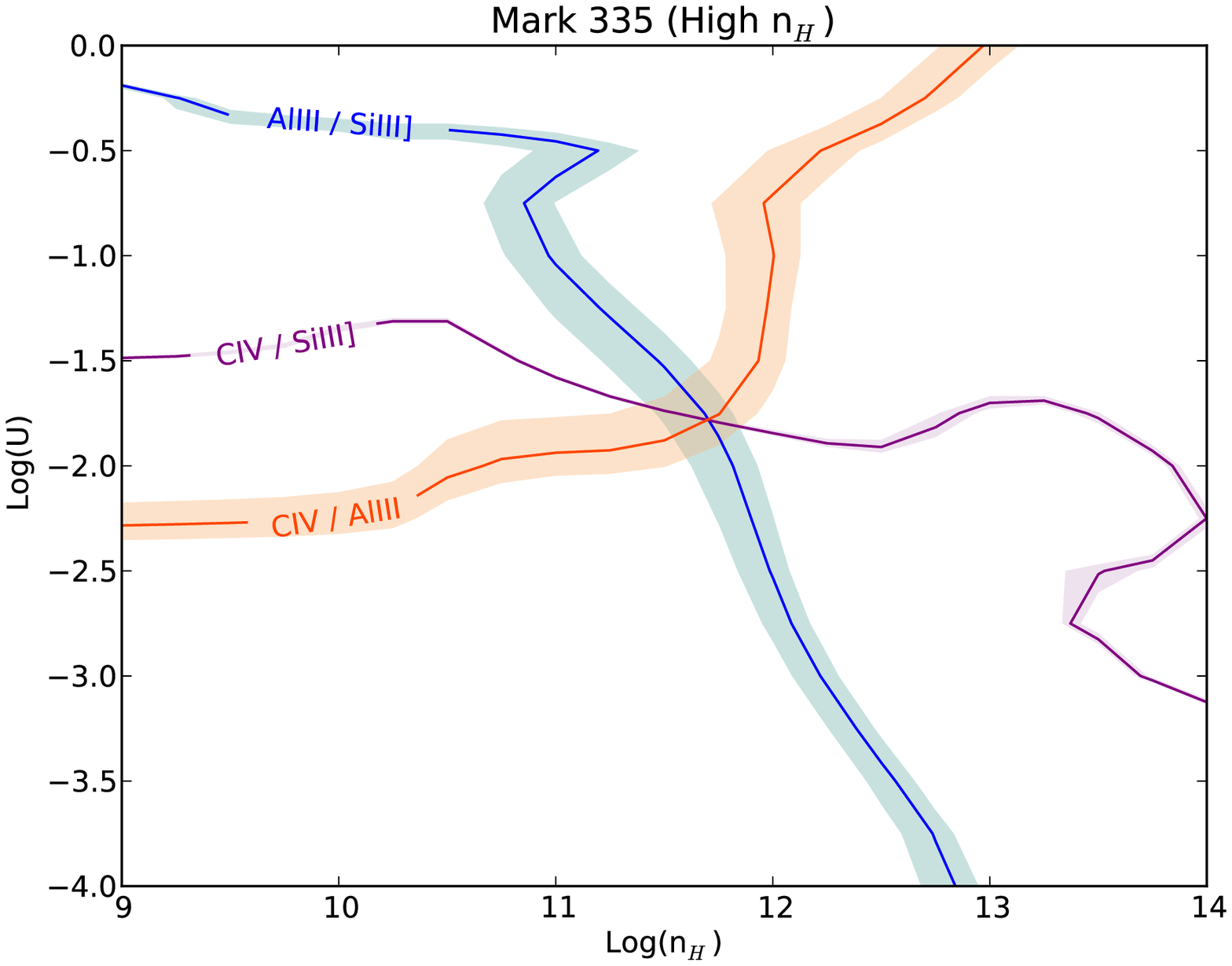}\includegraphics[scale=0.35]{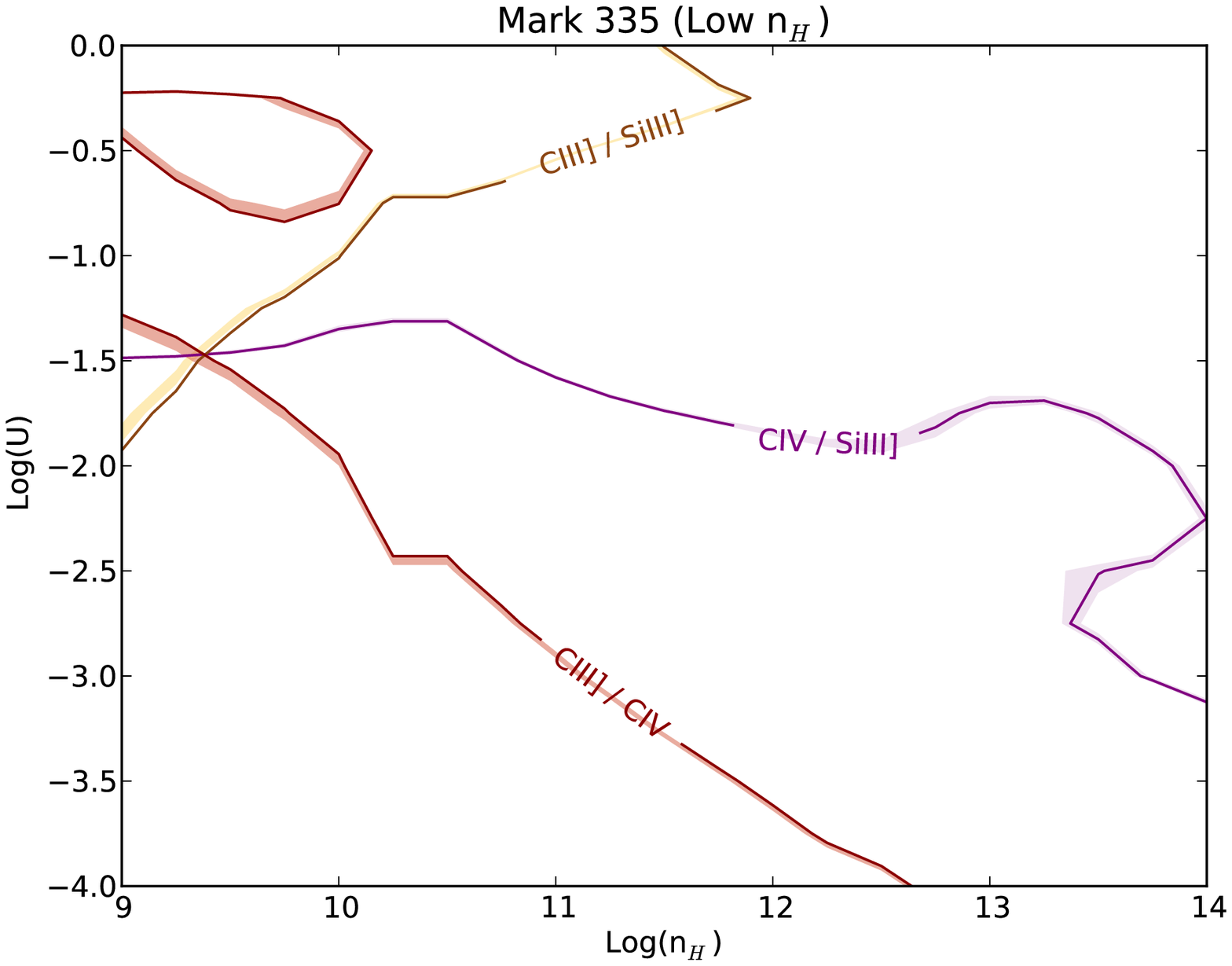}\\
\caption{Identification of the ``solutions'' in the plane (\nh, $U$).   Abscissa is \nh\ in cm$^{-3}$, ordinate is the ionization parameter, both in logarithm scale. Left panels are the high density solution. Right panels are the low density solution. The point where the isocontours cross determine the values of $\log$ \nh\ + $\log U$. The width of the isocontours are 1$\sigma$ confidence.
\label{fig:fit_neu}}
\end{figure}

Armed with line intensities of the BC components we can compute line ratios  \aliii/\siiii, \civ/\aliii\ (or \siiii) and \ciii/\siiii. These values  allow  us to draw isopleths i. e., curves representing measured values of these ratios in the \nh\ and $U$ plane.  The left panels of Figure \ref{fig:fit_neu}  show the isocontour maps based on line ratios for \siiii, \aliii\ and \civ\ emission lines. The right panels show isocontours maps based on line ratios for \ciii, \siiii\ and \civ.  The crossing  points give us best estimates of \nh\ and $U$\ in the BC of each source. Table \ref{tab:derived_nu} lists the estimated product \nhu\ for our sources.  Col. (2) is  the product derived from the reverberation mapped data (obtained extracting the product \nhu\ from Eq. \ref{eq:r} with the assumption \R = $c \tau$), Col. (3) is our estimation derived from the high density solution (left panels of Fig. \ref{fig:fit_neu}), while Col. (4) is our estimation derived from the low density solution (right panels of Fig. \ref{fig:fit_neu}). Col. (5) lists \nhu\  corrected because of systematic effects, as described in  \S \ref{subsec:comparison}.

\ciii\ is a semiforbidden line with critical electron density \n\ $\sim 10^{10}$ cm$^{-3}$ \citep{osterbrockferland06}. This density is usually lower than the value found from the diagnostic ratios based on BC components of the UV  lines. This means that either \ciii\ is weak or that it  is not produced in the same region. If \ciii\ is very weak there would be no ambiguity since there is only one solution that is possible: a low $U$, high \nh\ at the crossing points   \aliii/\siiii, \civ/\aliii. \citet{negreteetal12} have shown that in this case other line ratios (\siii/\siiii, \siiv/\civ, \siiii/\siiv) support this solution. This emitting region is expected to produce little or no \ciii. \ciii\ traces emission from lower density gas that can also emit \siiii, \civ, and other lines. All contribute to the BC line profile since the emitting region is unresolved. Even if \ciii\ $\approx$ \siiii\ it is possible to predict a correction to the line fluxes and compute ratios that are meant to be free of the low-density gas emission \citep{negrete11,marzianietal11}. If \ciii $\gtsim$ \siiii\ this approach is not possible. Fig. \ref{fig:fits} shows that \ciii\ is prominent in all of our sources.  From the crossing points between \aliii/\siiii\ and \civ/\siiii\ on the one hand, and between \civ/\siiii\ and \ciii/\siiii\ on the other we derive two mutually exclusive solutions. 

In Figure \ref{fig:fit_neu} we show isoplet diagrams in the $\log U$ vs. $\log$ \nh\ plane. In the right panels of Figure \ref{fig:fit_neu} we show the solution for a low density emitting region, involving the \ciii/\siiii\ ratio. In the left panels we present the high density emitting region solution that includes the \aliii/\siiii\ ratio. Table \ref{tab:derived_nu} compares  values obtained  from the \aliii/\siiii\ and \ciii/\siiii\ solution. The \aliii/\siiii\ solution closely corresponds to the one derived from RM extracting the product \nhu\ from Eq. \ref{eq:r}. The average $\log$ \nhu\ is different by only 0.1 dex in the two cases.    The \aliii/\siiii\ values tightly cluster around the average  (9.86) with a dispersion of just 0.23 dex, not much larger than the typical uncertainty in individual $\log$ \nhu\ measures.  The distribution of  $\log$ \nhu\  from the \ciii/\siiii\ is significantly offset, with average $\approx 8.0$.

We can therefore draw two conclusions: 1) the \ciii/\siiii\ ratio is not representative of gas responding to the continuum changes. As mentioned, the \ciii\ emitting gas must be  in a lower density region whose extent and location is, at present, a matter of guesswork (conceivable scenario may involve low density tails trailing dense clouds, \citealt{maiolinoetal10}, although a larger distance of the \ciii\ emitting gas seems more likely, as inferred below); 2) the high density solution derived from \aliii/\siiii\ is more representative of the reverberating gas. The uncorrected \aliii/\siiii\ solution offers a reasonable estimate of the \nhu\ value derived from RM. The \nhu(RM) and \aliii/\siiii\  \nhu\ solutions are uncorrelated.  The dispersion for \nhu(RM) values is significantly larger indicating however that there could be a dependence between \nhu\ estimates and additional parameters.

\subsection{Comparison Between \R\ Determinations}
\label{subsec:comparison}

As mentioned, the RM sample has the advantage that \R\ is independently known from reverberation. The \nhu\ and \R\ values are related rewriting Eq. \ref{eq:r} as   Eq. 8 of \citetalias{negreteetal12}:

\begin{equation}
r_{\rm BLR} \approx 93 \left[ \frac{f_{\lambda_0,-15} \tilde{Q}_\mathrm{H,0.01}}{(n_{\mathrm H} U)_{10}} \right]^\frac{1}{2} 
\zeta(z, 0.3, 0.7)  ~~\mathrm{lt- \, days.} 
\label{eq:rblr1}
\end{equation}

where $\lambda_0 = 1700$\AA, $ f_{\lambda_0,-15}$\ is the specific rest frame flux (measured on the spectra) in units of 10$^{-15}$ erg s$^{1}$ \cmq\ \AA$^{-1}$, the product \nhu\ is normalized to 10$^{10}$ cm$^{-3}$  and \R\ is now expressed in units of light days. Note that  $\int_{0}^{\lambda_{Ly}} f_\lambda \lambda d\lambda =  f_{\lambda0}  \cdot \tilde{Q}_\mathrm{H}$ with  $\tilde{Q}_\mathrm{H} = \int_{0}^{\lambda_\mathrm{Ly}} \tilde{s}_\lambda \lambda d\lambda$, where $\tilde{Q}_{H}$ depends  only on the shape of the ionizing  continuum for a given specific flux, and the integral is carried out from the Lyman limit to the shortest wavelengths.   $\zeta$\ is an interpolation function for radial comoving distance as a function of redshift \citep[given by][]{sulenticetal06} with $\Omega_M = 0.3$ and $\Omega_\Lambda = 0.7$. We use the Spectral Energy Distributions (SEDs) $\tilde{s}_\lambda$   by \citet{mathewsferland87} and  by \citet{laoretal97b}  that have been conveniently parameterized as a set of broken power-laws. $\tilde{Q}_{H}$\ is $\approx$ 0.00963 cm\AA\ and $\approx$0.02181 in the case  of the \citet{laoretal97b} and \citet{mathewsferland87} continuum respectively. We use an average $\tilde{s}_\lambda$\ value, since the derived \nhu\ through the photoionization maps is not sensitive to the two different shapes to a first approximation.\footnote{Since the Laor et al. (1997) continuum produces  fewer ionizing photons, the same value of $U$\ is obtained at a smaller distance. However, the  \nhu\  values are, to a first approximation, independent on the frequency distribution of the ionizing photons in the two SEDs considered. }  

Using Eq. \ref{eq:rblr1} to compute \R\ for the low- and high-ionization solution,  the differences in \R\  between the two cases confirm that the BLR is stratified, with \ciii\ likely emitted at a much larger distance (as  indicated in Cols. 7 and 8 of Table \ref{tab:derived_nu} and by reverberation studies; see also the recent analysis of NGC 5548 by \citealt{kollatschnyzetzl13}). The low-density BLR zone is therefore not responding to continuum changes on the same time-scale of \hb.  Hence, the \R\ derived for the high density region is the one that we shall use for any further comparison  with the \R\ derived from reverberation of the \hb\ emission line. The \R\ derived from the high-density photoionization solution based on the \aliii/\ciii\ ratio will be denoted as \R $_{,\Phi}$ in what follows.

\begin{figure}
\includegraphics[scale=0.4]{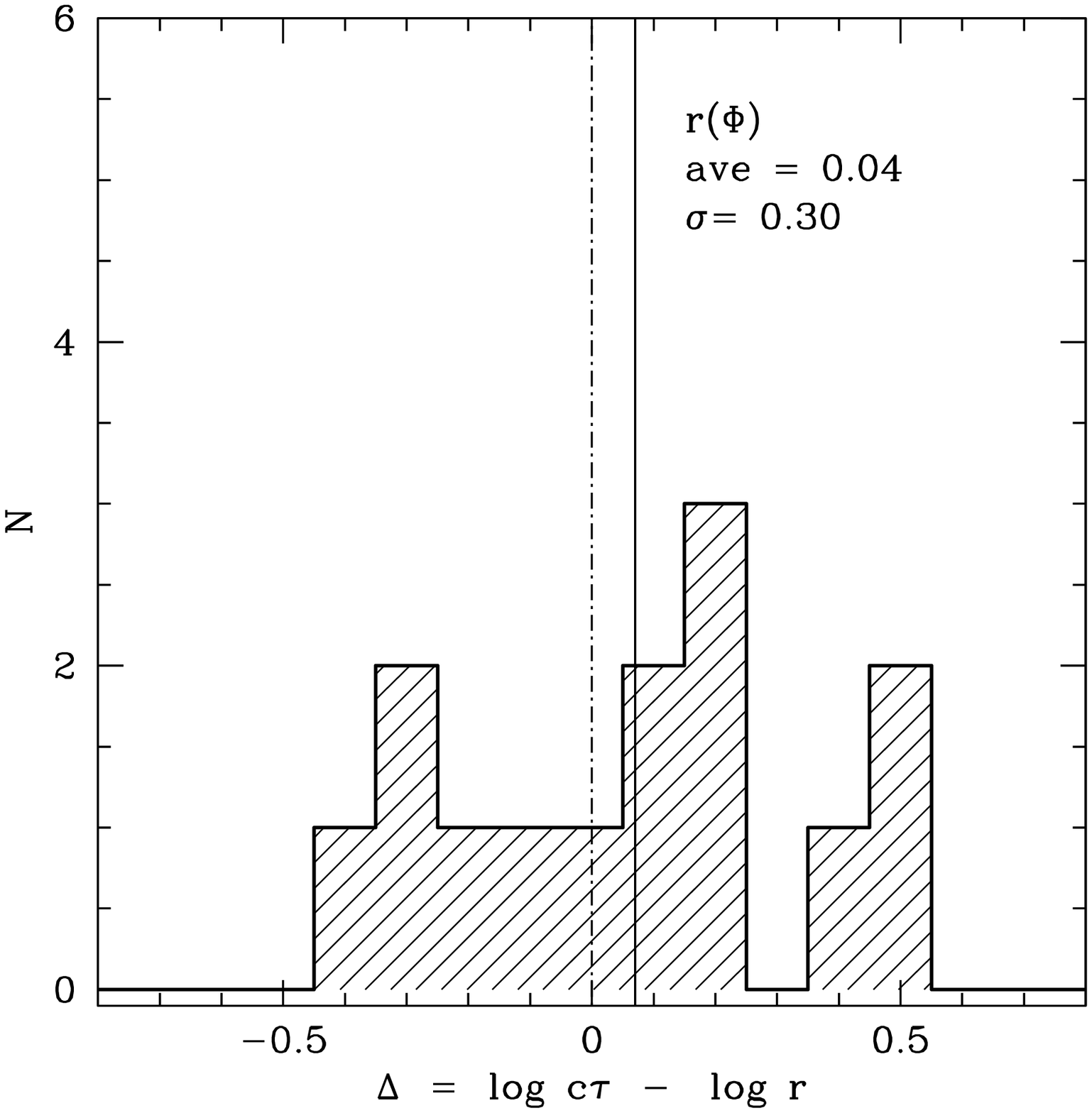}
\includegraphics[scale=0.4]{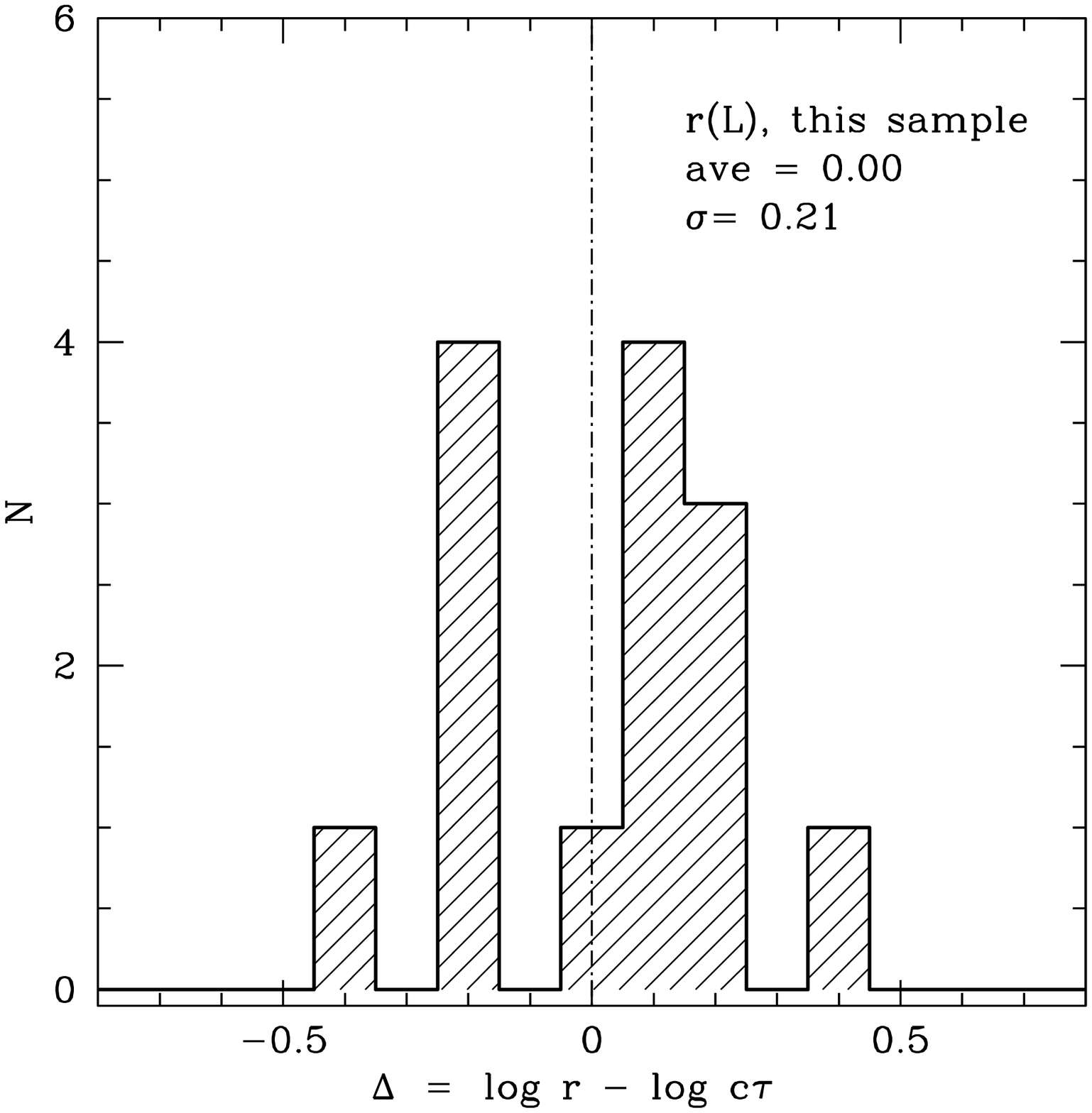}\\
\includegraphics[scale=0.4]{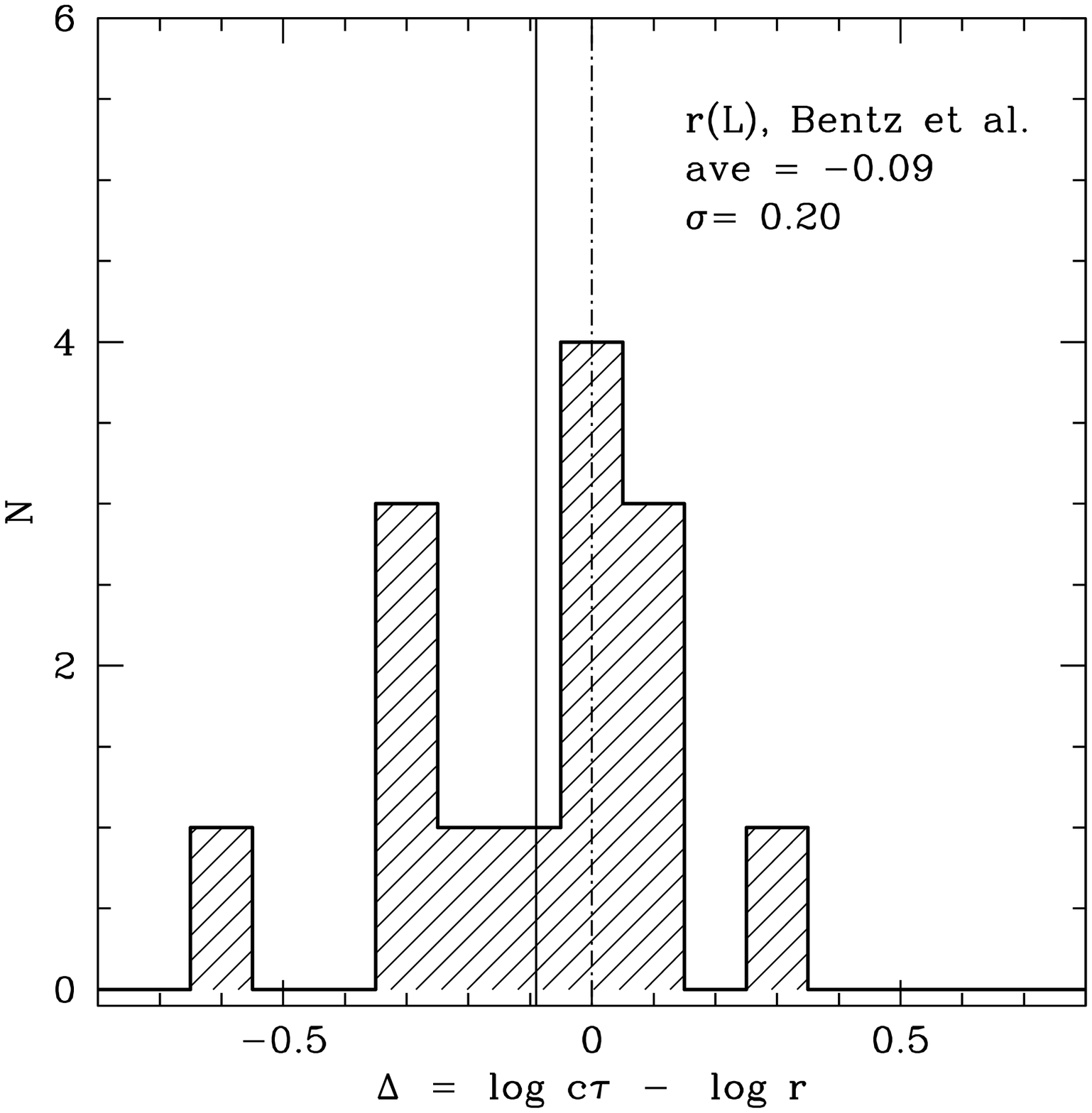}
\includegraphics[scale=0.4]{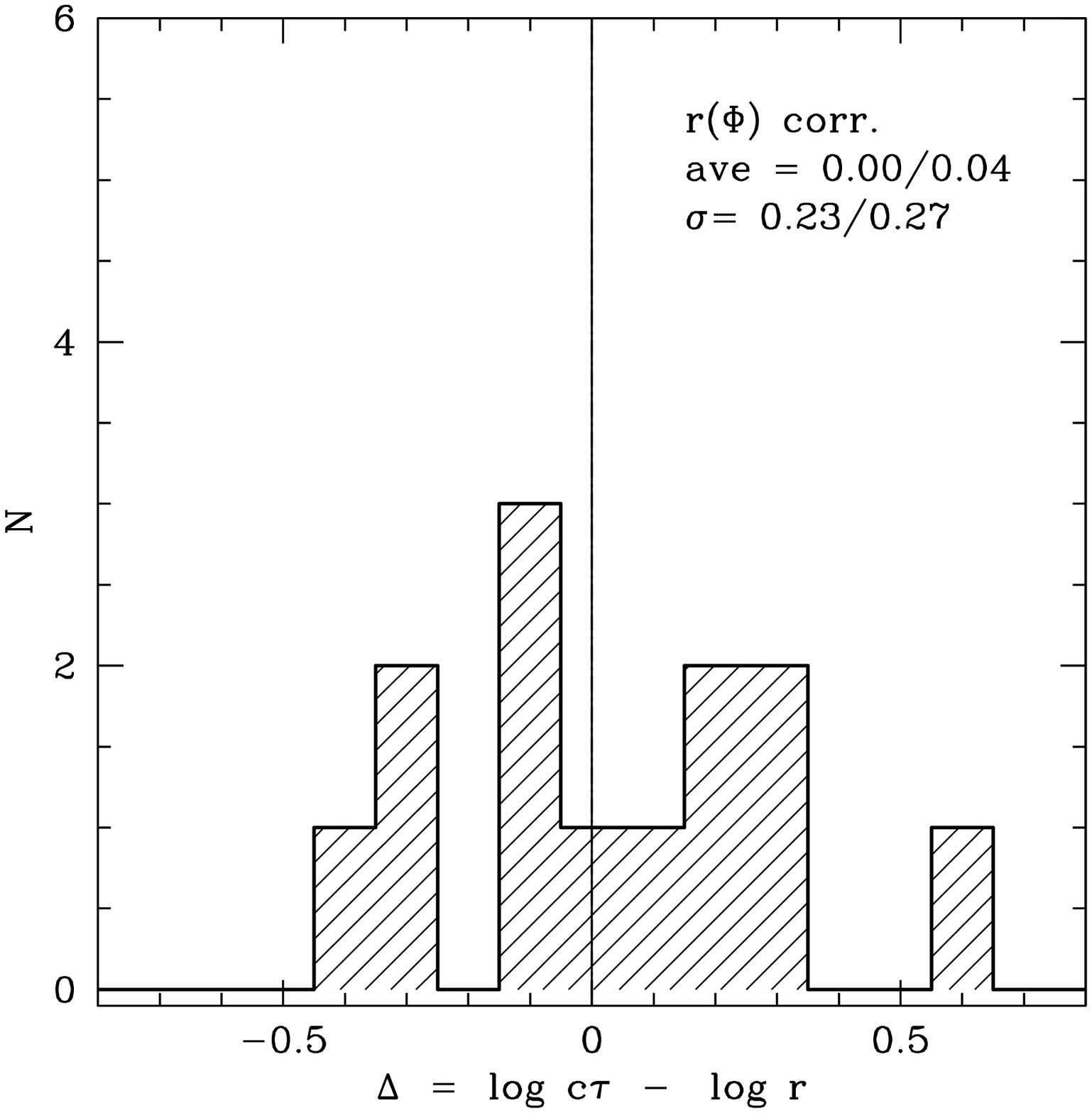}
\caption{Residuals $\Delta$\ between the reverberation based distance and \R\ with 4 different method. 
(upper left) \R\ with the photoionization method; 
(upper right) \R\ with the luminosity correlation defined on the present sample; 
(lower left) \R\ with the luminosity correlation of \citet{bentzetal09}; 
(lower right) \R\ with the photoionization method after correcting for a systematic effects dependent on the ratio W(\aliii)/W(\ciii). The two values in the lower right panel refer to the average and rms excluding/including PG 0953+414.   
\label{fig:histos}}
\end{figure}

In Fig \ref{fig:histos} we display the residuals \D$\log r_\mathrm{BLR} =  \log  r_\mathrm{BLR}  - \log c\tau$\ between the distance computed with four different methods and the reverberation based distance. In Fig. \ref{fig:histos} upper left we show the distribution of the \D$\log  r_\mathrm{BLR}$ difference between the photoionization and the reverberation distance, as reported in Cols. (6) and (7) of Table \ref{tab:derived_nu}. From this figure, we see that the agreement between $\log r_\mathrm{BLR,\Phi} $ and $\log c\tau$ is good with an average of $\overline{\Delta \log r_\mathrm{BLR}} \approx 0.07 \pm 0.29$ dex with a significant scatter.  In only two cases (Fairall 9 and NGC 7469) a $t$-test indicates a significant difference between the two methods. 

 We use the values of \l L$_\lambda$ (5100\AA) and the \hb\ time lags given by \citet{bentzetal09} for the 13 objects of this paper  which are included in their sample. We obtain:

\begin{equation}
\log r_\mathrm{BLR}(L) \approx -  (9.91 \pm 0.34) + (0.61 \pm 0.02)  \log \lambda L_{\lambda}(5100\mathrm{\AA})  
\label{eq:rLus}
\end{equation}

with residuals $-0.01 \pm$ 0.20 dex { (Fig. \ref{fig:histos} upper right)}. 

For the sample with RM data of $\approx$50 objects,  \citet{bentzetal09}   derive the equation
\begin{equation}
\label{eq:R_L}
\log R_\mathrm{BLR}(L) = -21.3 + 0.519 \, \log \lambda L_\lambda  (5100\mathrm{\AA})
\end{equation}

If we apply this correlation to our sample (Fig. \ref{fig:histos} lower left) we obtain a residual rms  $ \pm$ 0.20  dex but with a significant bias ($-0.09$ dex)  that makes the uncertainty again  $\approx$ 0.3 dex, as also indicated by the full sample of \citet{bentzetal09}.  Therefore \R$_{,\Phi}$ estimates ÊshowÊprecision and accuracy that are similar to Êthe ones based on the luminosity correlation of \citet{bentzetal09}. In our case the luminosity correlation of \citet{bentzetal09} systematically over-predicts \R\ by $\Delta \log r_\mathrm{BLR} \approx$ 0.1 dex, as shown also in Fig. \ref{fig:histos} lower left. 

The formal median uncertainty at $1 \sigma $\ confidence level is $\pm 0.1$ dex  in $R_\mathrm{BLR}$ for RM (Fig. \ref{fig:histos} lower left), and comparable to the photoionization method (Fig. \ref{fig:histos} upper right). If the RM values are assumed to be the ``true'' \R\ values, the photoionization method (in its simplest formulation)   yields information that is not accurate in 2 of 13 cases, as mentioned above.  Its precision  is comparable to  the luminosity correlation.

\subsection{Analysis of Systematic Differences}
\label{corr}

Emission line ratios used for the computation of the high density solution are most likely affected by lower density \ciii\ emission. Therefore, we expect that any disagreement between \nhu\ might be influenced by the equivalent widths and the relative strength of the emission lines considered, and especially by the prominence of \ciii.

\begin{figure}
\includegraphics[scale=0.55]{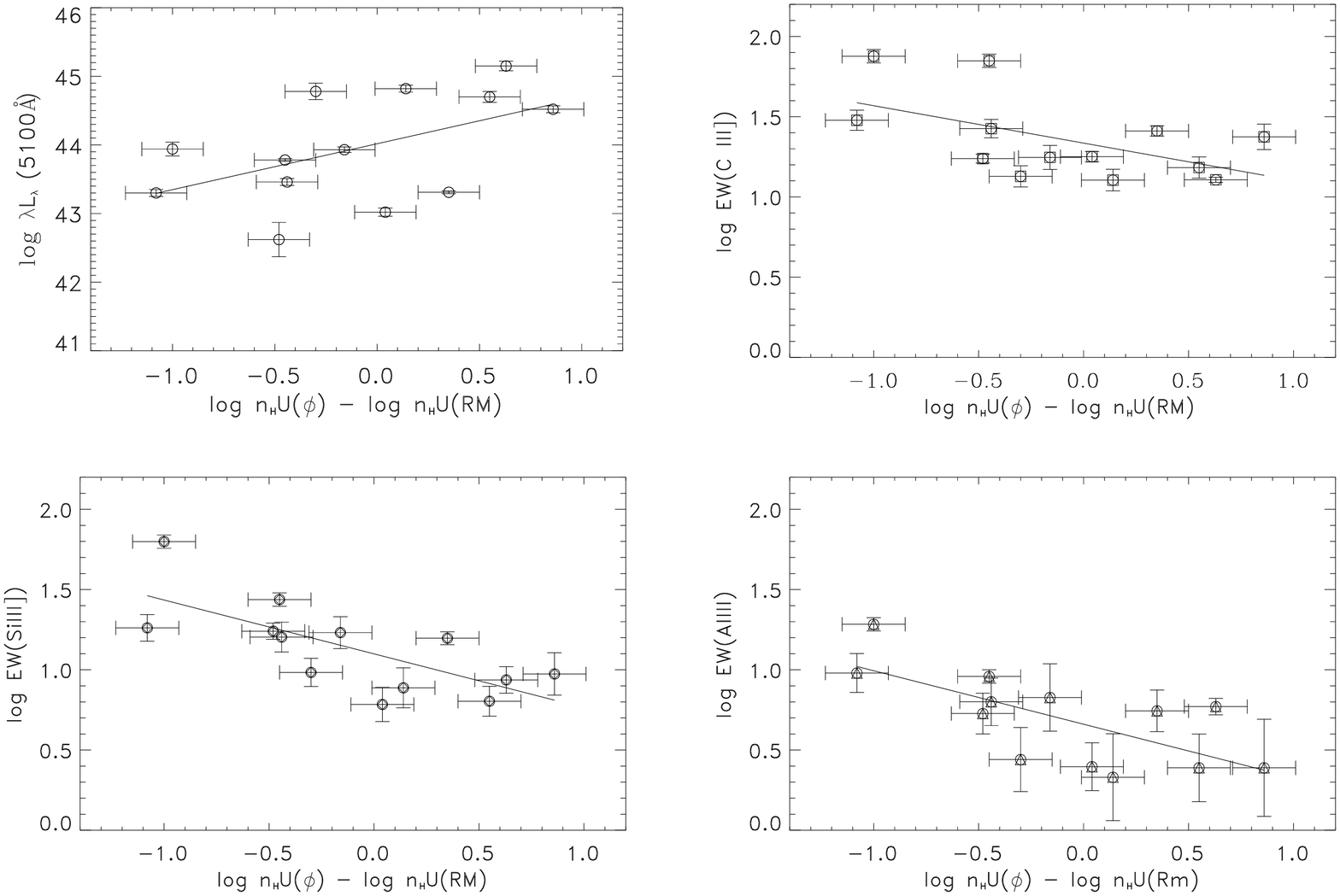}
\caption{Difference $\Delta n_H U$\ as a function of luminosity, $W$(\ciii), $W$(\aliii) and $W$(\siiii). The filled line  shows a least square fit.  
 \label{fig:delta}}
\end{figure}

 Three correlations emerge from the consideration of the residuals $\Delta \log n_\mathrm{H}U = \log n_\mathrm{H}U(\Phi) - \log n_\mathrm{H}U({\mathrm{RM}})$, where we have again conventionally indicated with \nhu $(\Phi)$\   the solution based on the \aliii/\siiii\  ratio. The \nhu(RM)  is derived from Eq. \ref{eq:rblr1} using c$\tau$  from reverberation. Figure \ref{fig:delta} presents  comparisons of $\Delta \log n_\mathrm{H}U$ as a function of luminosity, and the equivalent widths of \ciii, \siiii\ and \aliii. Errors on line intensity (computed considering continuum level uncertainty) are quadratically propagated to compute errors for  the diagnostic ratios and hence for the product \nhu.  Best fit lsq solutions are as follows:

\begin{equation}
\Delta \log n_\mathrm{H}U \approx (1.49 \pm  0.05)  \log \lambda L_\lambda(5100\mathrm{\AA})  -  (65.76 \pm  2.39).
\end{equation}

\begin{equation}
\Delta \log n_\mathrm{H}U \approx (-4.29 \pm 0.45)  \log W({\mathrm{C~III]}}\lambda1909)  +  (5.73 \pm 0.43)
\end{equation}

\begin{equation}
 \Delta \log n_\mathrm{H}U \approx (-2.98 \pm 0.27)  \log W({\mathrm{Si~III]}}\lambda1892)  +  (3.28 \pm 0.43)
\end{equation}

\begin{equation}
\Delta \log n_\mathrm{H}U \approx (-3.00 \pm 0.66)  \log W({\mathrm{Al~III}}\lambda1860)  +  (1.98 \pm 0.87)
\end{equation}

The equations of this section yielding a correction for $\log$ \nhu\ as a function of equivalent widths can be used to  recompute \R. This operation would eliminate any bias between \nhu($\Phi$) and \nhu(RM) estimates but the scatter in the \R\ residuals computed after applying a correction would remain large, $\approx $ 0.3 dex. A similar scatter is obtained if a correction is defined directly correlating $\Delta \log$\R\  against the equivalent widths of \ciii, \siiii\ and \aliii.  It is perhaps not surprising that the scatter is not reduced since equivalent widths are expected to be influenced by several factors affecting the gas physical conditions   (i.e., continuum luminosity, covering factor, etc.).  We can improve the correction if we use the ratios of line equivalent widths or fluxes, as shown below.  

\subsubsection{Improving the Agreement}

\begin{figure}
\includegraphics[scale=0.5]{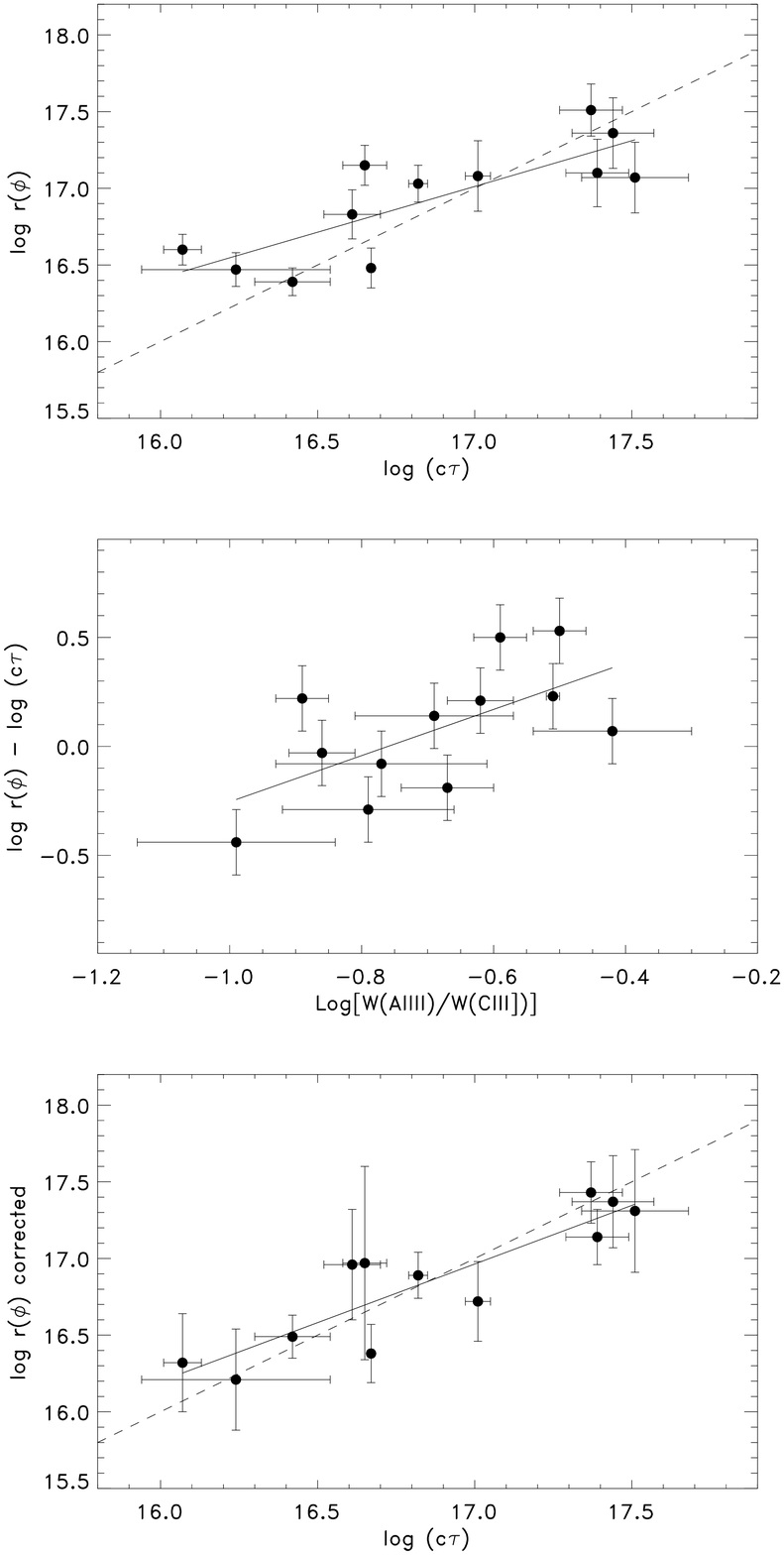}
\caption{\R\ comparison. Upper panel: $\log$\R$(\phi)$ using the method of the present paper {\it vs.} the product $\log (c\tau)$ reported by \citet{bentzetal09}. Middle panel: difference of distances $\Delta$\R\  {\it vs.} $\log$ W(\aliii)/W(\ciii). Lower panel: $\log$ \R$(\phi)$ corrected by the W(\aliii)/W(\ciii) ratio {\it vs.} $\log (c\tau)$. The dashed line shows equality; the filled line  shows the result of a least square fit.
\label{fig:comparison_al3_c3}}
\end{figure}

The relation between  $\log c\tau$  {\it vs.} $\log r_\mathrm{BLR,\Phi}$ (Figure \ref{fig:comparison_al3_c3} upper panel) is given by 

\begin{equation}
\log \mathrm{c \tau} \approx (1.16 \pm 0.07)   \log r_\mathrm{BLR,\Phi}  -  (2.74 \pm 0.40)
\label{eq:rRMF}
\end{equation}

The Pearson correlation coefficient is $R \approx$0.82, implying probability $P\approx 0.003$\ for the correlation to occur by chance. This equation yields an rmsÊ$\approx$Ê0.30 dex, a value similar to the scatter inÊrBLRÊvalues obtained through the correlation with luminosity.Ê

In order to obtain an even better correlation, we apply a correction using the equivalent width ratio  W(\aliii)/W(\ciii) that is more effective  (Fig. \ref{fig:comparison_al3_c3} middle panel):
\begin{equation}
\log r_\mathrm{BLR,\Phi} - \log  \mathrm{c \tau}  \approx (1.06 \pm 0.25)  \log \frac{\rm{W(Al\, III}\lambda 1860 )}{\rm{W(C \, III]}\lambda 1909)} +  (0.81 \pm 0.26)
\label{eq:correct}
\end{equation}
 
The correlation coefficient is 0.62 which for 12 objects implies a marginal significance slightly over a $2\sigma$\ confidence level.   Fig. \ref{fig:histos}(lower right) shows a somewhat more compact distribution in the $\Delta \log$\R\ values.  There is one outlying source that has been excluded from the analysis, PG 0953+414. This source shows considerable narrow absorption that appear to significantly eat away part of the \ciii\ line, making an estimate of the W(\aliii)/ W(\ciii) ratio rather difficult.  

The correlation of \R$_\phi$ corrected and $c\tau$ is (Fig. \ref{fig:comparison_al3_c3} lower panel):

\begin{equation}
\log r_\mathrm{BLR,\Phi corr}  \approx  (0.77 \pm 0.14)  \log \mathrm{c \tau}\  +  (3.94 \pm 0.75)
\end{equation}

with a scatter of $\approx$ 0.23 dex and a correlation coefficient of 0.89. The absence of a significant bias in the corrected \R$_{,\Phi}$ comes from the definition of the correlation on the present sample. However, since \R$_{,\Phi}$ is computed on an object by object basis it is reasonable to assume that no bias will be introduced in different samples. ÊThe photoionization method is therefore expected to be also somewhat more accurate than the luminosity correlation.

\section{Discussion} 
\label{sec:discussion}

The preceding sections showed that the photoionization method used in this paper is yielding physically meaningful values, that are consistent with Ê\R\ derived from reverberation mapping. Values of \R\ derived from photoionization arguments are known to be consistent with $c \tau$. ÊA similar method assuming a constant \nhu\ also provides consistent agreement (\citealt{padovanietal90,padovanirafanelli88,wandeletal99}; Êsee also Chapter 4 of \citealt{donofrioetal12}). A reassessment of the method of \citet{dibai77} based on the luminosity of Balmer line Êshows Êagreement with reverberation derived masses within $\pm$0.3 dex \citep{bochkarevgaskell09}. Ê

The agreement between photoionization results and the correlation with luminosity is expected since the diagnostic ratios measure the product \nh $U$\ that is the ionizing photo flux. If the correction provided by Eq. \ref{eq:correct} is valid in general, then the photoionization method can provide a significant improvement in precision for single epoch \R\ estimates, lowering the dispersion  around  RM- derived \R\ from more than a factor 2 (if the luminosity correlation is used) to $\approx$70\%. 
 
\subsection{Influence of Continuum}

We adopted a very simplified approach, that neglects (1) the diversity in the ionizing continua among sources, and  (2) the dependence of \nhu\ on ionizing continuum shape. It is unlikely that, with the chosen simplified approach, a better agreement between photoionization and RM \R\ estimates can be achieved. First, \R\ from RM is subject to a significant uncertainty, and shapes of the cross-correlation function are not always regular. Second, a significant part of the scatter is associated to the assumption of an average SED. Using the SEDs of each individual source, and repeating the photoionization simulation array that defines \nhu\ should lead to a significant improvement. The simplified  approach is meant to make the method easily applicable to high-$z$\ quasars for which SEDs data are  most often unavailable at present. 

\subsection{Interpretation of the Empirical Correction}

Eq. \ref{eq:correct} needs to be confirmed by more extended data. It is based on an heterogeneous sample of 12 objects only. Physical properties within the BC are not found to be identical across the E1 sequence. In principle, Eq. \ref{eq:correct} should be built separating the most populated spectral types along E1. This feat is however beyond the possibilities offered  by available data. Given the unclear role of continuum diversity, it is not easy to derive a unique physical interpretation beyond the following qualitative considerations. Eq. \ref{eq:correct} indicate that \R($\Phi$) for smaller \aliii/\ciii\ ratio sources significantly under-predicts $c\tau$, while the agreement is better for relatively large \aliii/\ciii\ ratios. This is consistent with the results of \citetalias{negreteetal12}. Stronger \ciii\ emitters (for example PG 1211+143 and PG 1411+442 of spectral type A1 of the E1) may  appreciably respond  to continuum changes at a systematically larger distance with respect to the denser, low ionization gas . The denser low ionization gas may account for a small fraction of the emitting gas if \ciii/\aliii$\gg$1. 

\subsection{Influence of Continuum Variability}

The previous results rely on the assumption that we can take an average AGN ionizing continuum. In fact, however, we know that the continuum is variable and in some cases very variable. When the ionizing continuum varies Eq. \ref{eq:rblr1} predicts a  variation in \R. The physical reasons for the variation in \R\ may be twofold: a) the ionizing continuum may penetrate farther among the BLR clouds and/or b) the effect of radiation pressure can push the clouds further away. Regardless of the ultimate physical interpretation, we can confront \R\ derived from reverberation mapping and from photoionization in different continuum states. We considered the case of NGC5548 which is a very well monitored object. The lowest and highest value for the flux at \l1700\AA\ were retrieved from the AGN watch website\footnote{http://www.astronomy.ohio-state.edu/\~agnwatch/n5548/spectra}. We then calculated \nh\ and $U$\ for the two states. For the lowest value ($0.94\, 10^{-14} \mathrm {erg\,  s}^{-1}$ cm$^{-2}$), we obtain  \nhu\ = 9.85, and for the highest value ($4.78 \, 10^{-14}$ erg s$^{-1}$ cm$^{-2}$) we obtain \nhu\ = 10.04. These products yield a variation from $\log$(\R) = 16.26 to $\log$(\R) = 16.52.  The isoplets indicate that the change in \nhu\ is driven by a change in $U$\ that is  affecting strongly \civ, remaining the \nh\ value almost constant. The \R\ change goes in the same sense of the ones derived from the AGN watch and reported by \citet{bentzetal09} although $c \tau$\ seems to be affected more strongly by continuum changes, with a 3-fold increase in \R\ for a 3-fold increase in continuum.

\subsection{\mbh\ Computation}

Knowing \R\  enables us to estimate the black hole  mass (\M) assuming virial motions of the gas using:

\begin{equation}
\label{eq:vir}
M_\mathrm{BH} = f \frac{\Delta v^2 r_\mathrm{BLR}}G = \frac 3{ 4G} f_{0.75} \mathrm{FWHM}^2 r_{\mathrm{BLR}} 
\end{equation}

were  $G$\ is the gravitational constant. If $\Delta v = FWHM$ of a line, the geometry factor $f = \sqrt{3}/2$ if the orbits of the BLR clouds are randomly oriented. We use $f_{0.75} = 1.4$ \citep{grahametal11}.

As mentioned earlier, the BC of \civ, \siiii, and \aliii\  isolates emission that is believed to come from the same low-ionization region emitting the core of \hb, and LILs like \mgii, \feii, \siii.  It is believed that the BC broadening is due to Keplerian motions since the BC does not present strong asymmetries or centroid shifts with respect to rest frame. Therefore, use of \aliii\ or \siiii\  BC FWHM derived from the multicomponent fits should be regarded as ``safe'' as the use FWHM(H$\beta_{\mathrm{BC}}$)  for obtaining a BLR velocity dispersion indicator. This is not true for the blue-shifted component and for the VBC.  The \civ\ blue-shifted asymmetry found in many quasars is read as the signature of an outflowing wind \citep[][and references therein]{marzianisulentic12}.  The large  shift of the VBC similarly suggests that non-virial motions play a significant role.   The low-ionization part of the BLR  that should emit the BC we isolated  is still prominent in high luminosity quasars   \citep{marzianietal09}, and this makes the photoionization method discussed in this paper straightforwardly applicable to high redshift quasars (Negrete et al. in preparation).   The RM sample offered the possibility to check that the BC FWHM of \aliii\ and \siiii\ is indeed consistent with the BC FWHM of \hb\ (last columns of Table \ref{tab:obs}). \aliii, \siiii\ may offer the most consistent FWHM estimators. \civ\ should be avoided unless a detailed analysis as in Figure \ref{fig:fits} can be carried since it is often blueward asymmetric. \ciii\ may be significantly narrower than \siiii\ and \aliii, and, as stressed, is not associated to the high-density solution. Therefore also \ciii\ FWHM should be avoided as a virial broadening estimator. 

\begin{figure}
\includegraphics[scale=0.5]{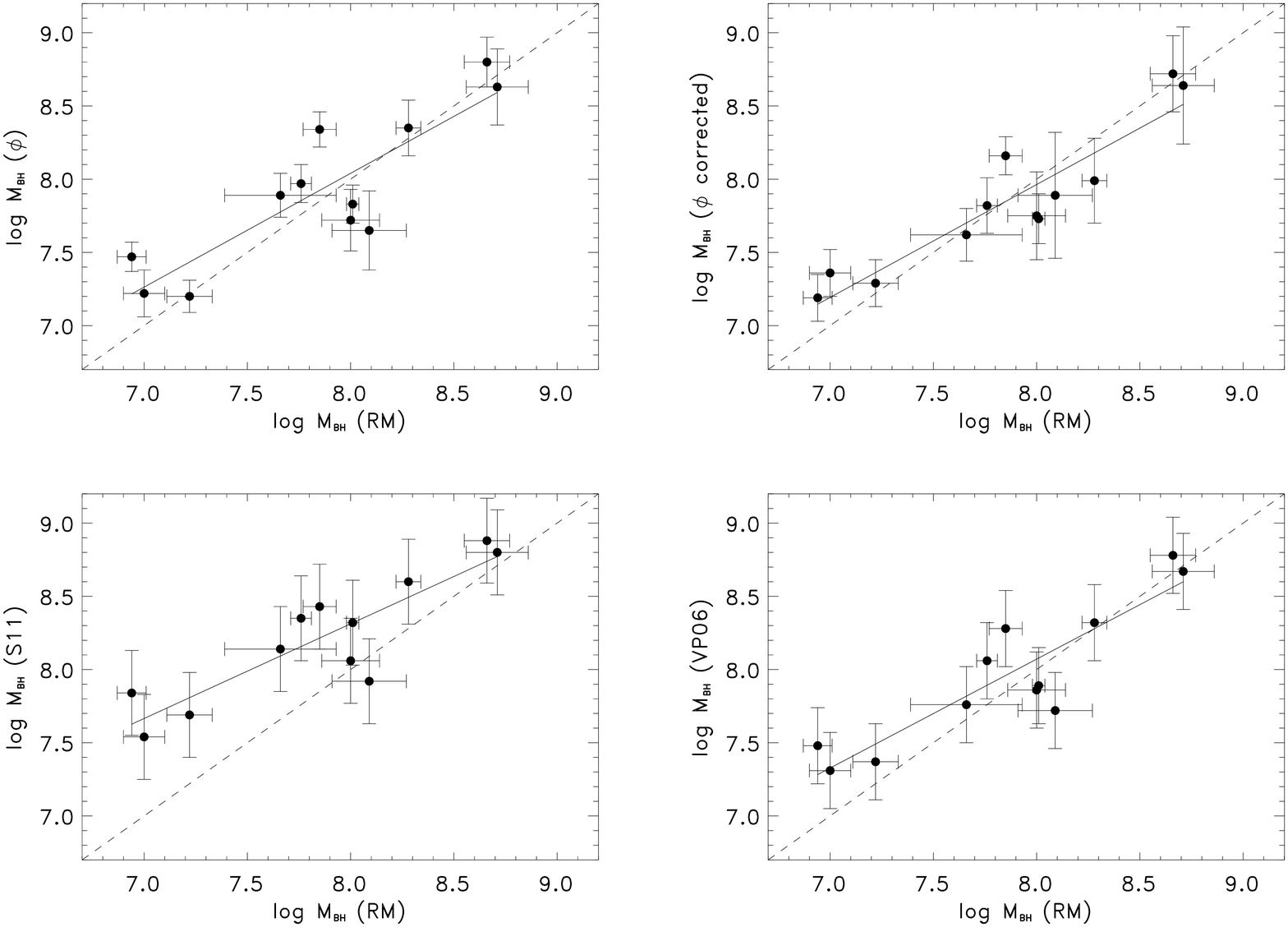}
\caption{Comparison between masses derived using the reverberation mapping distance, and derived from the photoionization method ($\Phi$, upper panels).  Lower panels show two luminosity correlations. We consider  12 sources of this work. Upper right: uncorrected $\Phi$; upper left:    $\Phi$\ corrected for systematic effects on the basis of the W(\aliii)/W(\ciii) ratio; lower left: relation with  \mbh\ computed following \citet{shenetal11}; lower right: elation with  \mbh\ computed following \citet{vestergaardpeterson06}.  The dashed line shows equality; the filled line shows the result of a least square fit.
 \label{fig:mass}}
\end{figure}

Fig. \ref{fig:mass} compares the mass computed from Eq. \ref{eq:vir} using \R\ estimated from RM,  our photoionization method and two luminosity correlations. As seen for \R\ the agreement is improved if systematic effects are corrected with the W(\aliii)/W(\ciii) relation. The scatter and the bias in the \civ\ luminosity derived masses of \citet{shenetal11} is probably related to significant broadening of the CIV line by non-virial motion \citep[e.g., ][]{netzeretal07,sulenticetal07}.  On the contrary, the \citet{vestergaardpeterson06} relationship provides more accurate values since it has been calibrated on a dataset that includes the sources considered in this paper.

\subsection{Further Considerations}

The assumption of RM values as the true values is a working hypothesis. RM based masses may be accurate to within a factor $\approx$3 \citep{vestergaardpeterson06}, if they are compared to the masses derived from the \mbh\ -- bulge velocity dispersion. However, the origin of this dispersion may include statistical (i.e., orientation) and systematic effects (as the geometry factor $f$) that do not enter in the \R\ measures. The determination of \R\ on an individual source basis allows  an immediate comparison with RM values and to independently consider  other systematic and statistical effects involved in the \mbh\ estimate. An average $f$\ value for all AGNs is unlikely to be appropriate (as derived from the scaling with the the \mbh\ -- bulge velocity dispersion) since the line profiles of the strongest emission lines suggest structural and dynamical changes along the so-called ``Eigenvector 1'' sequence. Therefore the photoionization method has the potential advantage (unlike methods based on \mbh\ -- FWHM -- luminosity correlation) to ``reproduce'' RM \R\ values at high redshift, leaving the possibility to consider  $f$\ and orientation effects on an individual basis. 

\section{Conclusion}
 
In summary, we are able to estimate BLR distances using an independent photoionization method that yields results consistent with reverberation values for 13 sources in common \citep{bentzetal09}.  Although we cannot constrain BLR physical conditions as well as we were able to do for extreme Population A sources \citepalias{negreteetal12}, we are nonetheless able to derive empirical relations that further improve  the agreement between photoionization and RM \R\ determinations. We suggest that the derived \R\ values can significantly improve black hole mass estimation especially at $z \gtsim 2$ when the intermediate ionization lines are shifted into the wavelength range accessible to optical spectrometers. The width  of the broad intermediate ionization lines likely provides a reliable virial estimator leaving the geometry factor $f$\  and poorly understood orientation effects as the main sources of uncertainty.

\acknowledgments
A. Negrete and D. Dultzin acknowledge  support form grant IN111610 PAPIIT UNAM, and  Yair Krongold for fruitful discussions. JWS acknowledges support under a Proyecto de Excelencia contract from La Junta de Andaluc'a.

\bibliographystyle{apj}


\begin{deluxetable}{lccccccccccccccc}
\tabletypesize{\scriptsize}
\setlength{\tabcolsep}{3pt}
\tablecaption{Measured quantities \label{tab:obs}}
\tablewidth{0cm}
\tablehead{
\colhead{}  &  \colhead{} &  \colhead{} & \multicolumn{4}{c}{Line Flux\tablenotemark{c}}  &  \colhead{} & \multicolumn{3}{c}{FWHM\tablenotemark{d}}\\
\cline{4-7}
\cline{9-11}
\colhead{Object}  & \colhead{Pop. type\tablenotemark{a}}  & \colhead{$f_{\lambda}$(1700 \AA)\tablenotemark{b}} & 
\colhead{\civ} &  \colhead{\aliii} & \colhead{\siiii} &\colhead{\ciii} & \colhead{ } & 
\colhead{\hb} &  \colhead{UV} & \colhead{$\sigma$} \\
\colhead{(1)} & \colhead{(2)} & \colhead{(3)} & \colhead{(4)} & \colhead{(5)} & \colhead{(6)} & \colhead{(7)}  & \colhead{ } & \colhead{(8)}  & \colhead{(9)} & \colhead{(10)} }
\startdata
AKN120	&	B	&	8.1	$\pm$	0.5	&	82.8	$^{+	4.1	}_{-	1.0	}$ &	4.7	$^{+	2.6	}_{-	1.3	}$ &	13.0	$^{+	1.2	}_{-	1.3	}$ &	13.2	$^{+	0.6	}_{-	1.4	}$ & &	5480	&	4990	&	420	\\
Fairall 9	&	B	&	3.6	$\pm$	0.4	&	25.3	$^{+	2.5	}_{-	2.5	}$ &	1.4	$^{+	0.4	}_{-	0.4	}$ &	4.5	$^{+	0.5	}_{-	0.5	}$ &	5.3	$^{+	0.5	}_{-	0.5	}$ & &	4540	&	4550	&	50	\\
MRK 335	&	A	&	7.1	$\pm$	0.7	&	54.8	$^{+	5.5	}_{-	5.5	}$ &	1.4	$^{+	0.4	}_{-	0.4	}$ &	4.2	$^{+	0.5	}_{-	0.5	}$ &	10.6	$^{+	1.1	}_{-	1.1	}$ & &	1960	&	1870	&	200	\\
MRK 509	&	A	&	8.9	$\pm$	0.6	&	116.5	$^{+	2.5	}_{-	3.5	}$ &	5.1	$^{+	1.6	}_{-	0.9	}$ &	12.5	$^{+	2.0	}_{-	0.5	}$ &	20.6	$^{+	1.4	}_{-	1.6	}$ & &	3390	&	3290	&	300	\\
NGC 3516	&	B	&	4.7	$\pm$	0.2	&	49.1	$^{+	2.4	}_{-	2.6	}$ &	2.3	$^{+	0.6	}_{-	0.4	}$ &	7.5	$^{+	0.5	}_{-	0.3	}$ &	7.4	$^{+	0.1	}_{-	0.4	}$ & &	6530	&	5270	&	700	\\
NGC 3783	&	A	&	10.9	$\pm$	0.1	&	106.9	$^{+	4.1	}_{-	2.9	}$ &	2.5	$^{+	0.9	}_{-	0.3	}$ &	5.9	$^{+	1.3	}_{-	0.1	}$ &	17.3	$^{+	0.5	}_{-	0.3	}$ & &	2870	&	2860	&	100	\\
NGC 5548	&	B	&	3.1	$\pm$	0.1	&	51.1	$^{+	2.4	}_{-	1.4	}$ &	1.6	$^{+	0.5	}_{-	0.3	}$ &	4.3	$^{+	0.2	}_{-	0.2	}$ &	7.0	$^{+	0.2	}_{-	0.2	}$ & &	5820	&	5390	&	330	\\
NGC 7469	&	A	&	4.7	$\pm$	0.4	&	57.0	$^{+	1.0	}_{-	1.5	}$ &	4.2	$^{+	0.8	}_{-	0.6	}$ &	7.8	$^{+	0.7	}_{-	0.4	}$ &	12.7	$^{+	0.5	}_{-	0.7	}$ & &	2850	&	3090	&	210	\\
PG 0052+251	&	B	&	2.4	$\pm$	0.1	&	15.0	$^{+	0.5	}_{-	0.2	}$ &	0.6	$^{+	0.3	}_{-	0.2	}$ &	2.0	$^{+	0.3	}_{-	0.2	}$ &	2.7	$^{+	0.2	}_{-	0.1	}$ & &	5340	&	5240	&	530	\\
PG 0953+414	&	A	&	1.9	$\pm$	0.1	&	16.8	$^{+	0.4	}_{-	0.5	}$ &	1.0	$^{+	0.1	}_{-	0.2	}$ &	1.0	$^{+	0.3	}_{-	0.2	}$ &	2.8	$^{+	0.2	}_{-	0.6	}$ & &	3390	&	3520	&	210	\\
PG 1211+143	&	A	&	2.9	$\pm$	0.1	&	25.0	$^{+	0.8	}_{-	0.7	}$ &	0.6	$^{+	0.3	}_{-	0.2	}$ &	1.6	$^{+	0.2	}_{-	0.3	}$ &	3.9	$^{+	0.3	}_{-	0.2	}$ & &	2440	&	2350	&	280	\\
PG 1307+085	&	B	&	1.6	$\pm$	0.1	&	111.8	$^{+	7.2	}_{-	4.7	}$ &	3.4	$^{+	2.1	}_{-	1.2	}$ &	10.8	$^{+	2.3	}_{-	0.9	}$ &	17.7	$^{+	1.1	}_{-	0.9	}$ & &	5290	&	4970	&	330	\\
PG 1411+442	&	A	&	1.5	$\pm$	0.1	&	11.9	$^{+	0.5	}_{-	0.5	}$ &	0.4	$^{+	0.3	}_{-	0.1	}$ &	1.2	$^{+	0.3	}_{-	0.1	}$ &	3.4	$^{+	0.2	}_{-	0.1	}$ & &	2540	&	2270	&	220	\\
\enddata
\tablenotetext{a}{As defined in Paper I and \citet{sulenticetal00a}}
\tablenotetext{b}{Rest-frame specific continuum flux at 1700 \AA\ in units of 10$^{-14}$ erg \, s$^{-1}$ cm$^{-2}$ \AA$^{-1}$.}
\tablenotetext{c}{Rest-frame line flux of \ciii, \siiii, \aliii\  and of the \civ\ BC in units of 10$^{-13}$ erg \, s$^{-1}$ cm$^{-2}$.}
\tablenotetext{d}{Rest frame FWHM of the \hb\ BC. UV is for \siiii, \aliii\ and of the \civ\ BC in units of \kms.}
\end{deluxetable}

\begin{rotate}
\begin{deluxetable}{lccccccccc}
\tabletypesize{\scriptsize}
\tablecaption{Derived Products\tablenotemark{*}}
\tablewidth{0pt}
\tablehead{
\colhead{}  & \multicolumn{4}{c}{\nhu}  &  \colhead{} & \multicolumn{4}{c}{\R} \\
\cline{2-5}
\cline{7-10}
\colhead{Object name} & \colhead{c\.$\tau$} &\colhead{$Al{\sc iii}/Si{\sc iii]}$} &\colhead{$C{\sc iii]}/Si{\sc iii]}$} &\colhead{corr.}  &  \colhead{} & \colhead{c\.$\tau$} &\colhead{$Al{\sc iii}/Si{\sc iii]}$} &\colhead{$C{\sc iii]}/Si{\sc iii]}$} &\colhead{corr.} \\
\colhead{(1)}&\colhead{(2)}&\colhead{(3)}&\colhead{(4)}&\colhead{(5)} &  \colhead{} &\colhead{(6)}&\colhead{(7)}&\colhead{(8)}&\colhead{(9)}
}
\startdata
AKN 120	&	9.86	$^{+	0.11	}_{-	0.08	}$ &	9.70	$^{+	0.22	}_{-	0.12	}$ &	8.17	$^{+	0.07	}_{-	0.03	}$ &	10.44	$^{+	0.36	}_{-	0.21	}$ & &	17.01	&	17.08	$^{+	0.22	}_{-	0.13	}$ &	17.85	$^{+	0.08	}_{-	0.05	}$ &	16.72	$^{+	0.36	}_{-	0.22	}$ \\
Fairall 9	&	10.55	$^{+	0.19	}_{-	0.15	}$ &	9.55	$^{+	0.11	}_{-	0.10	}$ &	7.99	$^{+	0.03	}_{-	0.06	}$ &	9.92	$^{+	0.12	}_{-	0.12	}$ & &	16.65	&	17.15	$^{+	0.12	}_{-	0.11	}$ &	17.93	$^{+	0.06	}_{-	0.08	}$ &	16.97	$^{+	0.13	}_{-	0.13	}$ \\
MRK 335	&	10.39	$^{+	0.20	}_{-	0.17	}$ &	9.94	$^{+	0.14	}_{-	0.13	}$ &	7.91	$^{+	0.08	}_{-	0.09	}$ &	9.68	$^{+	0.15	}_{-	0.14	}$ & &	16.61	&	16.83	$^{+	0.15	}_{-	0.14	}$ &	17.84	$^{+	0.09	}_{-	0.10	}$ &	16.96	$^{+	0.16	}_{-	0.15	}$ \\
MRK 509	&	10.34	$^{+	0.07	}_{-	0.07	}$ &	9.90	$^{+	0.10	}_{-	0.11	}$ &	8.09	$^{+	0.10	}_{-	0.07	}$ &	10.21	$^{+	0.21	}_{-	0.16	}$ & &	16.82	&	17.03	$^{+	0.10	}_{-	0.12	}$ &	17.94	$^{+	0.11	}_{-	0.08	}$ &	16.89	$^{+	0.21	}_{-	0.17	}$ \\
NGC 3516	&	10.08	$^{+	0.39	}_{-	0.61	}$ &	9.60	$^{+	0.10	}_{-	0.07	}$ &	8.20	$^{+	0.04	}_{-	0.02	}$ &	10.15	$^{+	0.18	}_{-	0.10	}$ & &	16.24	&	16.47	$^{+	0.10	}_{-	0.08	}$ &	17.17	$^{+	0.05	}_{-	0.04	}$ &	16.21	$^{+	0.18	}_{-	0.10	}$ \\
NGC 3783	&	10.11	$^{+	0.18	}_{-	0.24	}$ &	10.15	$^{+	0.07	}_{-	0.11	}$ &	8.07	$^{+	0.26	}_{-	0.11	}$ &	9.97	$^{+	0.19	}_{-	0.13	}$ & &	16.42	&	16.39	$^{+	0.07	}_{-	0.12	}$ &	17.43	$^{+	0.26	}_{-	0.11	}$ &	16.49	$^{+	0.20	}_{-	0.13	}$ \\
NGC 5548	&	9.58	$^{+	0.03	}_{-	0.03	}$ &	9.93	$^{+	0.12	}_{-	0.10	}$ &	8.30	$^{+	0.03	}_{-	0.02	}$ &	10.15	$^{+	0.19	}_{-	0.15	}$ & &	16.67	&	16.48	$^{+	0.12	}_{-	0.11	}$ &	17.30	$^{+	0.05	}_{-	0.04	}$ &	16.38	$^{+	0.20	}_{-	0.15	}$ \\
NGC 7469	&	10.91	$^{+	0.14	}_{-	0.13	}$ &	9.83	$^{+	0.08	}_{-	0.06	}$ &	7.90	$^{+	0.05	}_{-	0.03	}$ &	10.41	$^{+	0.17	}_{-	0.12	}$ & &	16.07	&	16.60	$^{+	0.09	}_{-	0.07	}$ &	17.56	$^{+	0.07	}_{-	0.06	}$ &	16.32	$^{+	0.18	}_{-	0.13	}$ \\
PG 0052+251	&	9.94	$^{+	0.21	}_{-	0.21	}$ &	9.64	$^{+	0.16	}_{-	0.14	}$ &	8.07	$^{+	0.06	}_{-	0.09	}$ &	9.81	$^{+	0.31	}_{-	0.21	}$ & &	17.37	&	17.51	$^{+	0.17	}_{-	0.15	}$ &	18.29	$^{+	0.07	}_{-	0.09	}$ &	17.43	$^{+	0.31	}_{-	0.21	}$ \\
PG 0953+414	&	9.73	$^{+	0.12	}_{-	0.12	}$ &	10.49	$^{+	0.25	}_{-	0.17	}$ &	8.24	$^{+	0.39	}_{-	0.28	}$ &			\ldots			& &	17.59	&	17.20	$^{+	0.26	}_{-	0.17	}$ &	18.33	$^{+	0.40	}_{-	0.28	}$ &			\ldots			\\
PG 1211+143	&	9.45	$^{+	0.32	}_{-	0.21	}$ &	10.00	$^{+	0.21	}_{-	0.17	}$ &	8.17	$^{+	0.20	}_{-	0.21	}$ &	9.95	$^{+	0.35	}_{-	0.24	}$ & &	17.39	&	17.10	$^{+	0.22	}_{-	0.18	}$ &	18.02	$^{+	0.21	}_{-	0.22	}$ &	17.14	$^{+	0.35	}_{-	0.24	}$ \\
PG 1307+085	&	9.63	$^{+	0.32	}_{-	0.25	}$ &	9.77	$^{+	0.22	}_{-	0.29	}$ &	8.17	$^{+	0.13	}_{-	0.07	}$ &	9.76	$^{+	0.44	}_{-	0.35	}$ & &	17.44	&	17.36	$^{+	0.22	}_{-	0.29	}$ &	18.16	$^{+	0.13	}_{-	0.08	}$ &	17.37	$^{+	0.44	}_{-	0.35	}$ \\
PG 1411+442	&	9.01	$^{+	0.35	}_{-	0.35	}$ &	9.87	$^{+	0.22	}_{-	0.31	}$ &	7.53	$^{+	0.21	}_{-	0.09	}$ &	9.41	$^{+	0.50	}_{-	0.36	}$ & &	17.51	&	17.07	$^{+	0.22	}_{-	0.31	}$ &	18.24	$^{+	0.22	}_{-	0.10	}$ &	17.31	$^{+	0.50	}_{-	0.36	}$ \\
\enddata
\tablenotetext{(*)}{ c\.$\tau$ is derived from RM data. The $Al{\sc iii}/Si{\sc iii]}$ ratio represents the high density solution, while the $C{\sc iii]}/Si{\sc iii]}$ ratio is the low density solution (see section \ref{sec:results}). {\it corr.} is the correction due systematic effects (see section \ref{subsec:comparison})}
\label{tab:derived_nu}
\end{deluxetable}
\end{rotate}

\clearpage

\clearpage
{\includegraphics[scale=0.35]{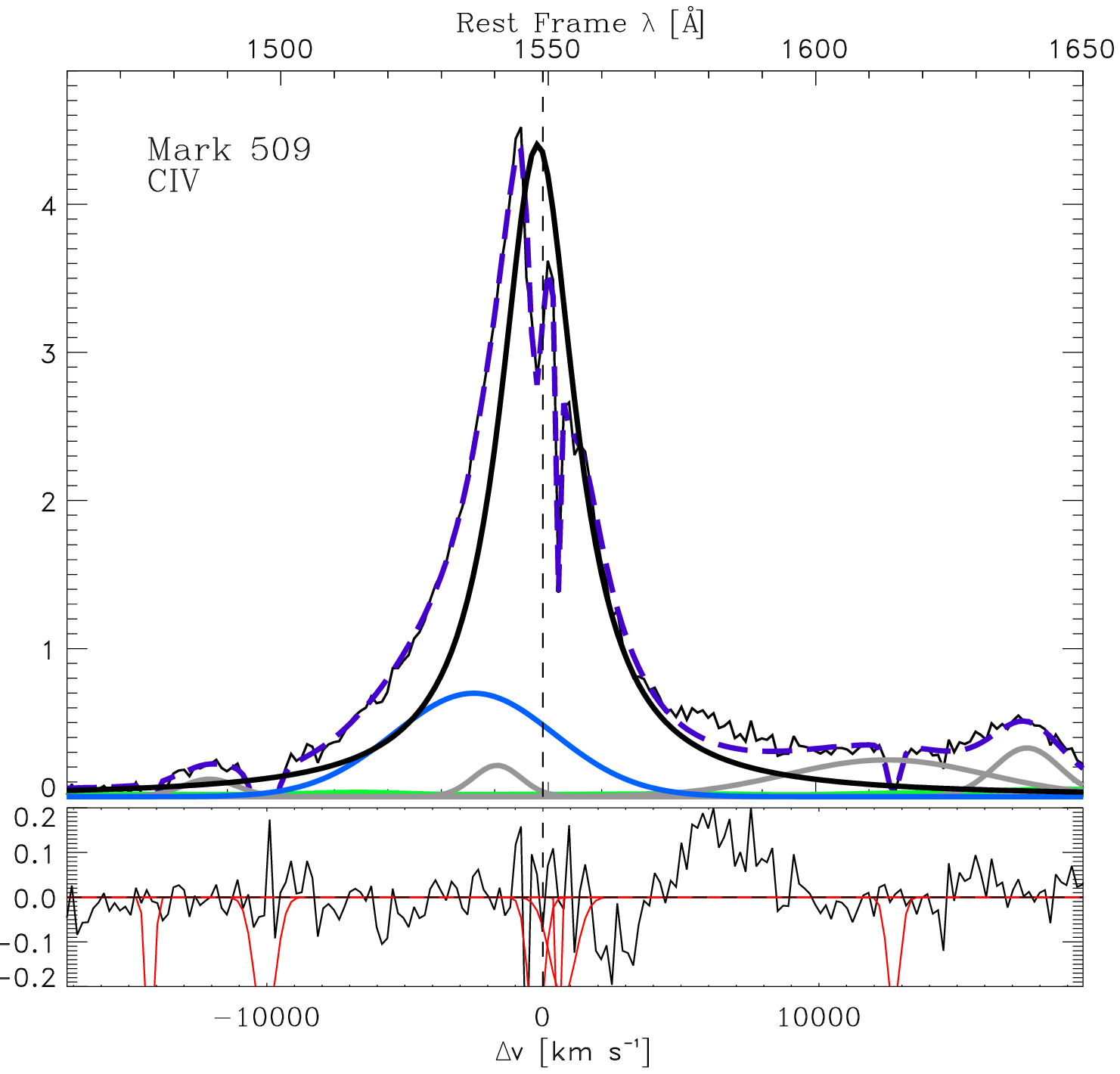}\includegraphics[scale=0.35]{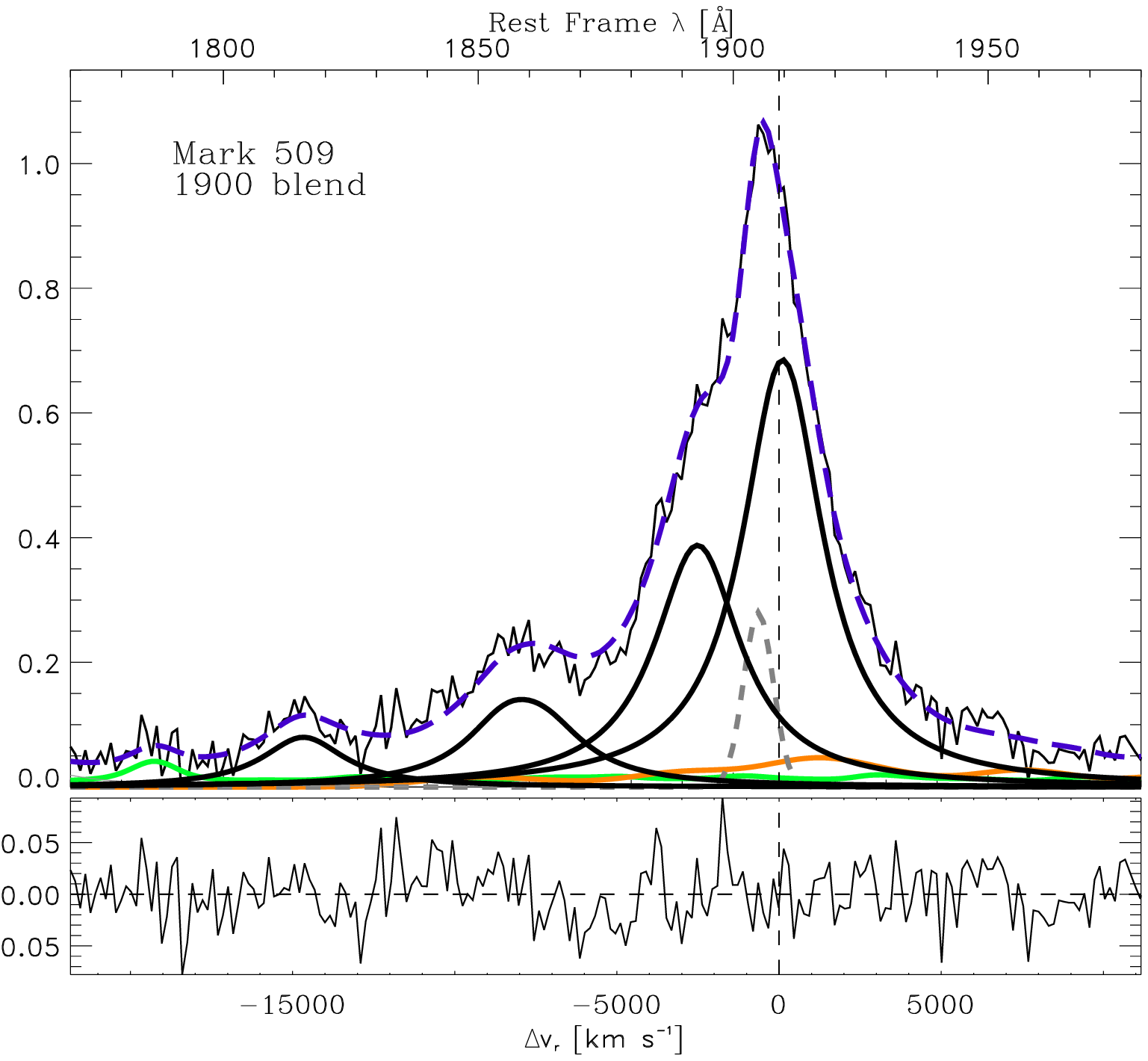}\includegraphics[scale=0.35]{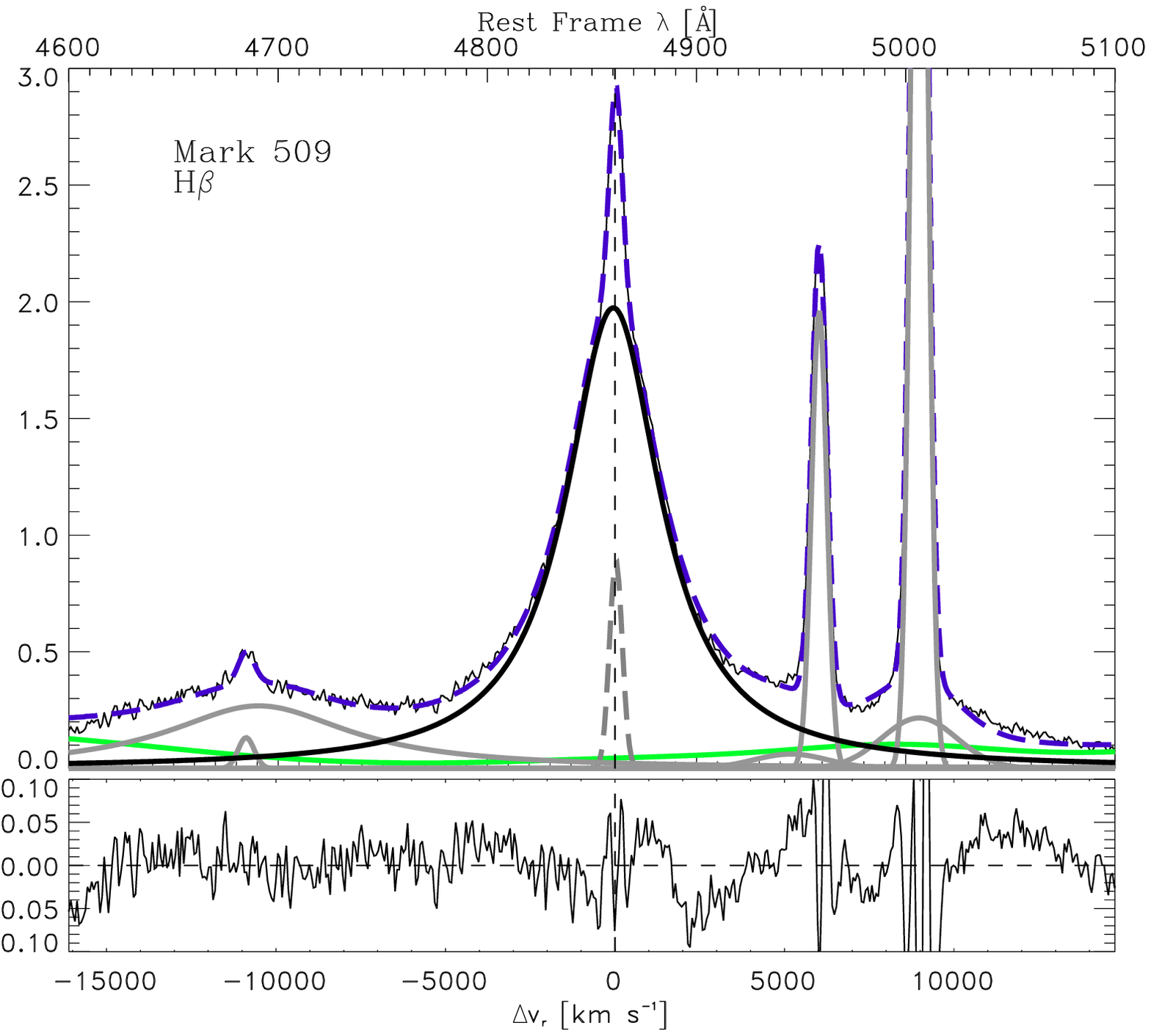}\\
\includegraphics[scale=0.35]{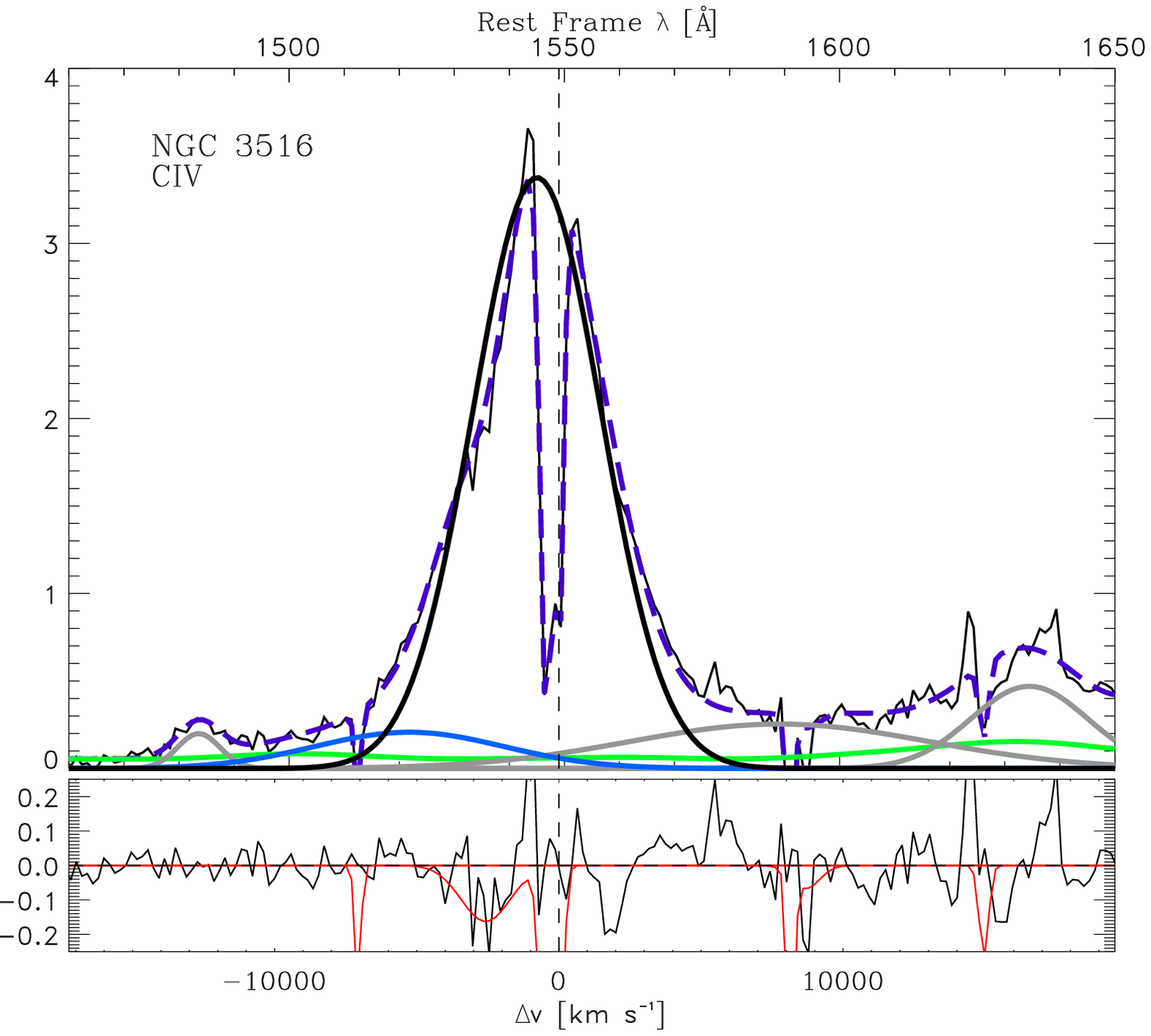}\includegraphics[scale=0.35]{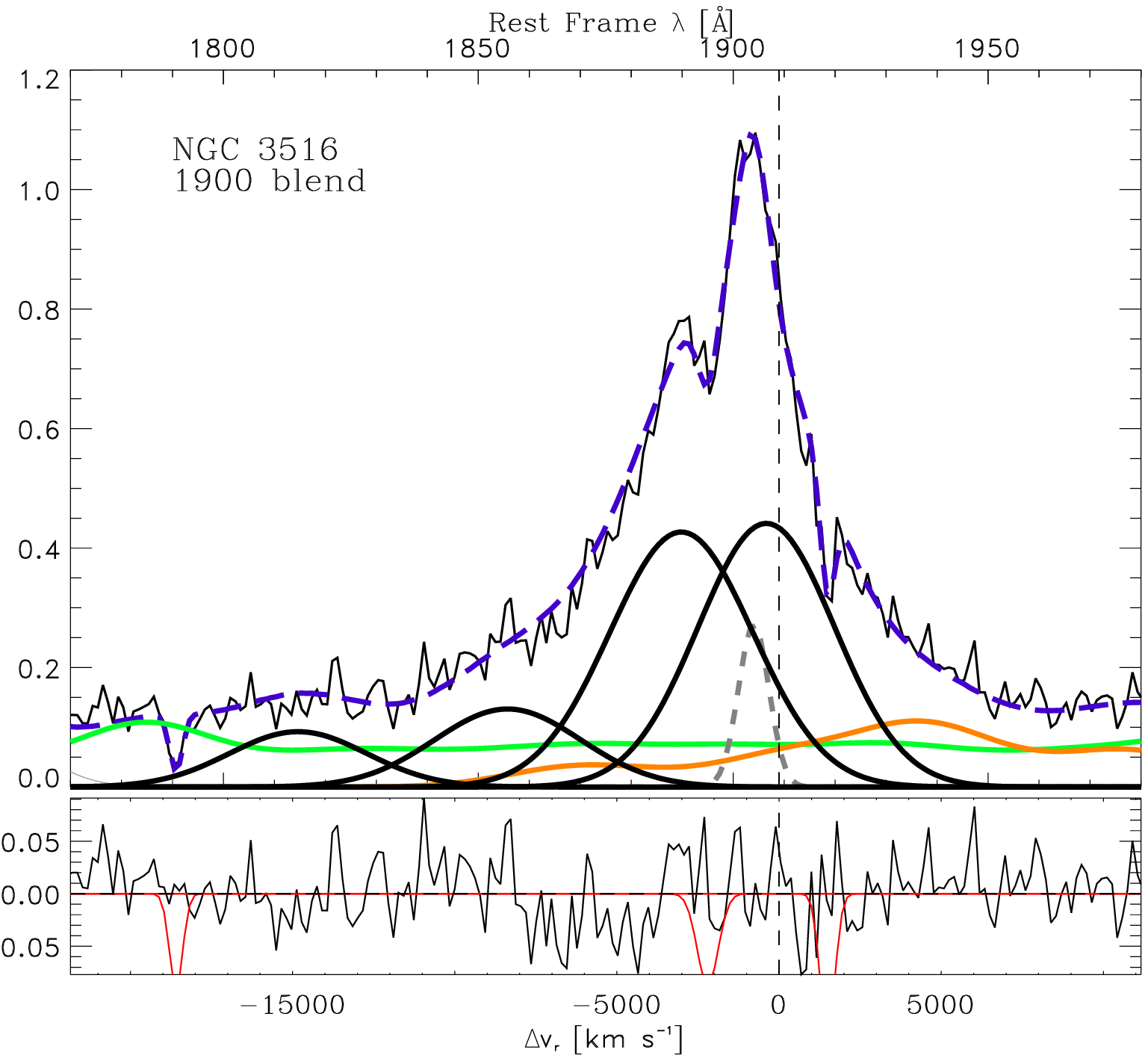}\includegraphics[scale=0.35]{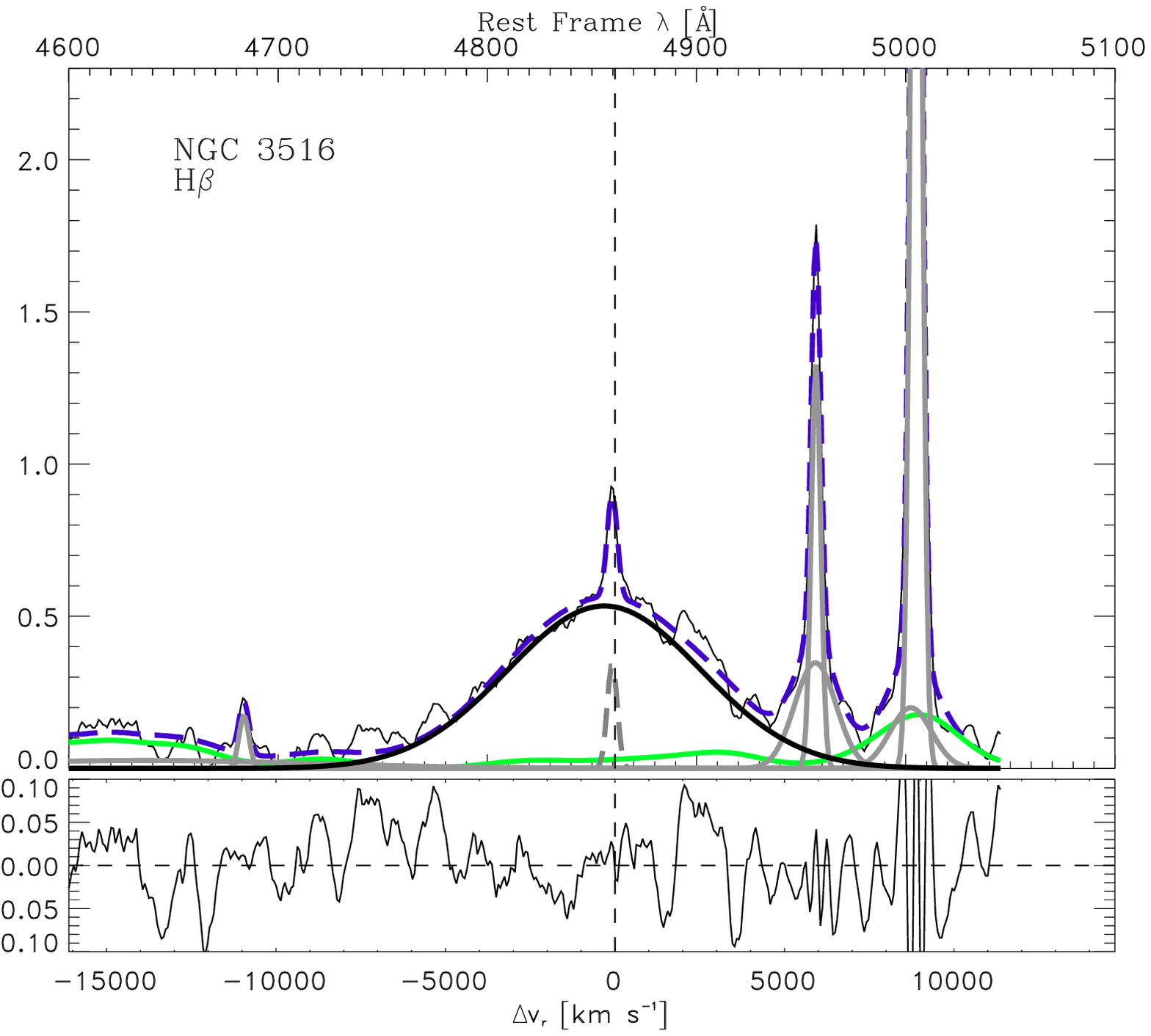}\\
\includegraphics[scale=0.35]{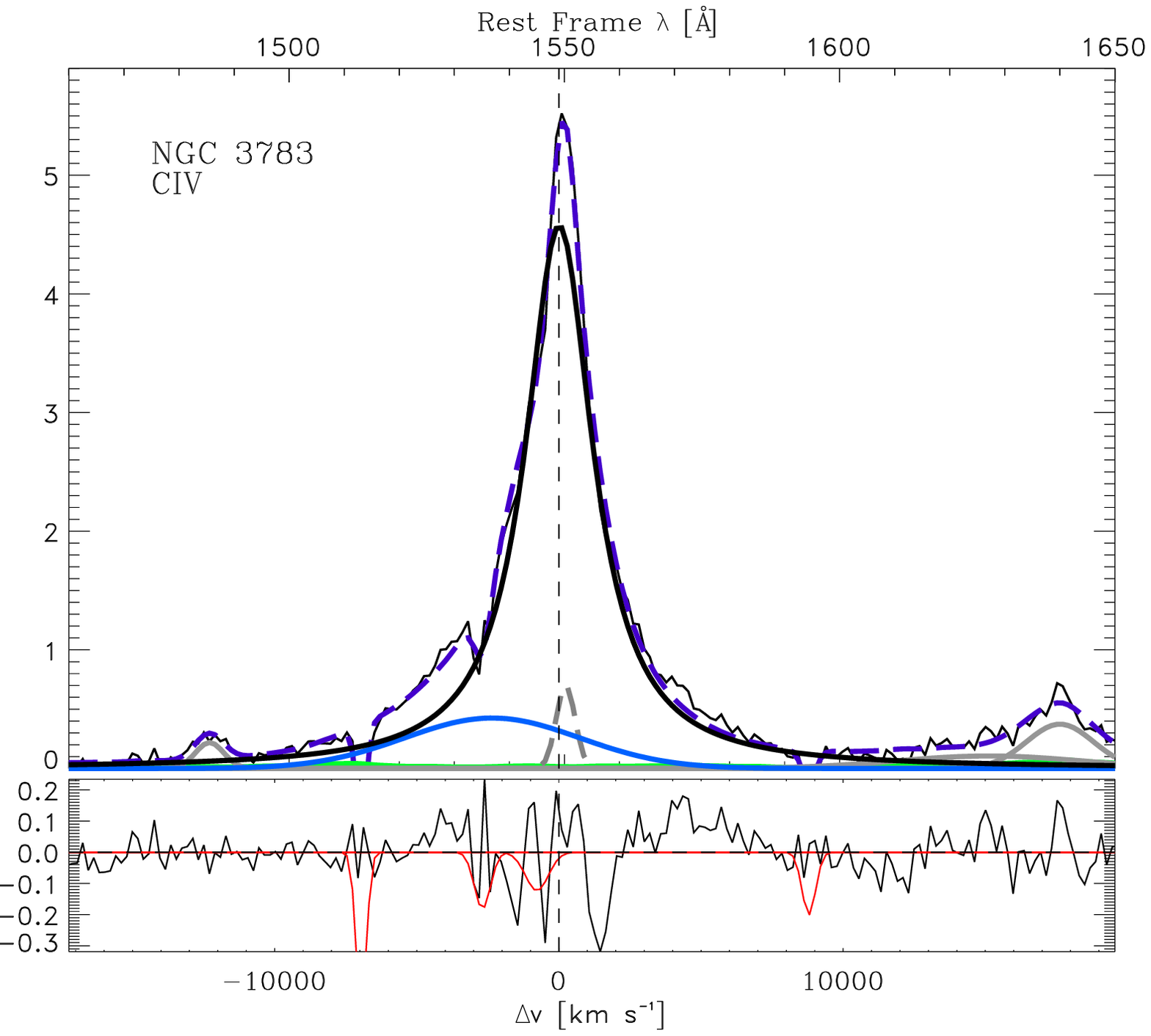}\includegraphics[scale=0.35]{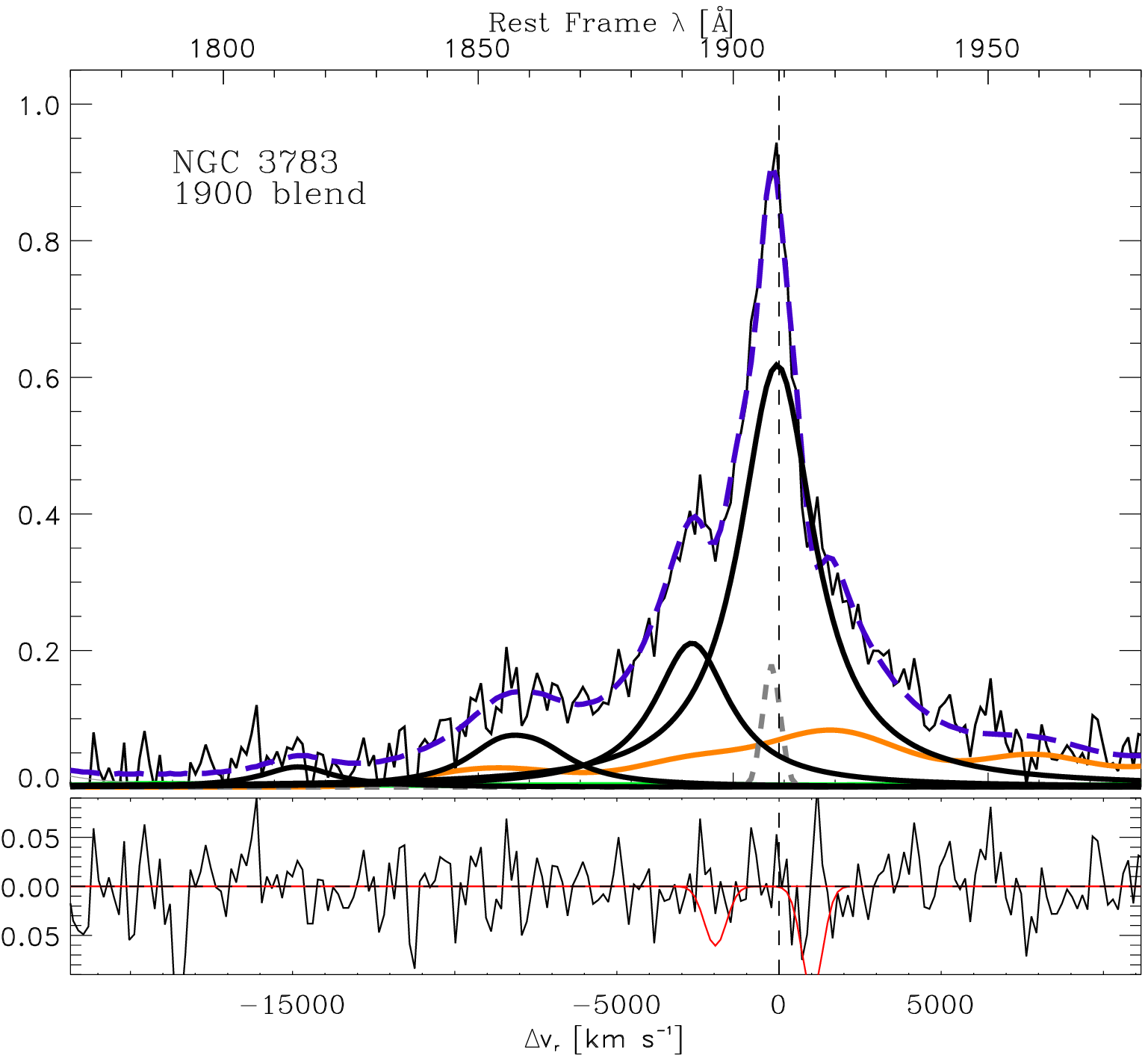}\includegraphics[scale=0.35]{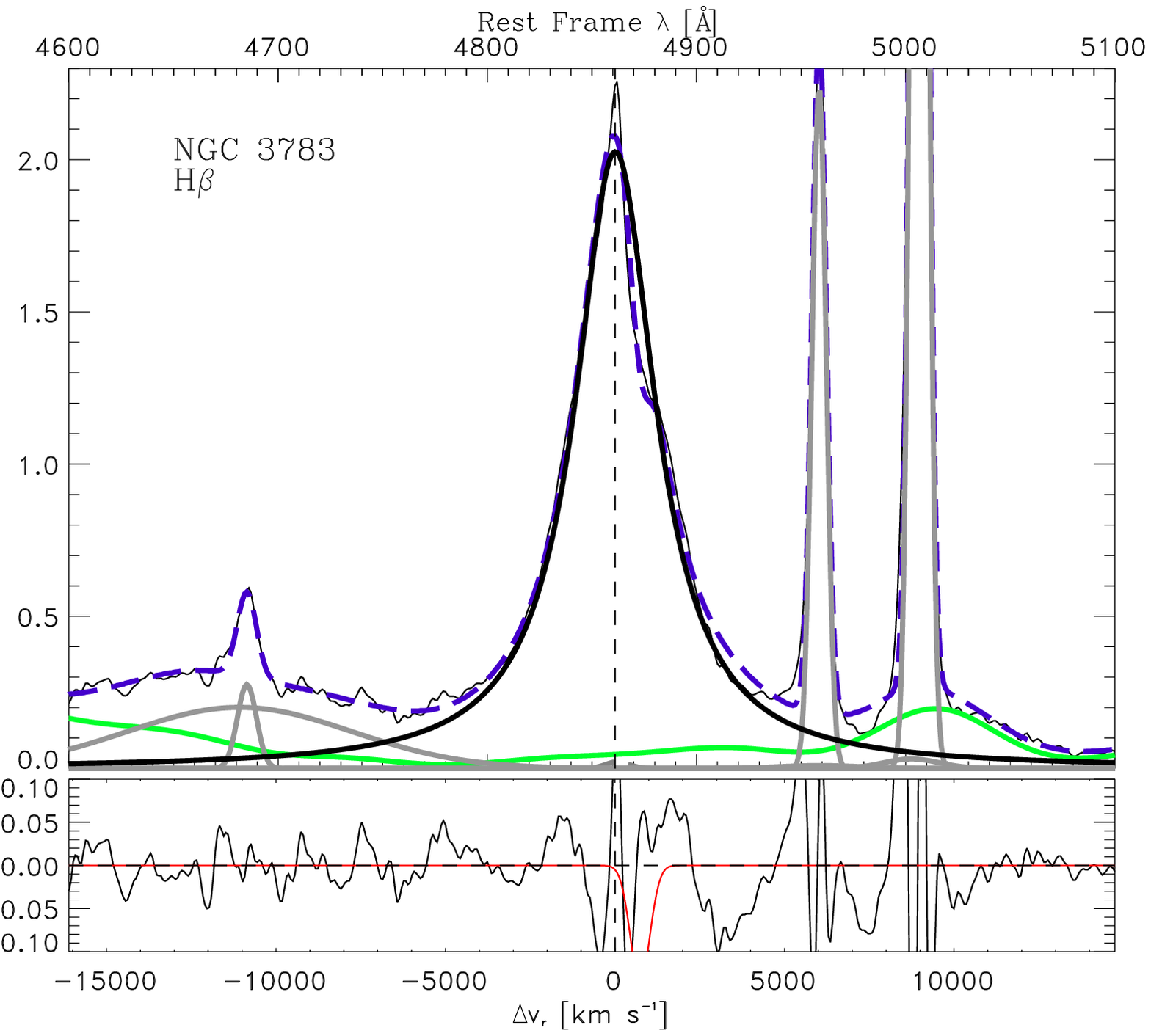}}
\centerline{Fig. 1. --- Continued.}
\clearpage
\includegraphics[scale=0.35]{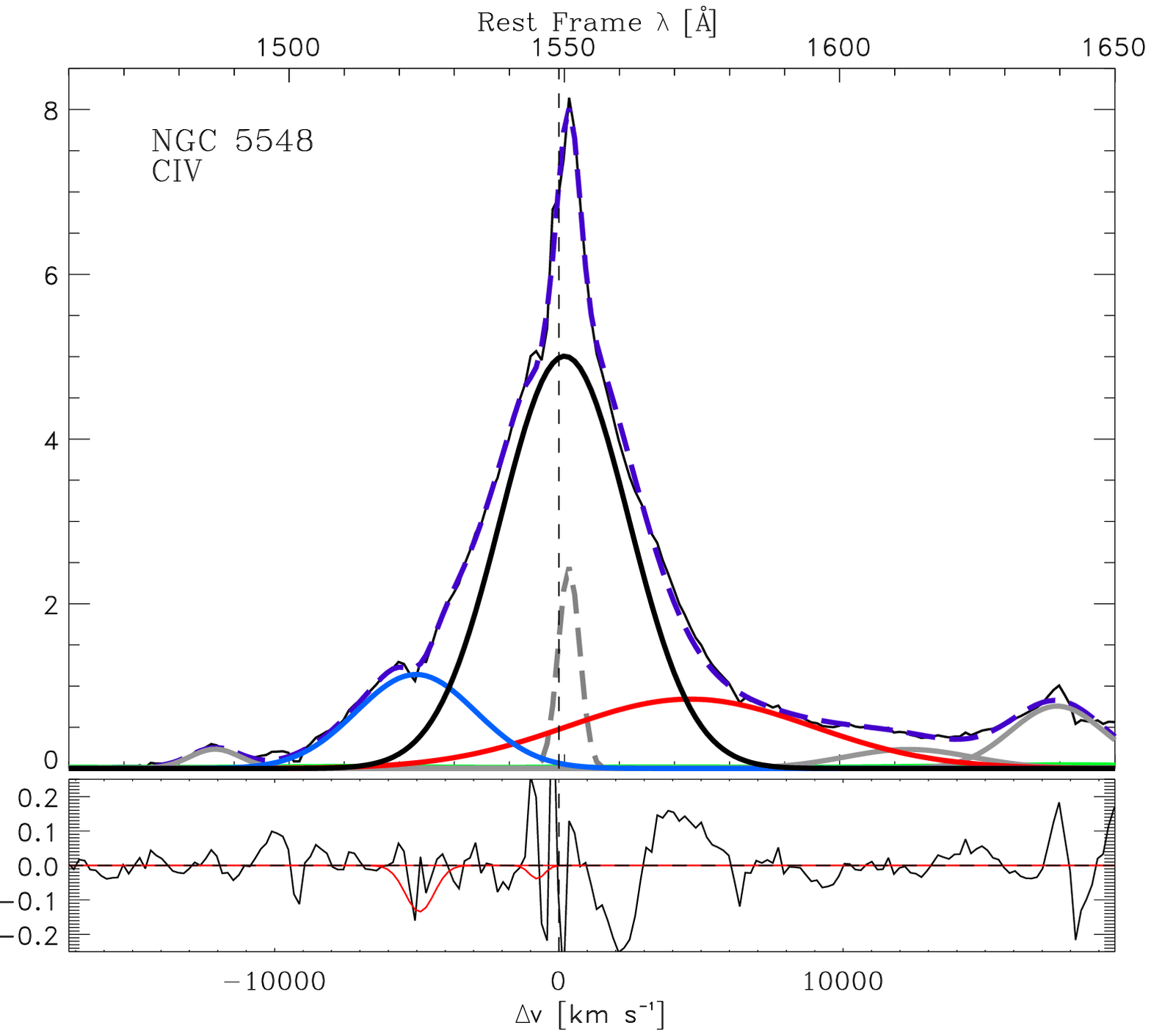}\includegraphics[scale=0.35]{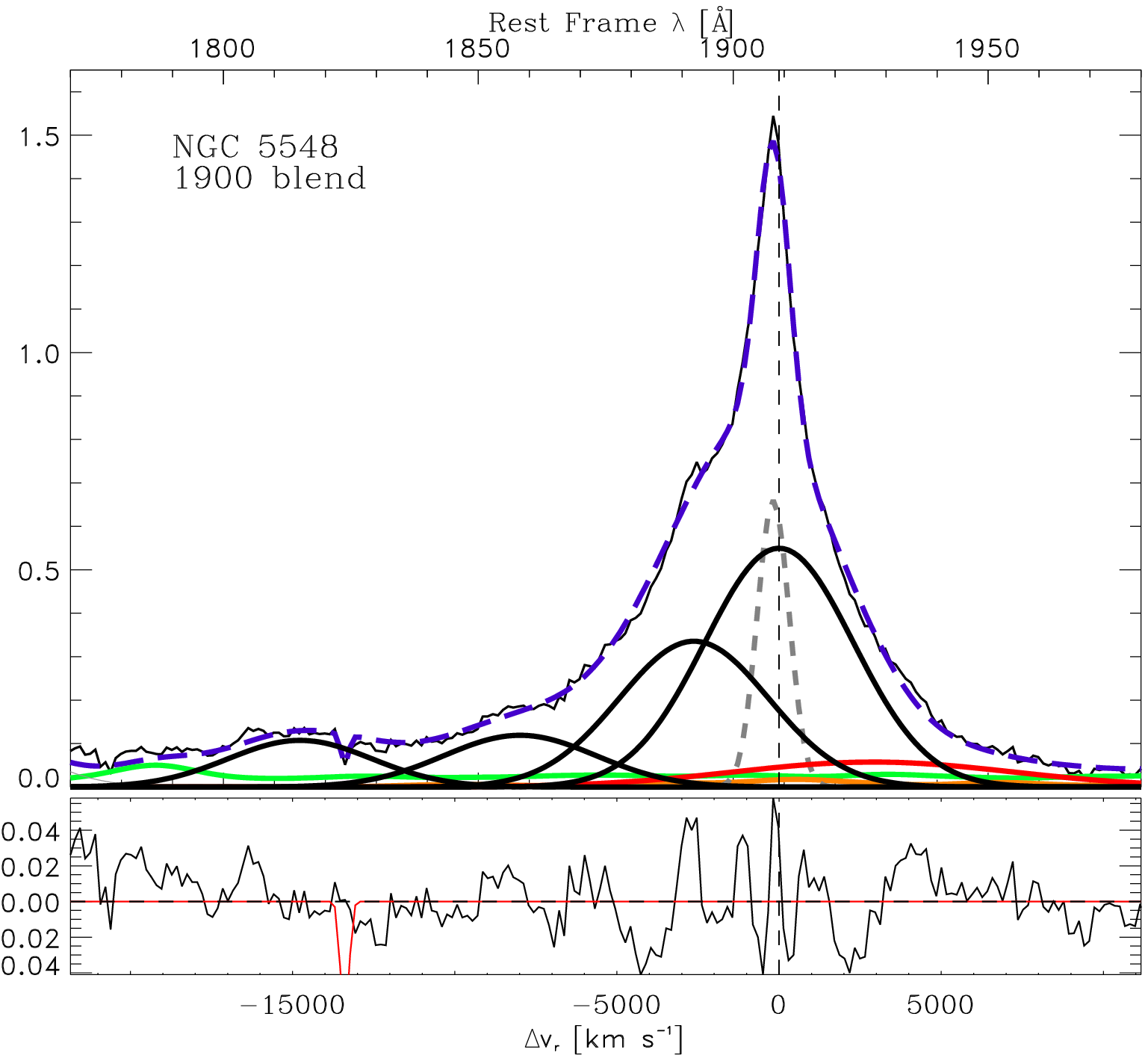}\includegraphics[scale=0.35]{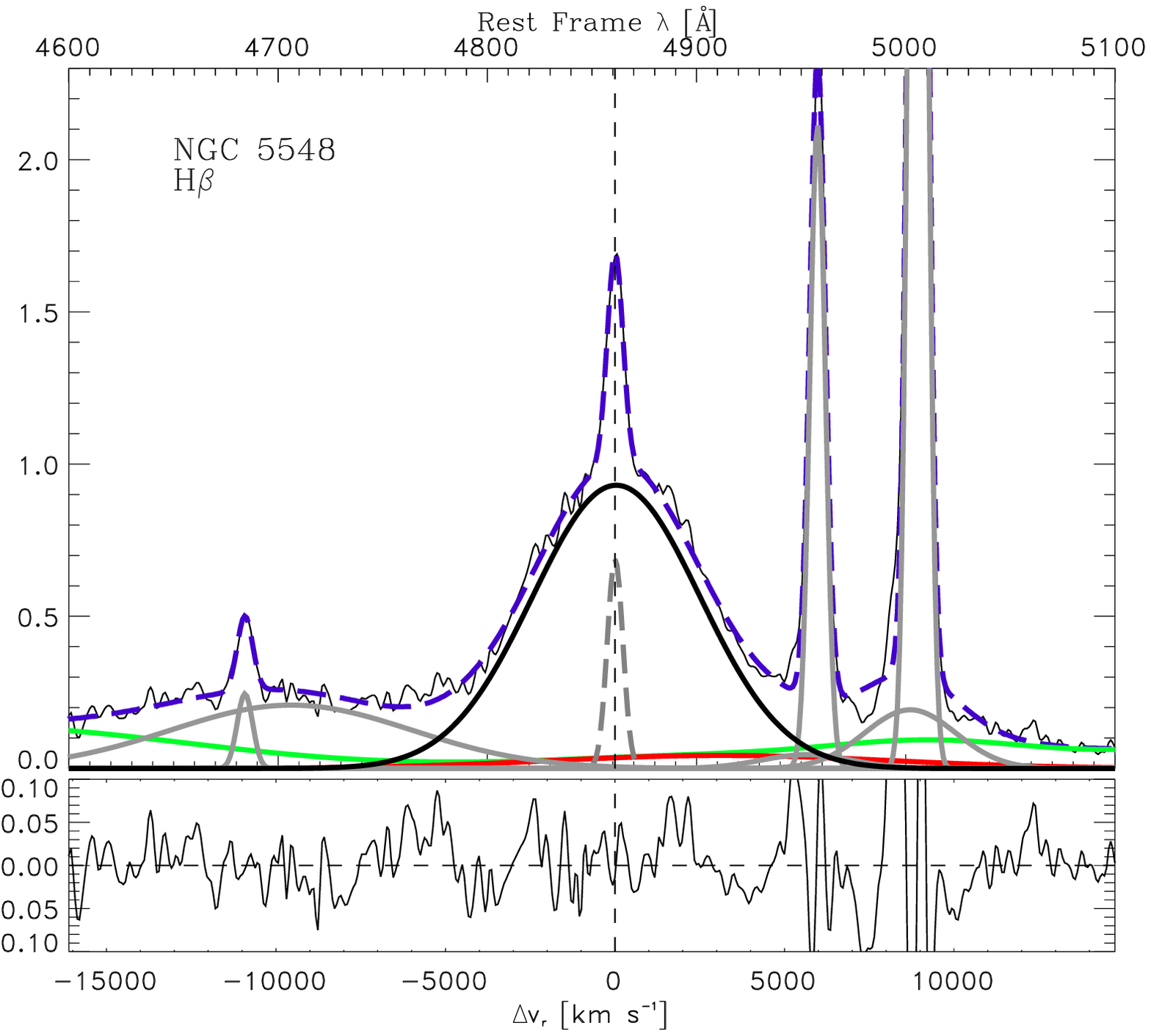}\\
\includegraphics[scale=0.35]{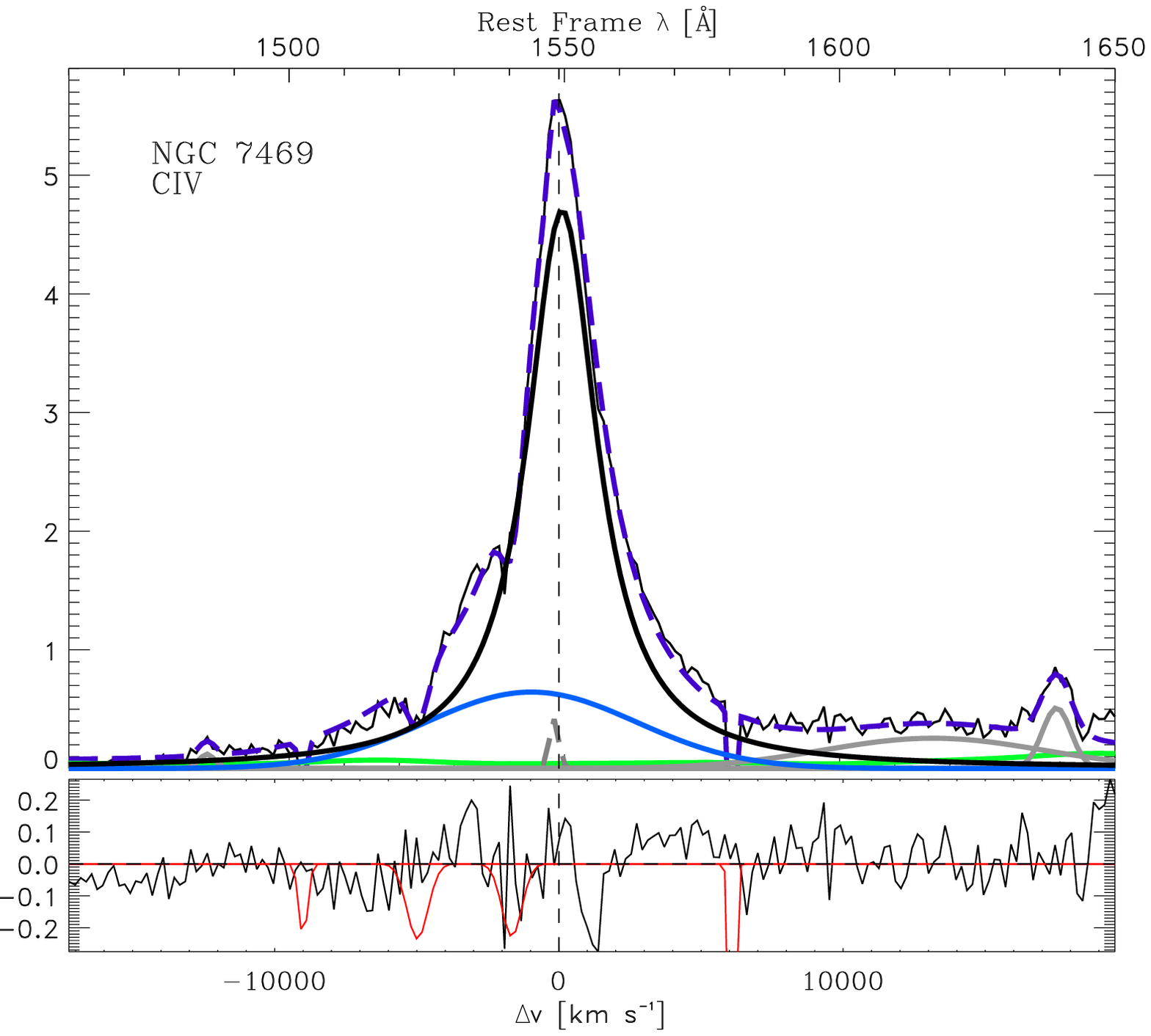}\includegraphics[scale=0.35]{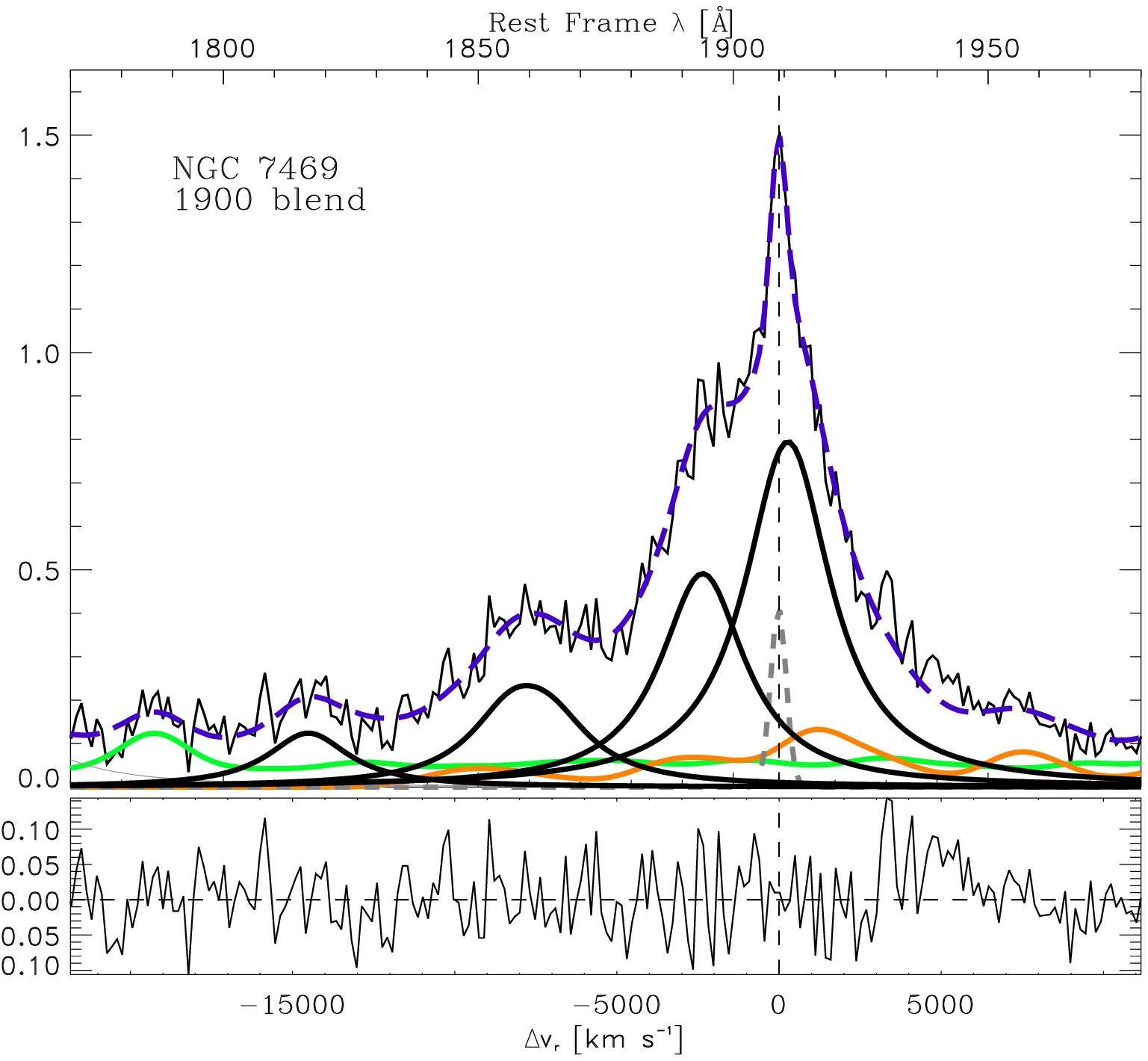}\includegraphics[scale=0.35]{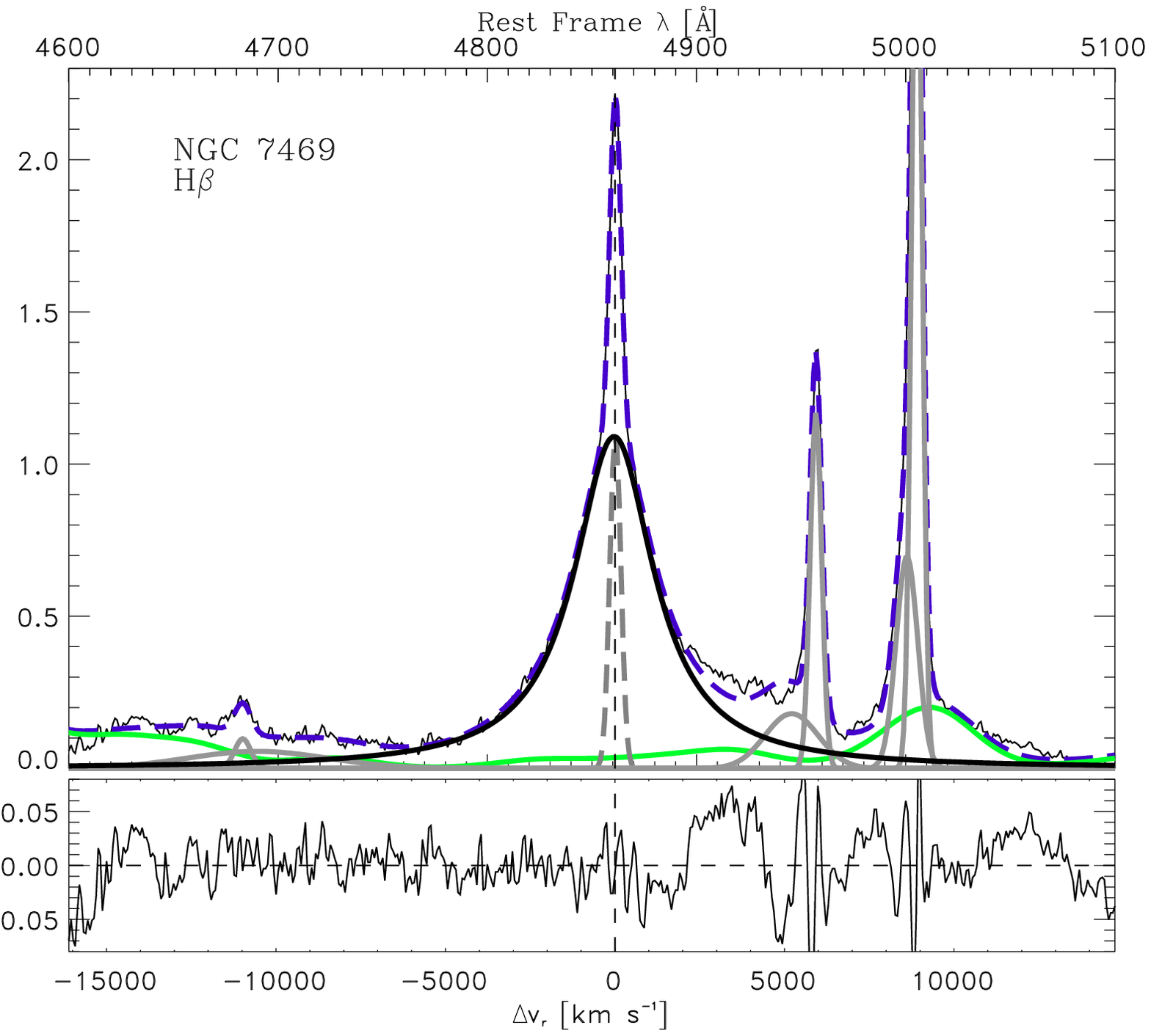}\\
\includegraphics[scale=0.35]{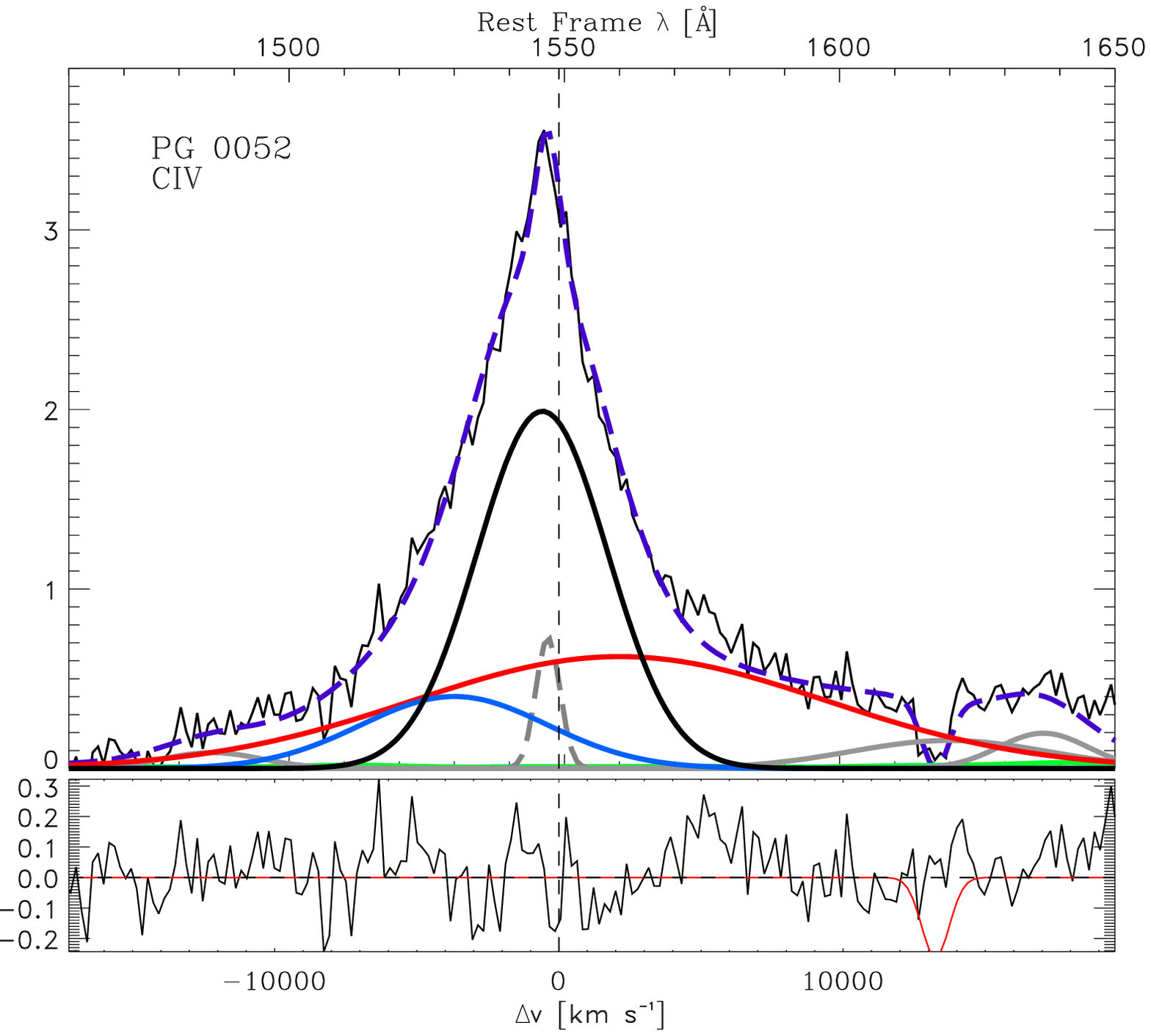}\includegraphics[scale=0.35]{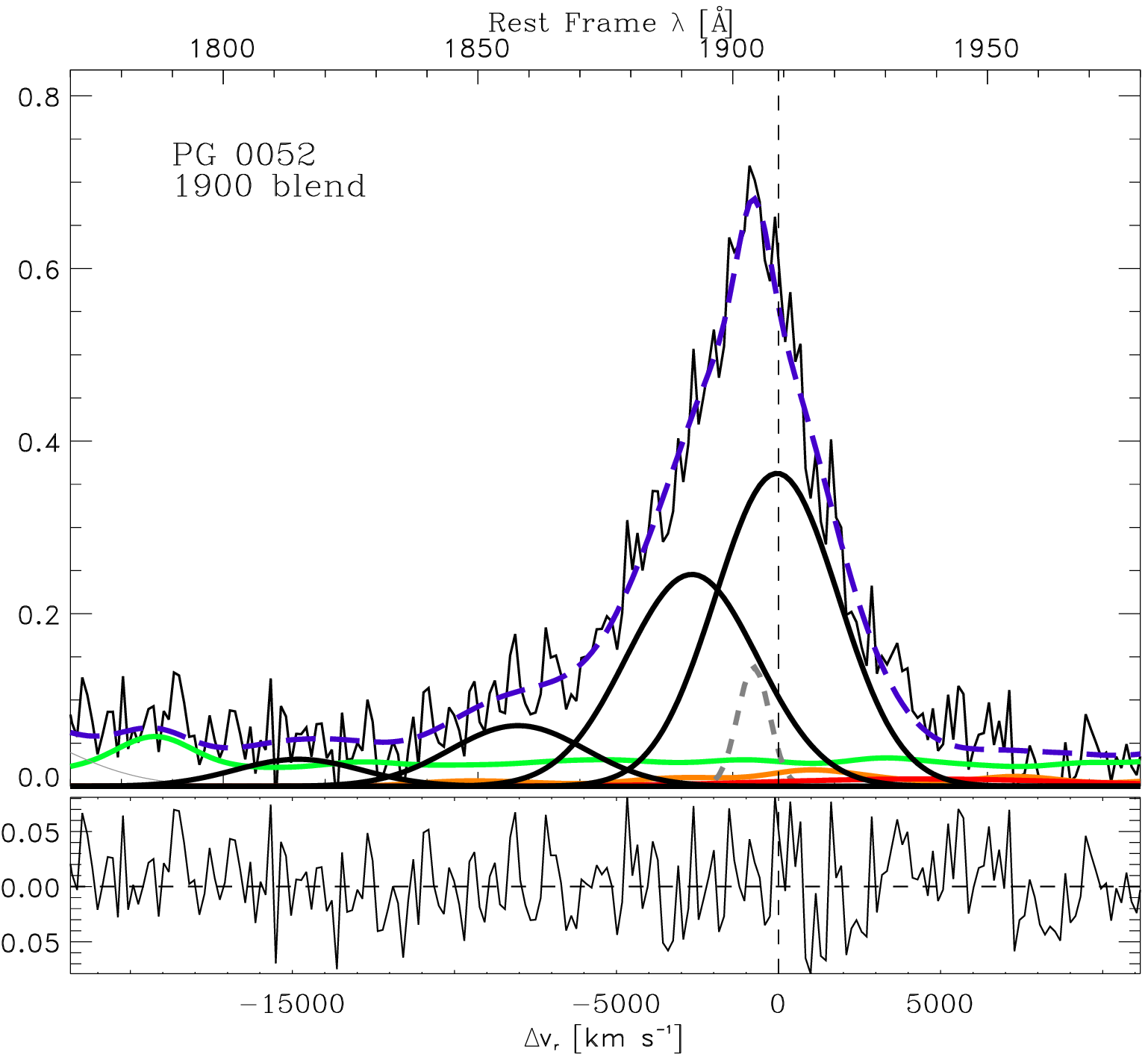}\includegraphics[scale=0.35]{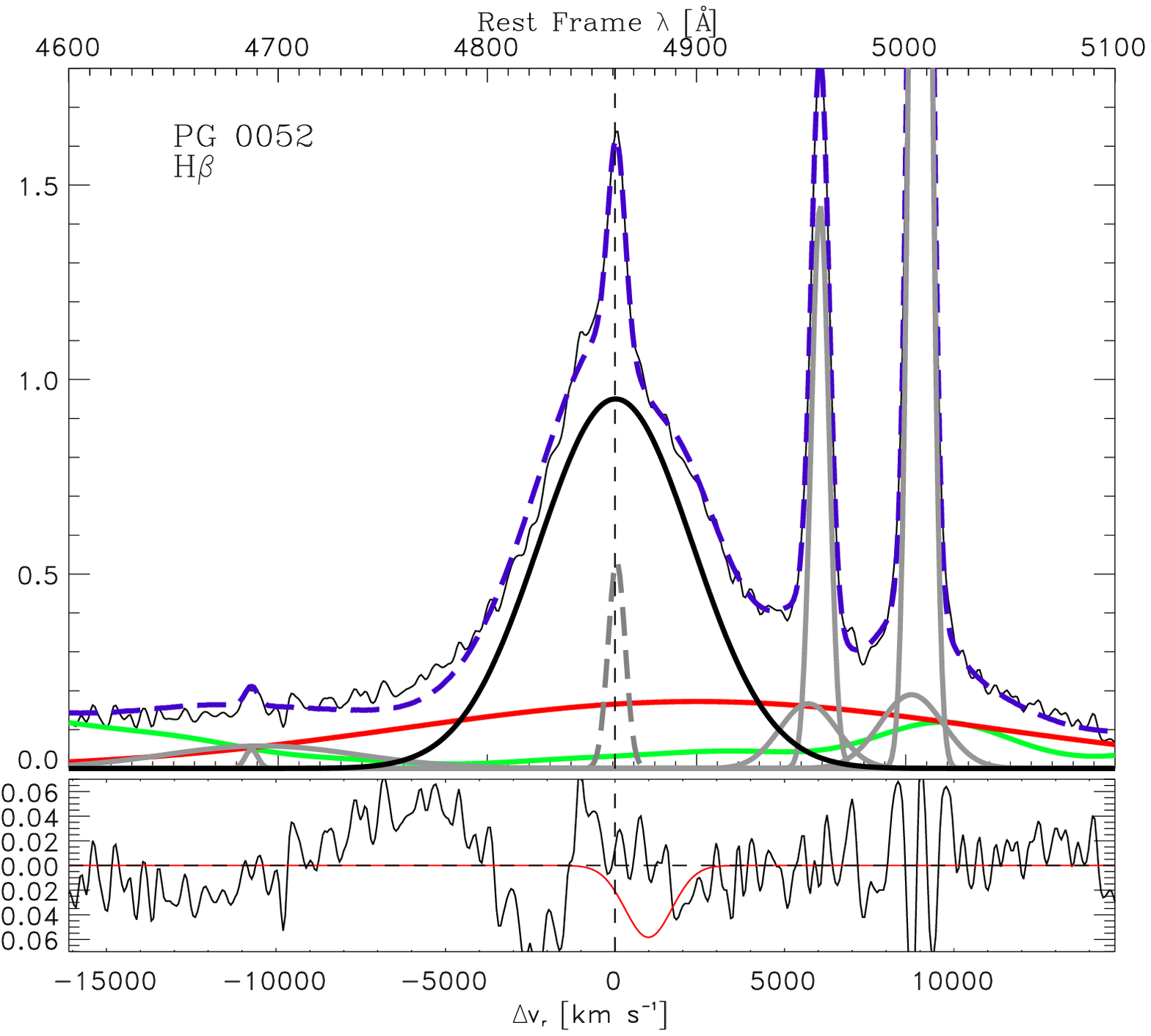}
\centerline{Fig. 1. --- Continued.}
\clearpage
\includegraphics[scale=0.35]{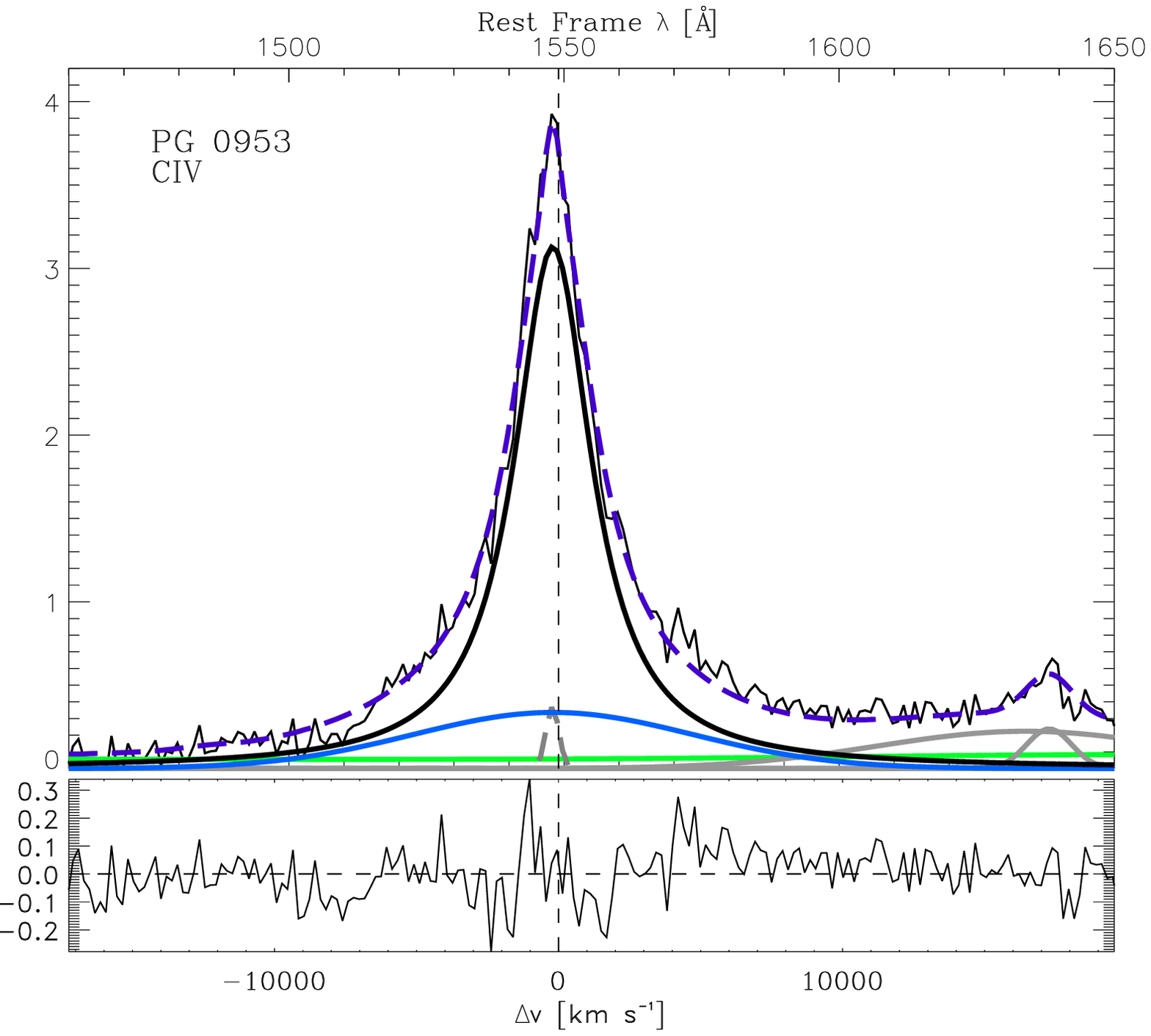}\includegraphics[scale=0.35]{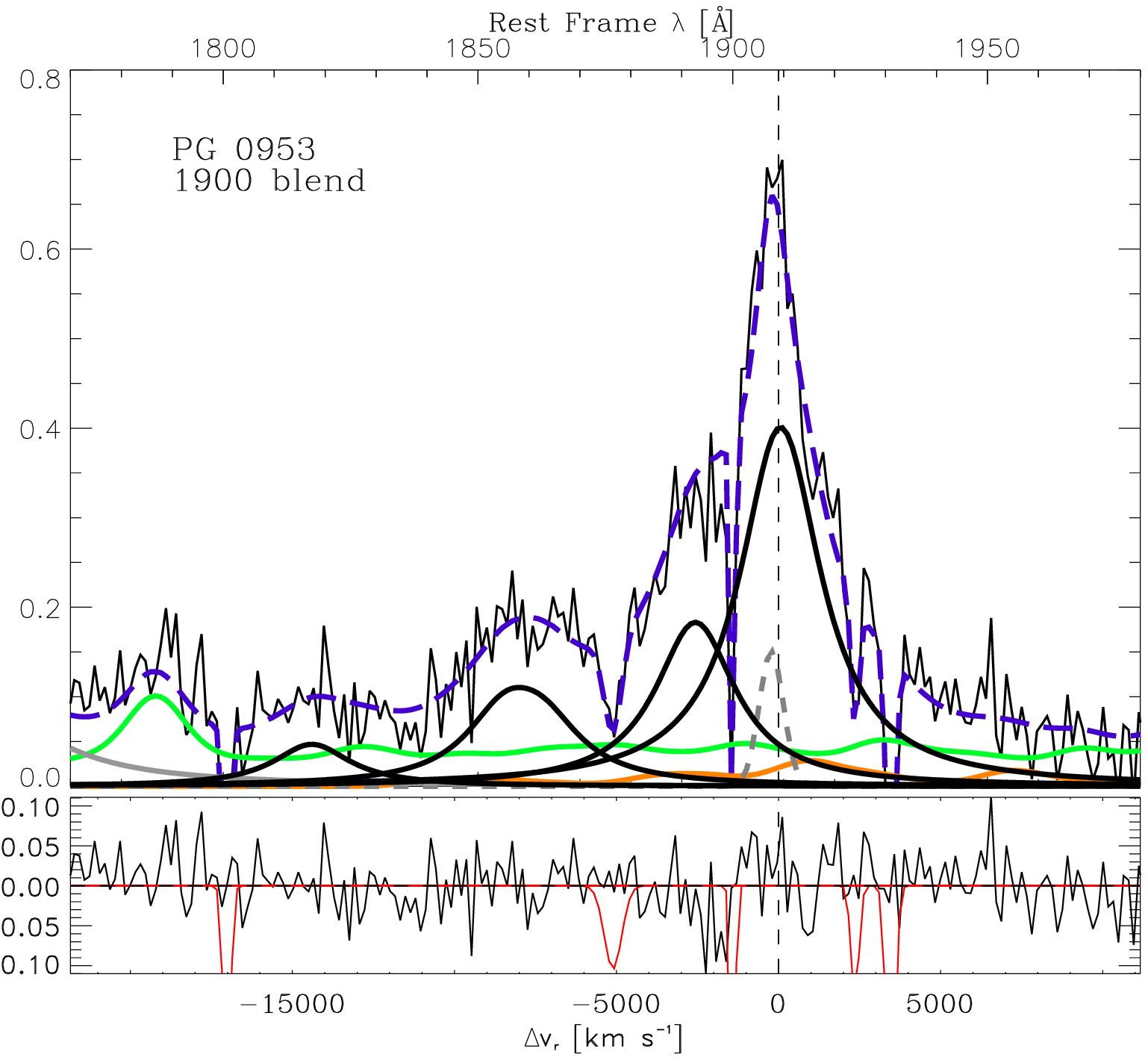}\includegraphics[scale=0.35]{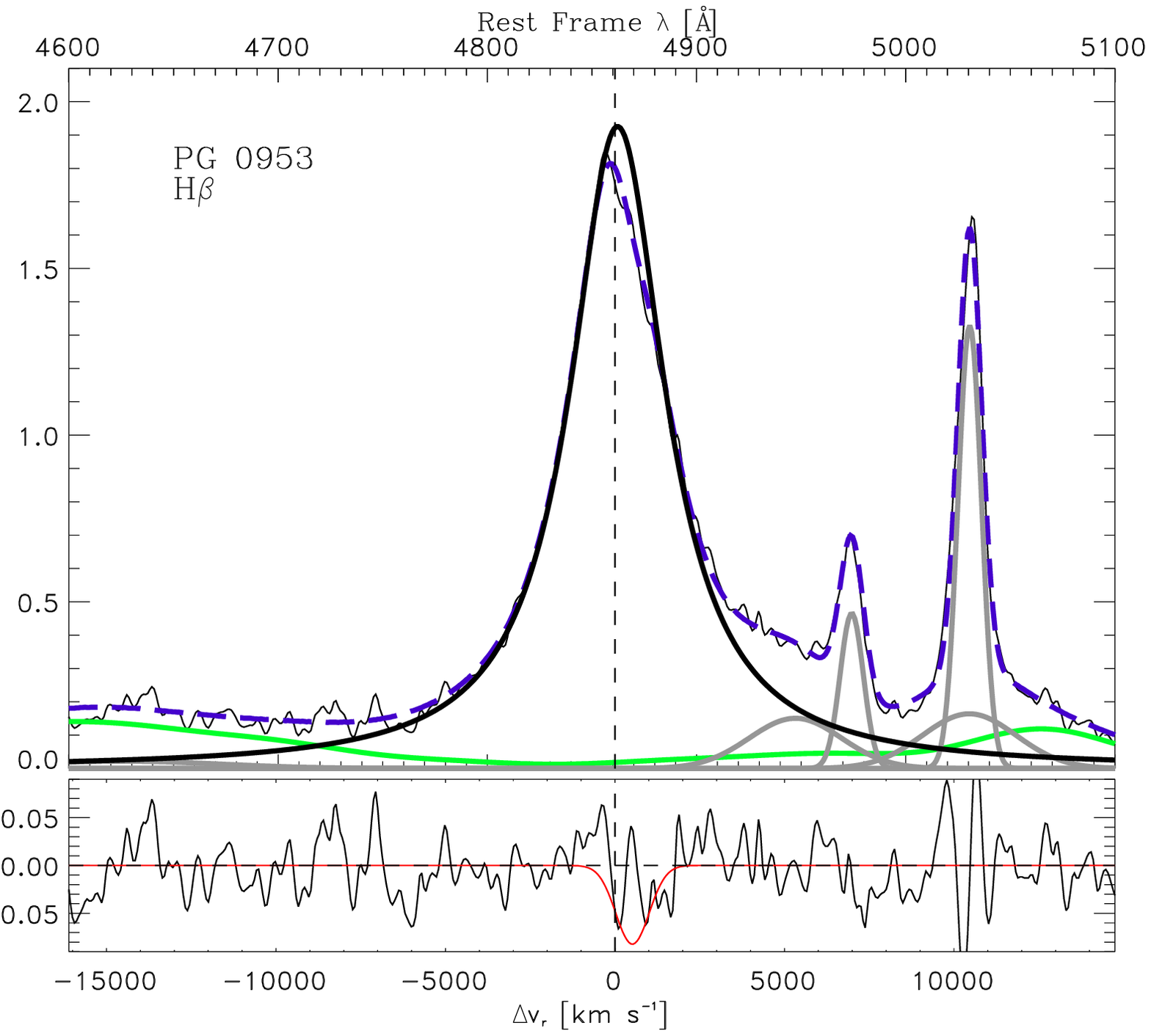}\\
\includegraphics[scale=0.35]{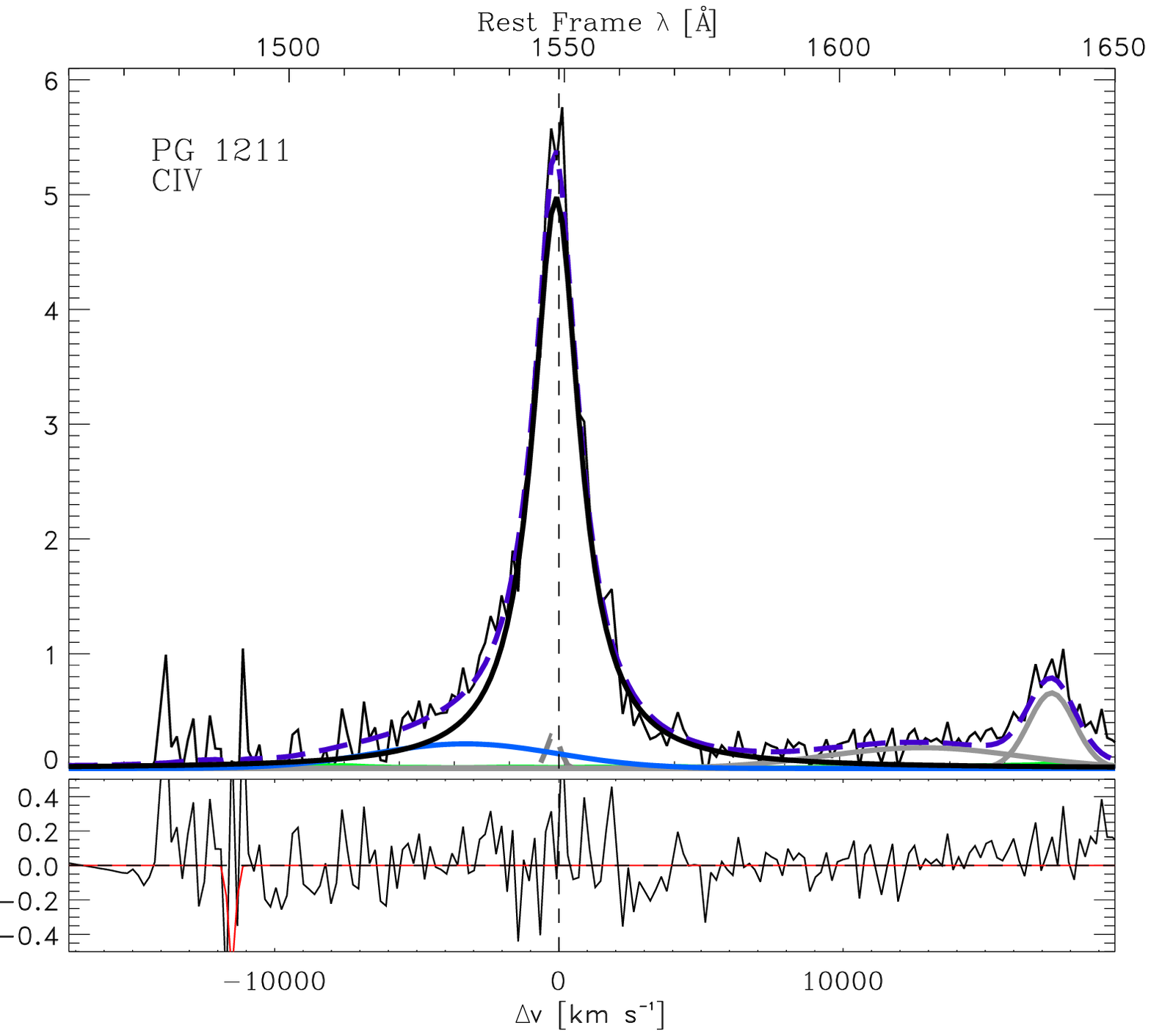}\includegraphics[scale=0.35]{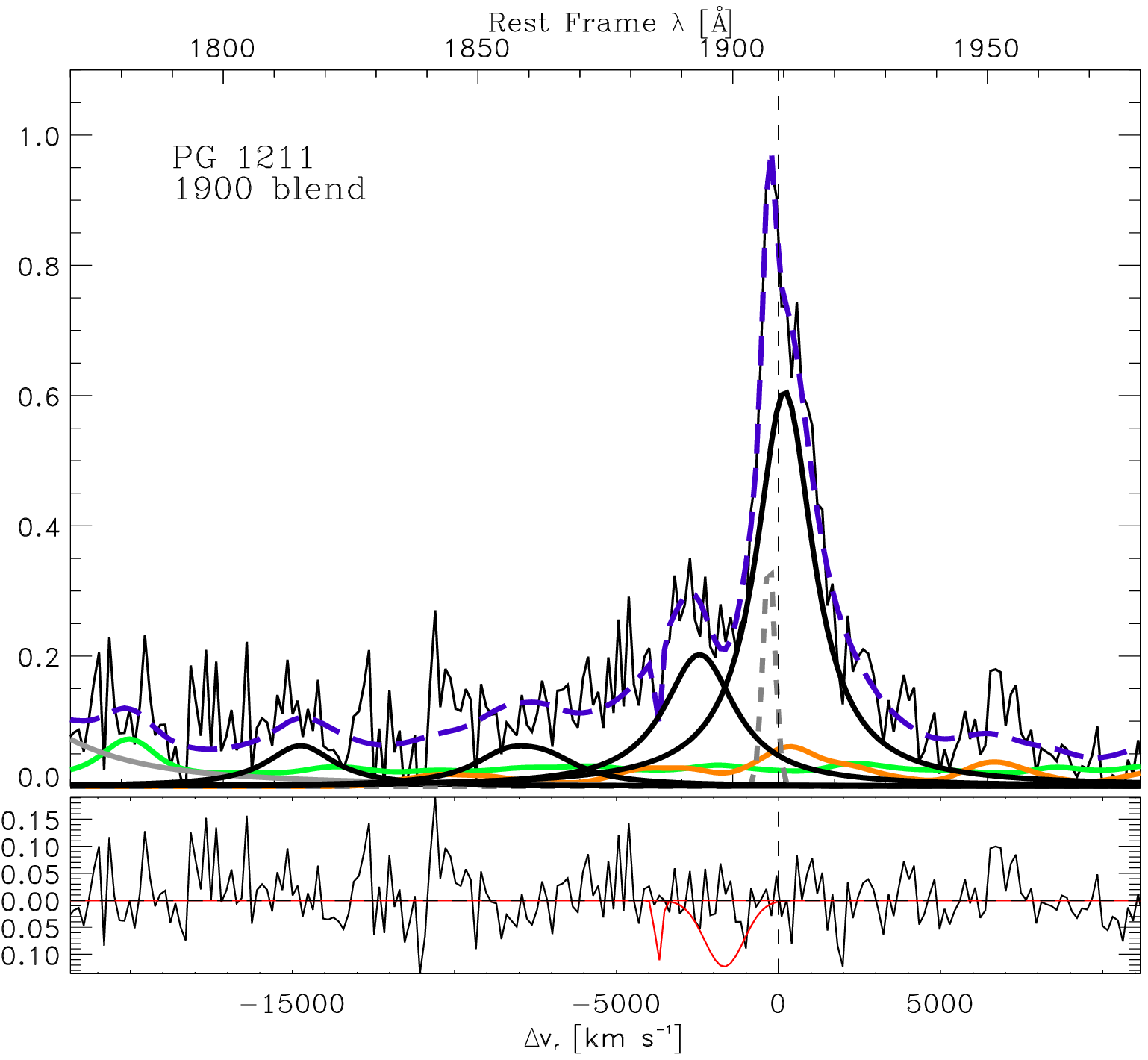}\includegraphics[scale=0.35]{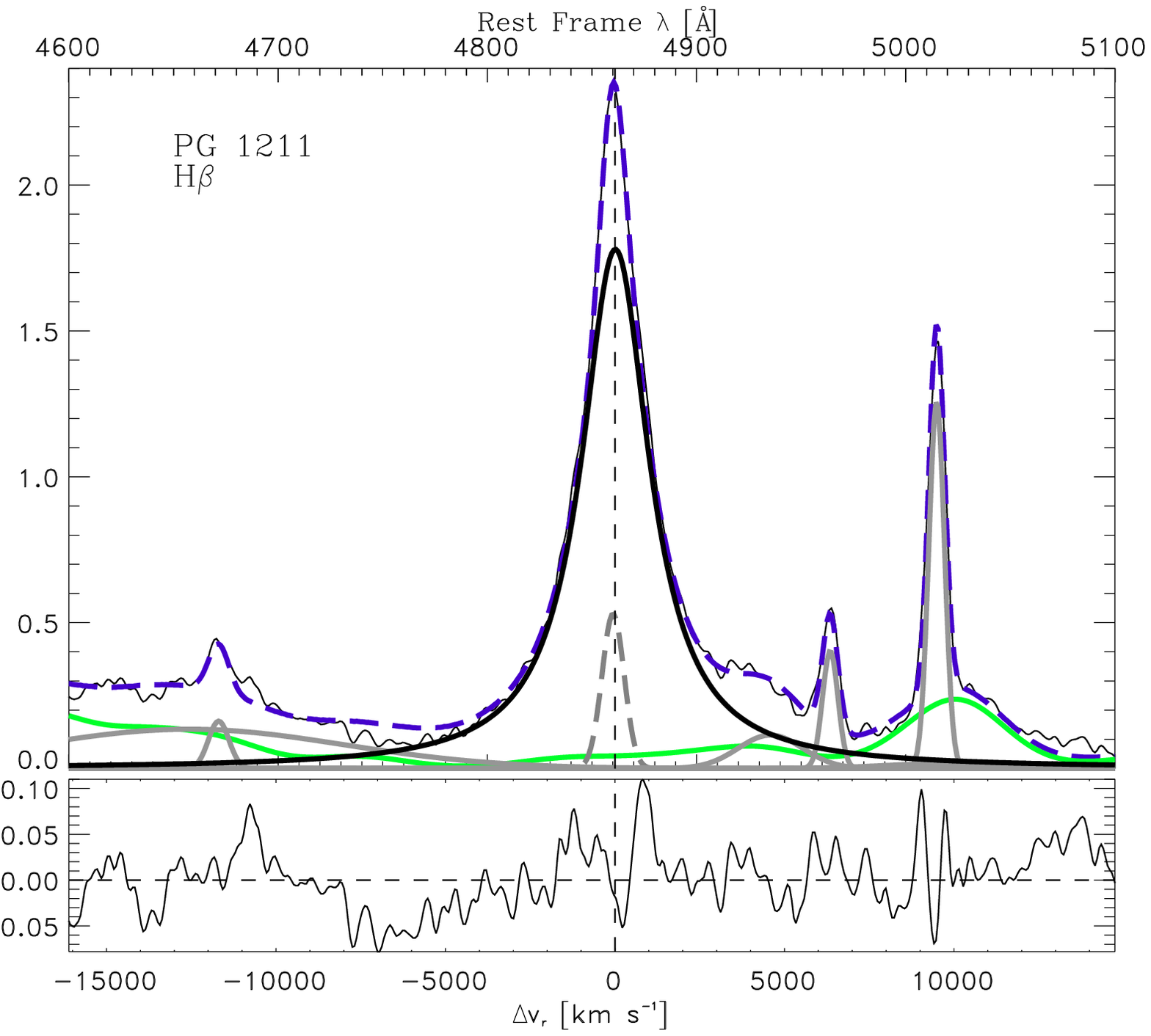}\\
\includegraphics[scale=0.35]{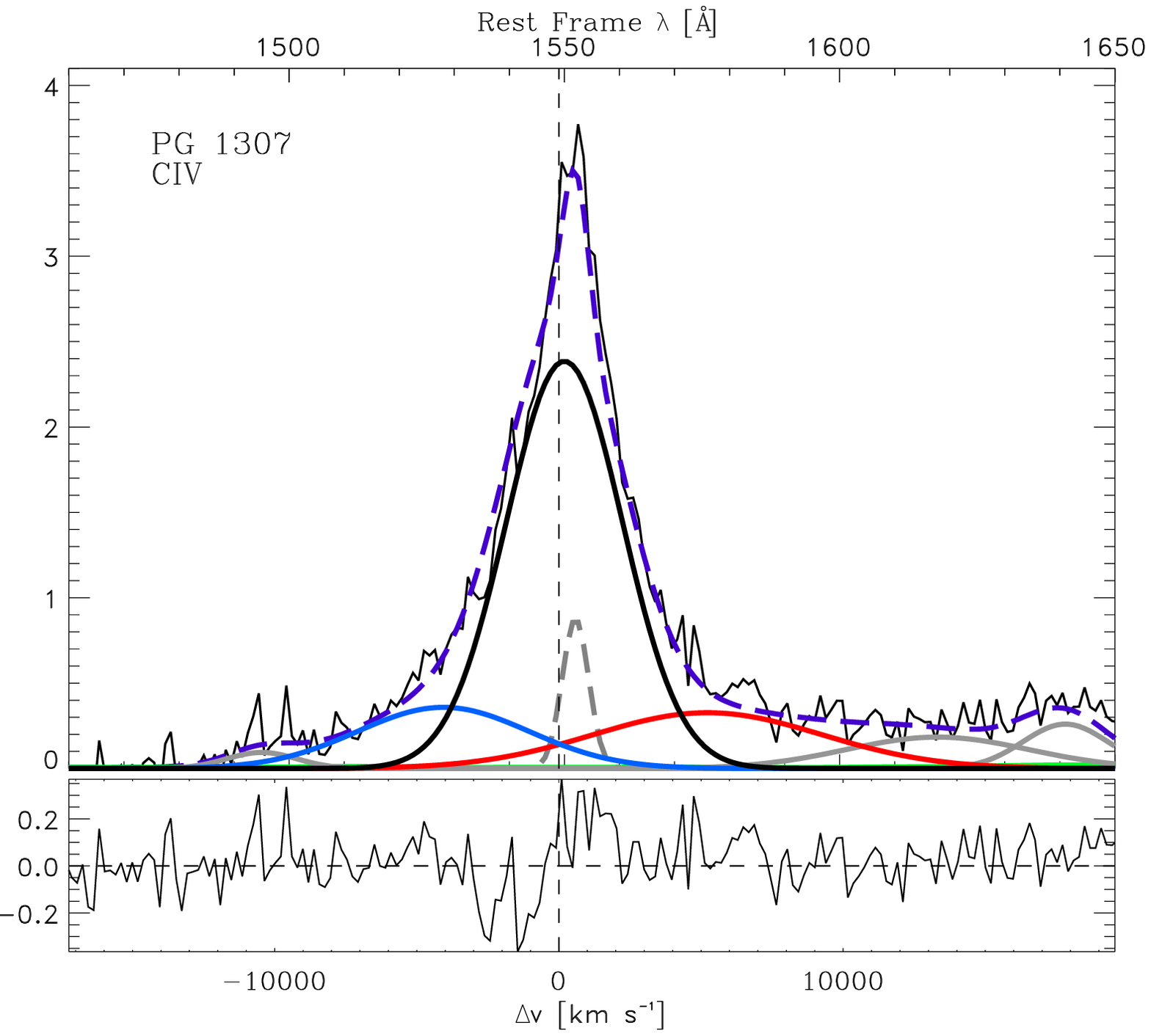}\includegraphics[scale=0.35]{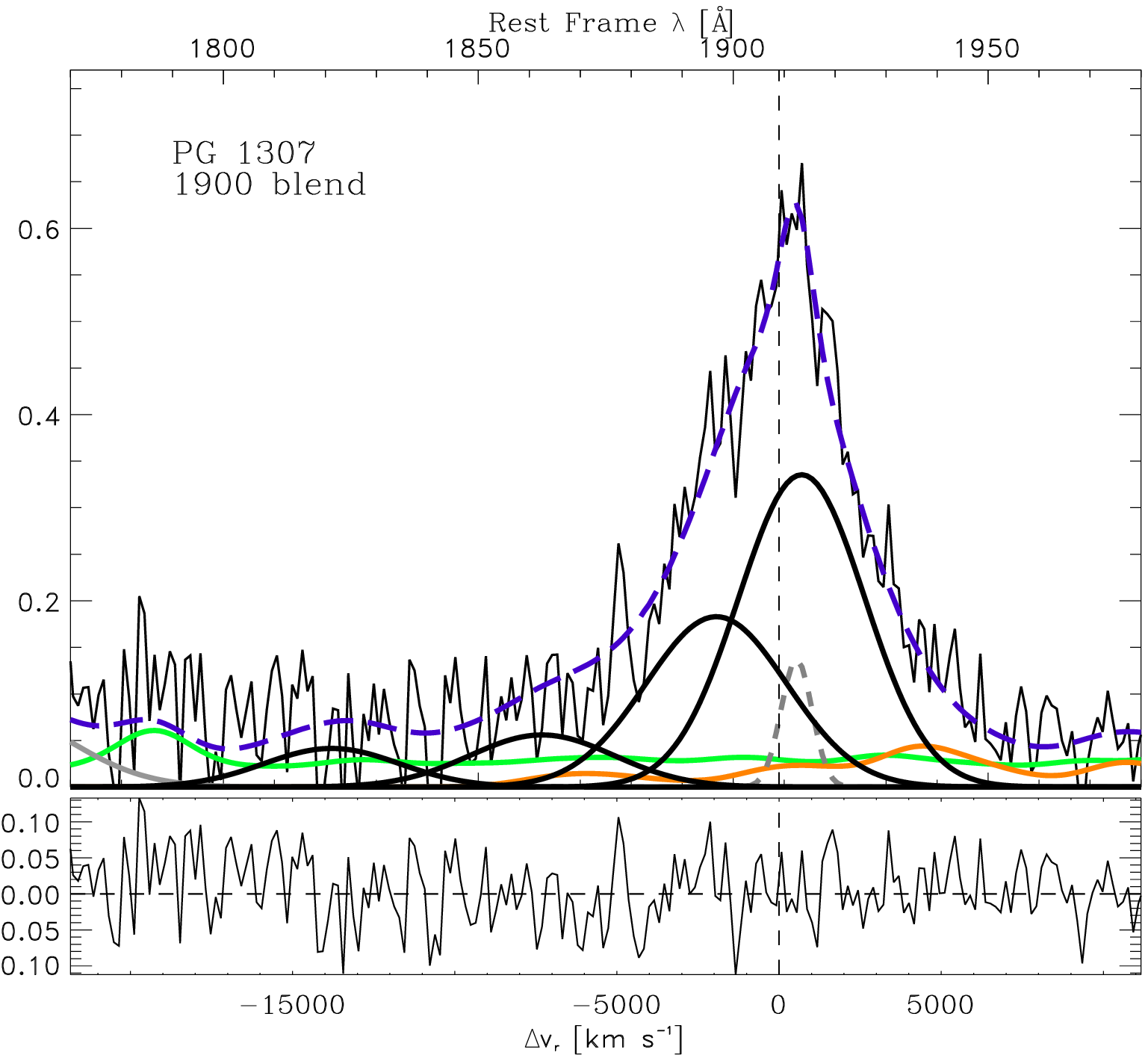}\includegraphics[scale=0.35]{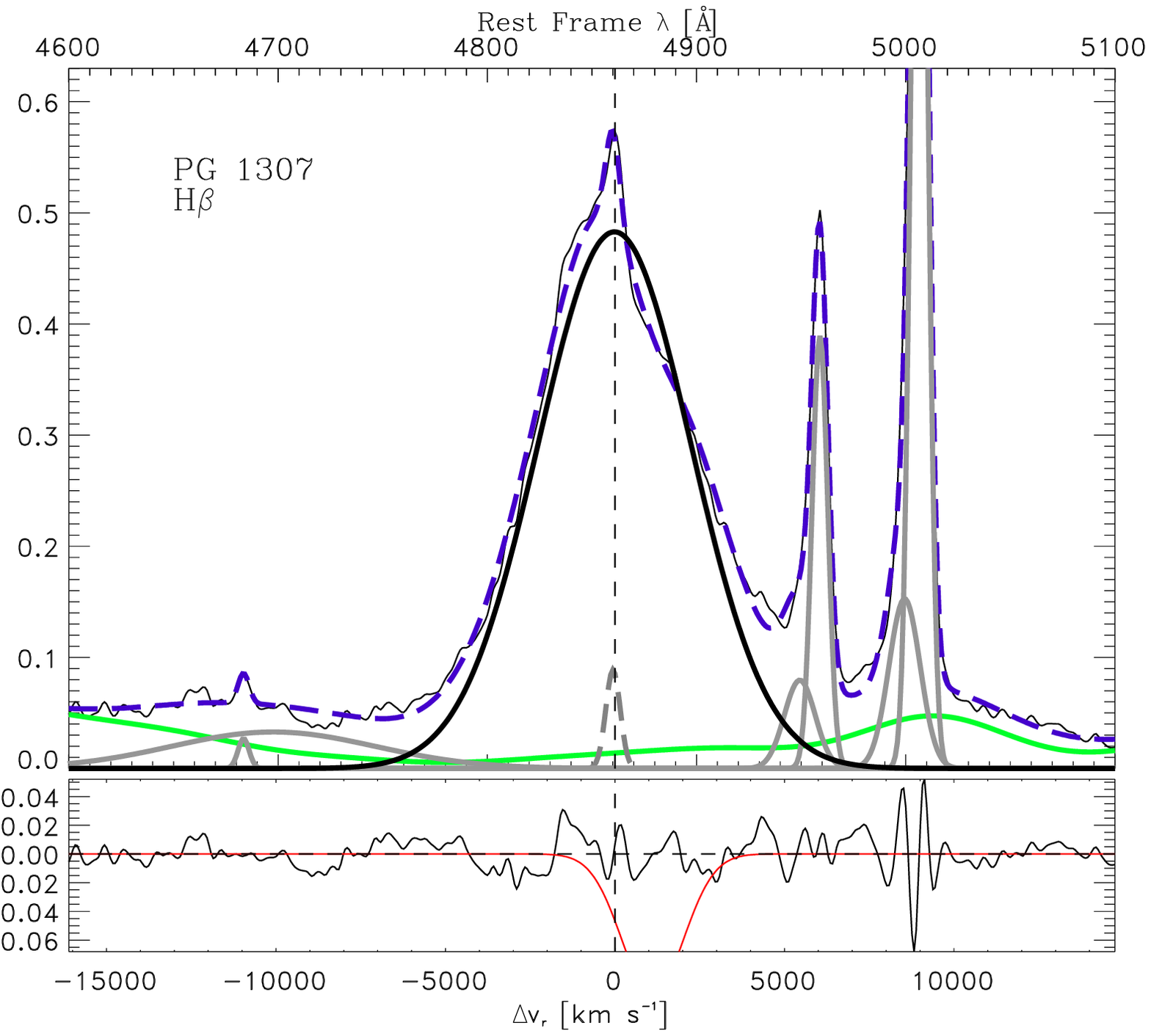}
\centerline{Fig. 1. --- Continued.}
\clearpage
\includegraphics[scale=0.35]{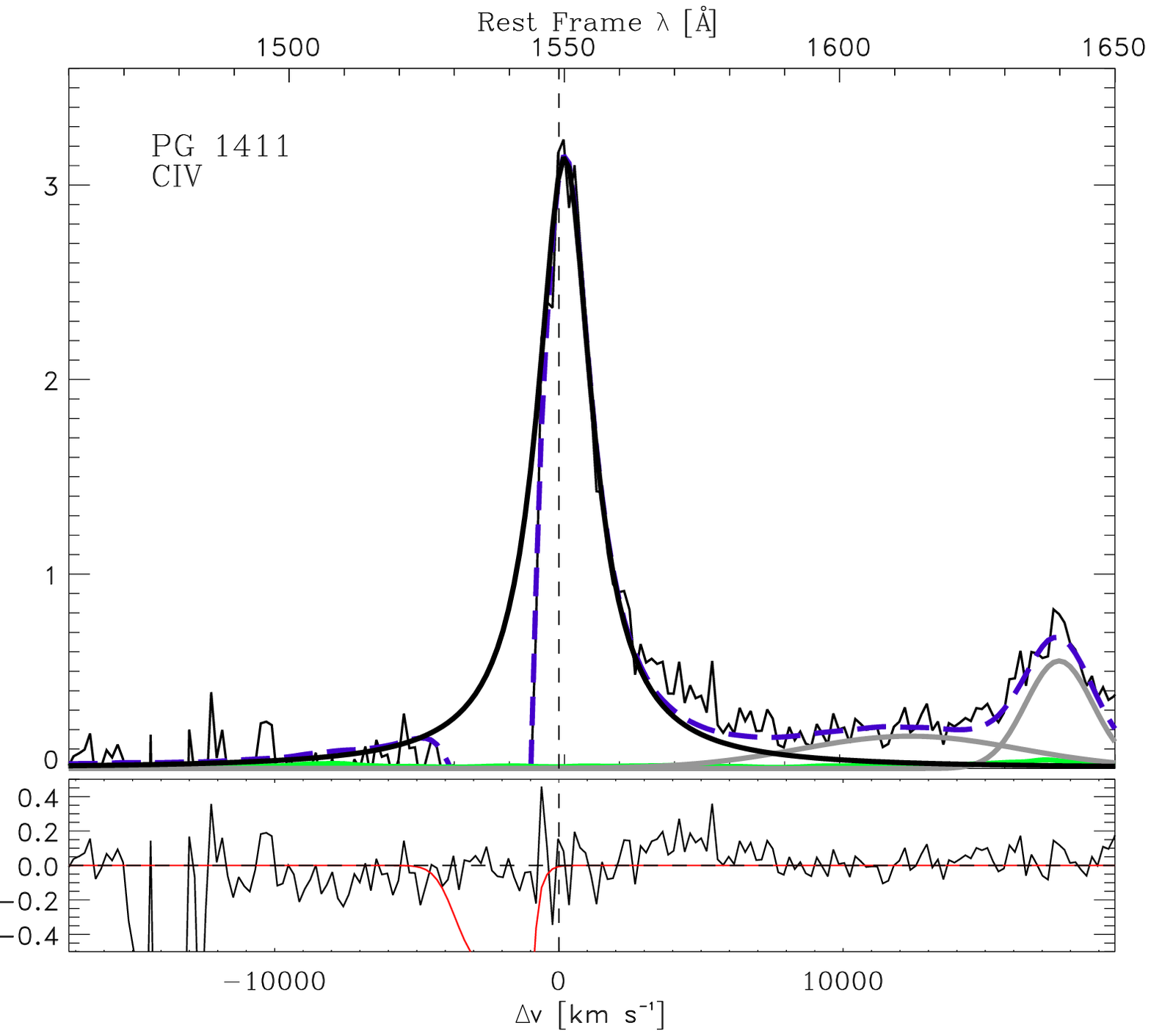}\includegraphics[scale=0.35]{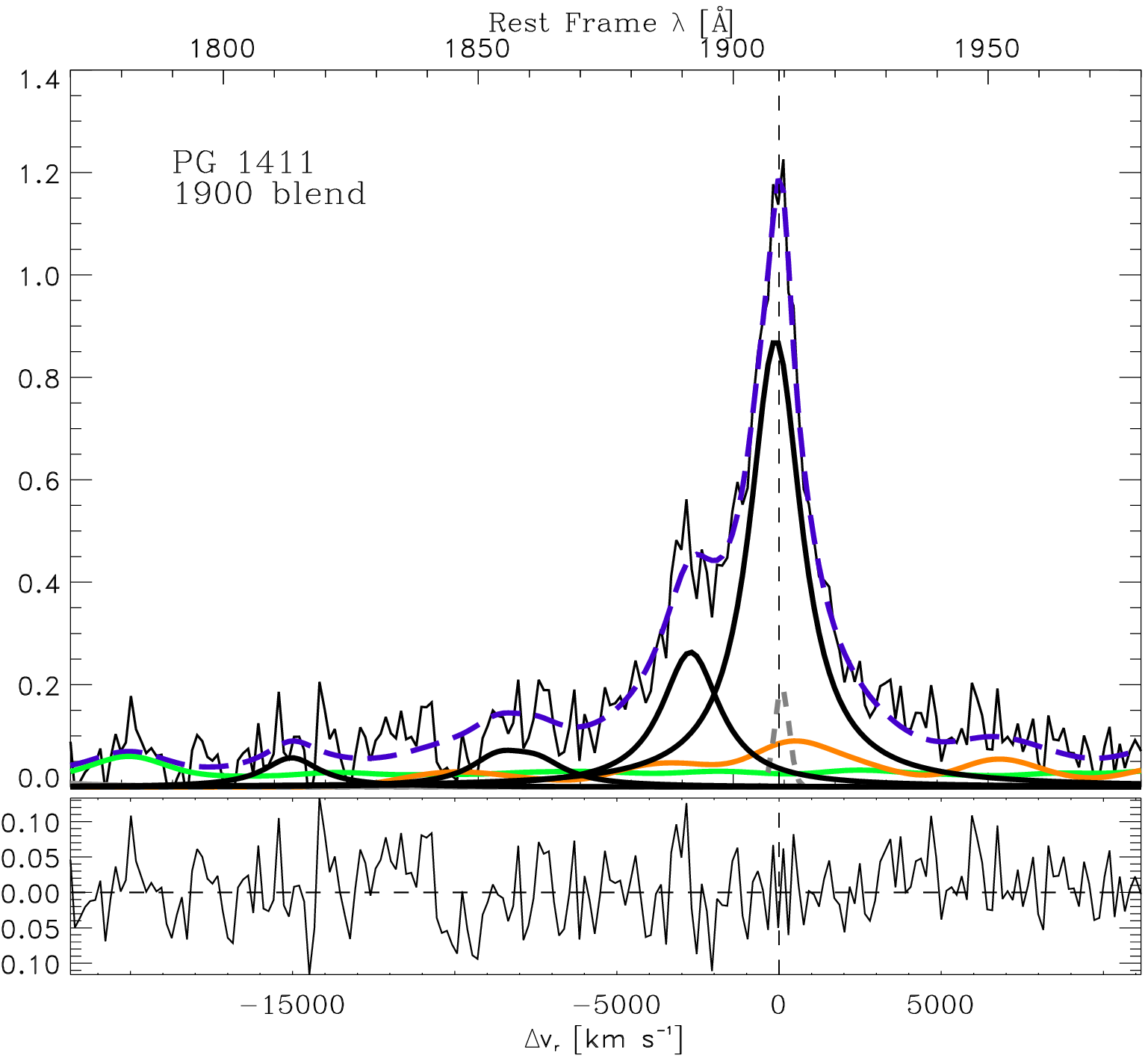}\includegraphics[scale=0.35]{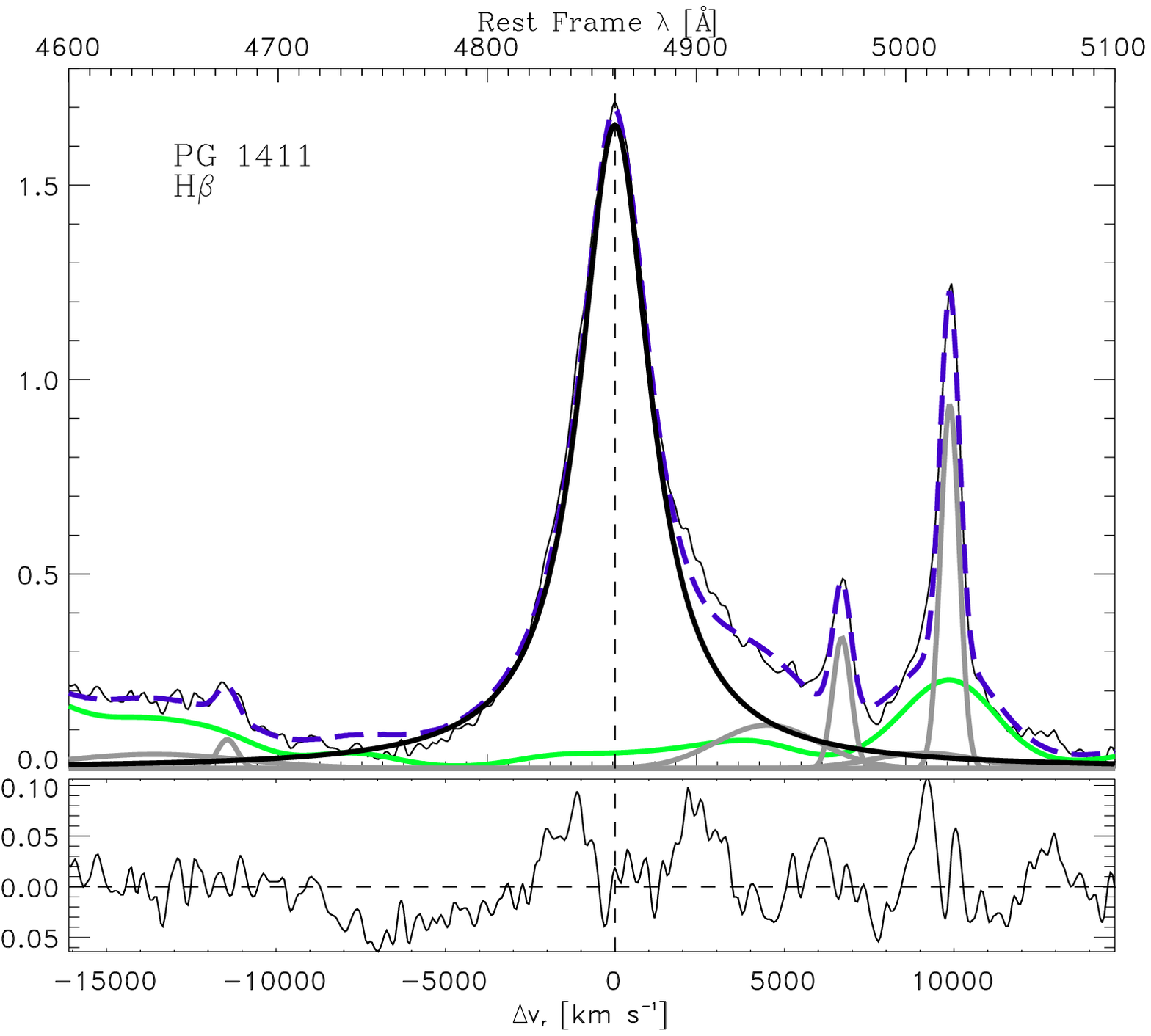}\\
\centerline{Fig. 1. --- Continued.}

\clearpage
\includegraphics[scale=0.35]{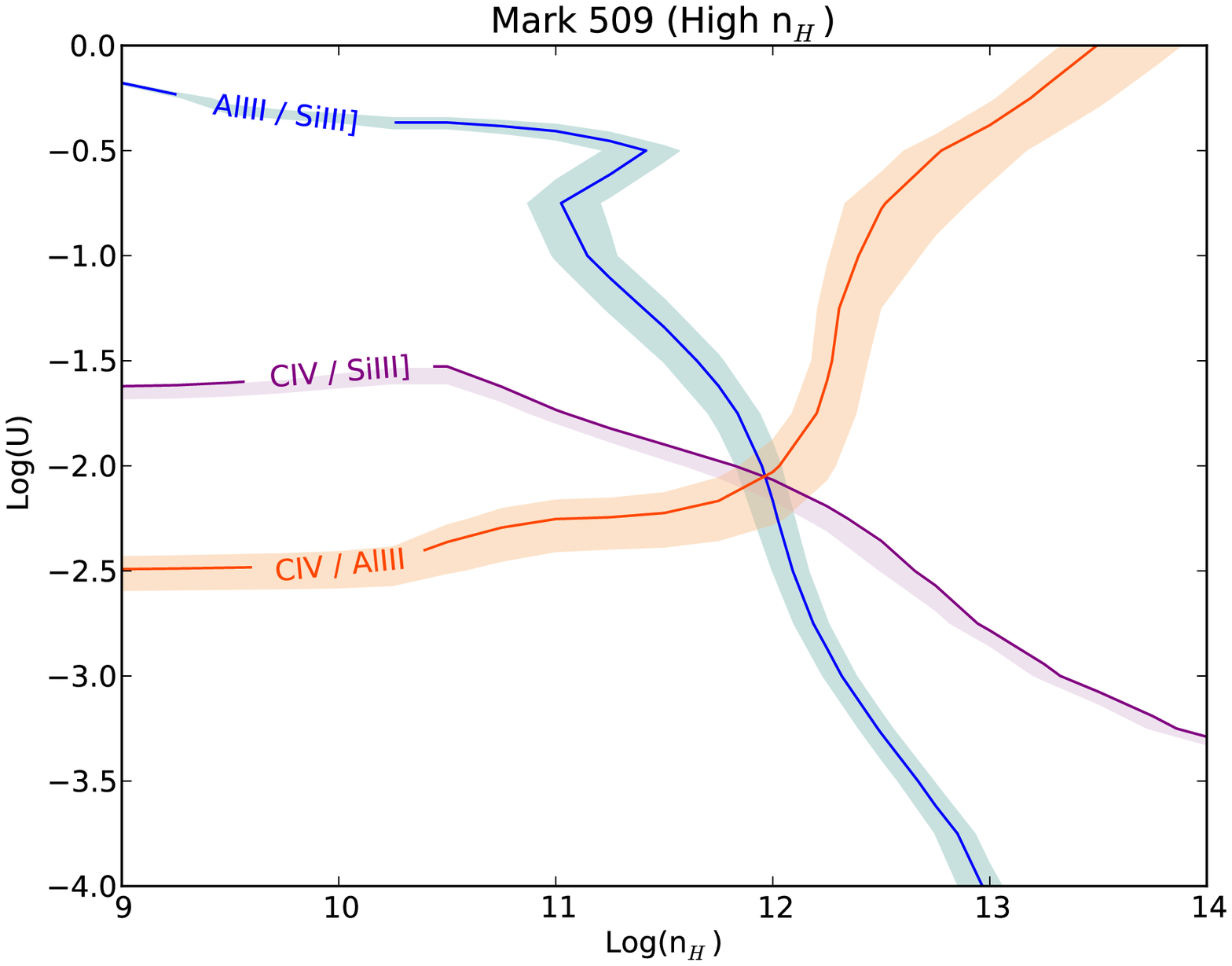}\includegraphics[scale=0.35]{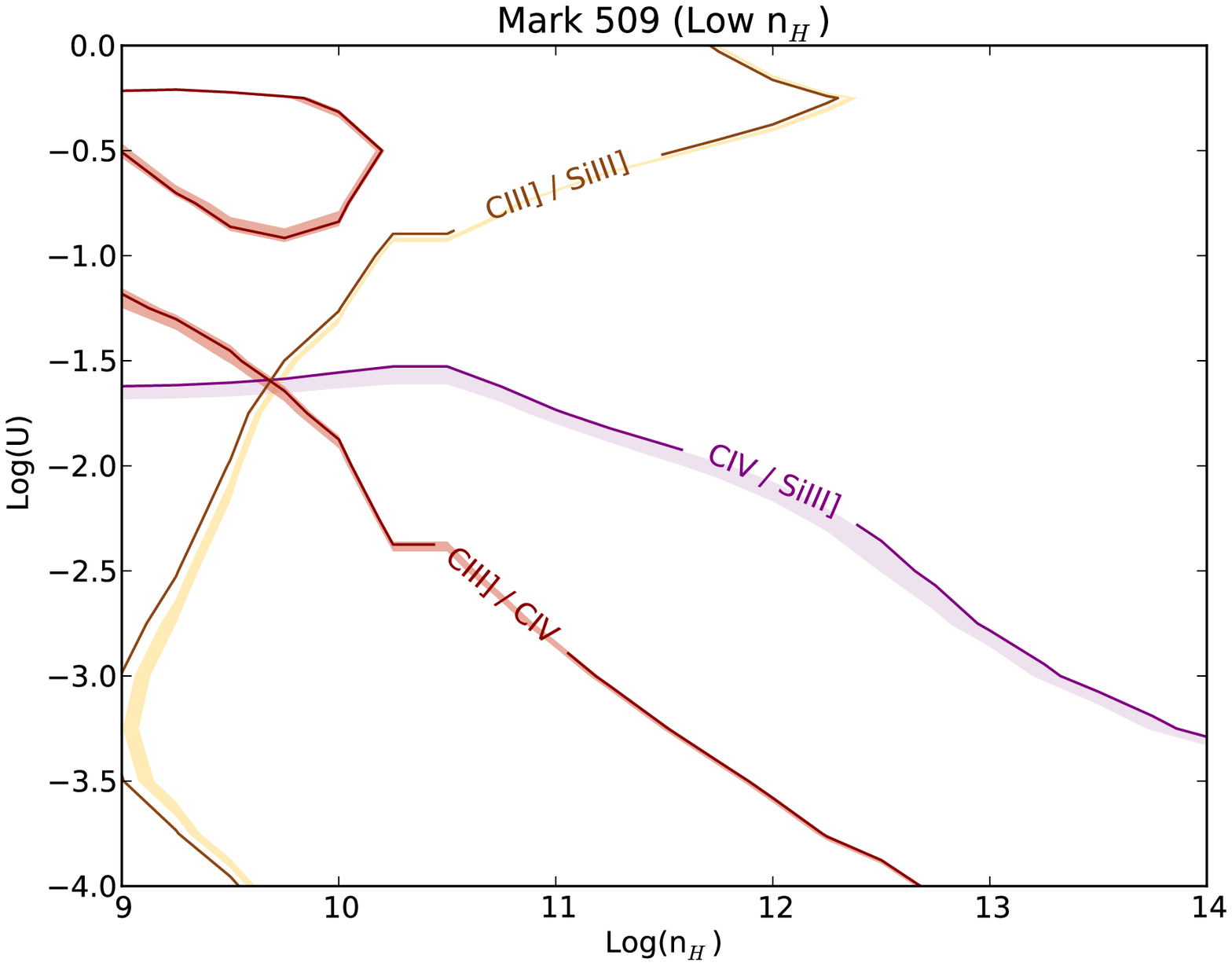}\\
\includegraphics[scale=0.35]{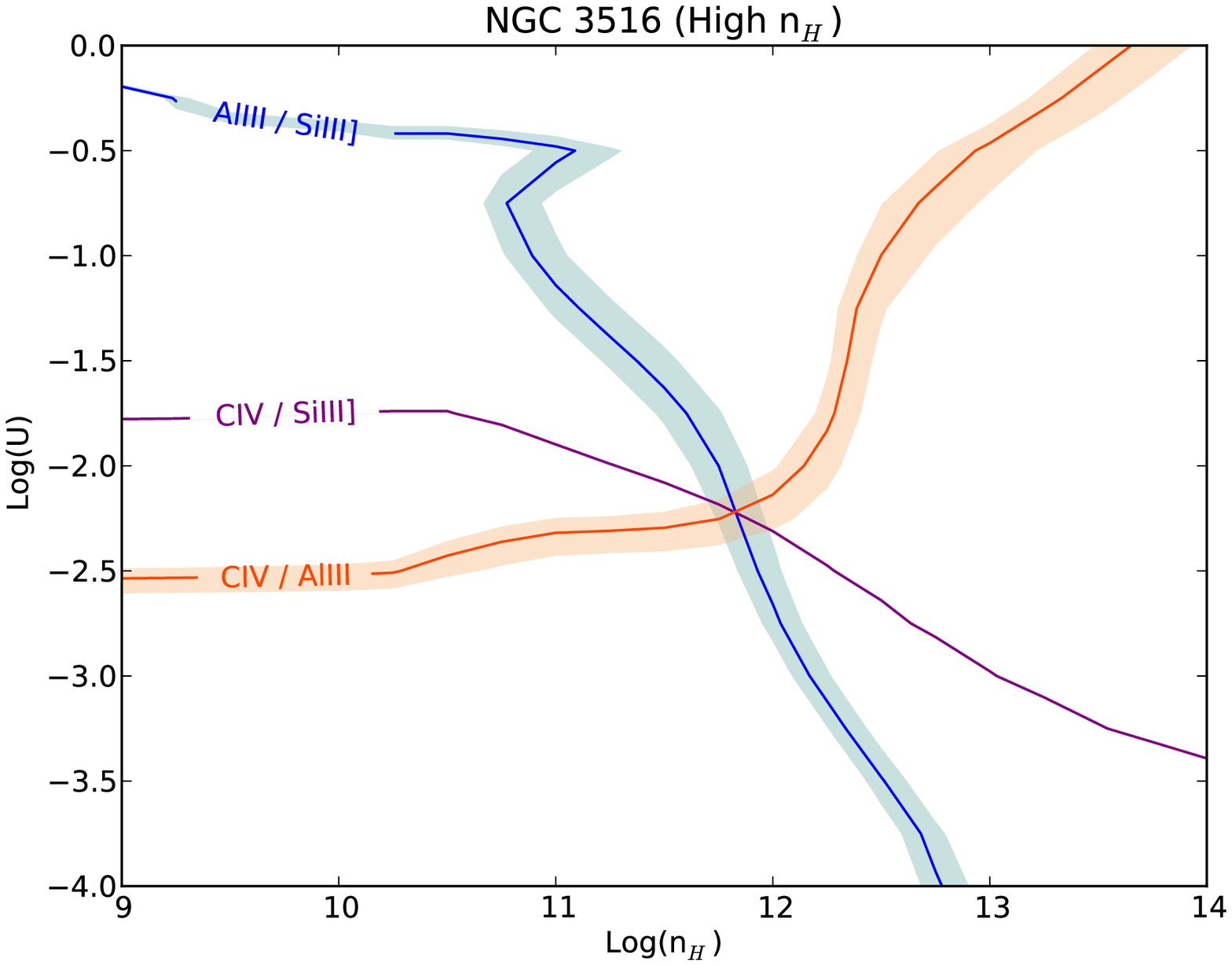}\includegraphics[scale=0.35]{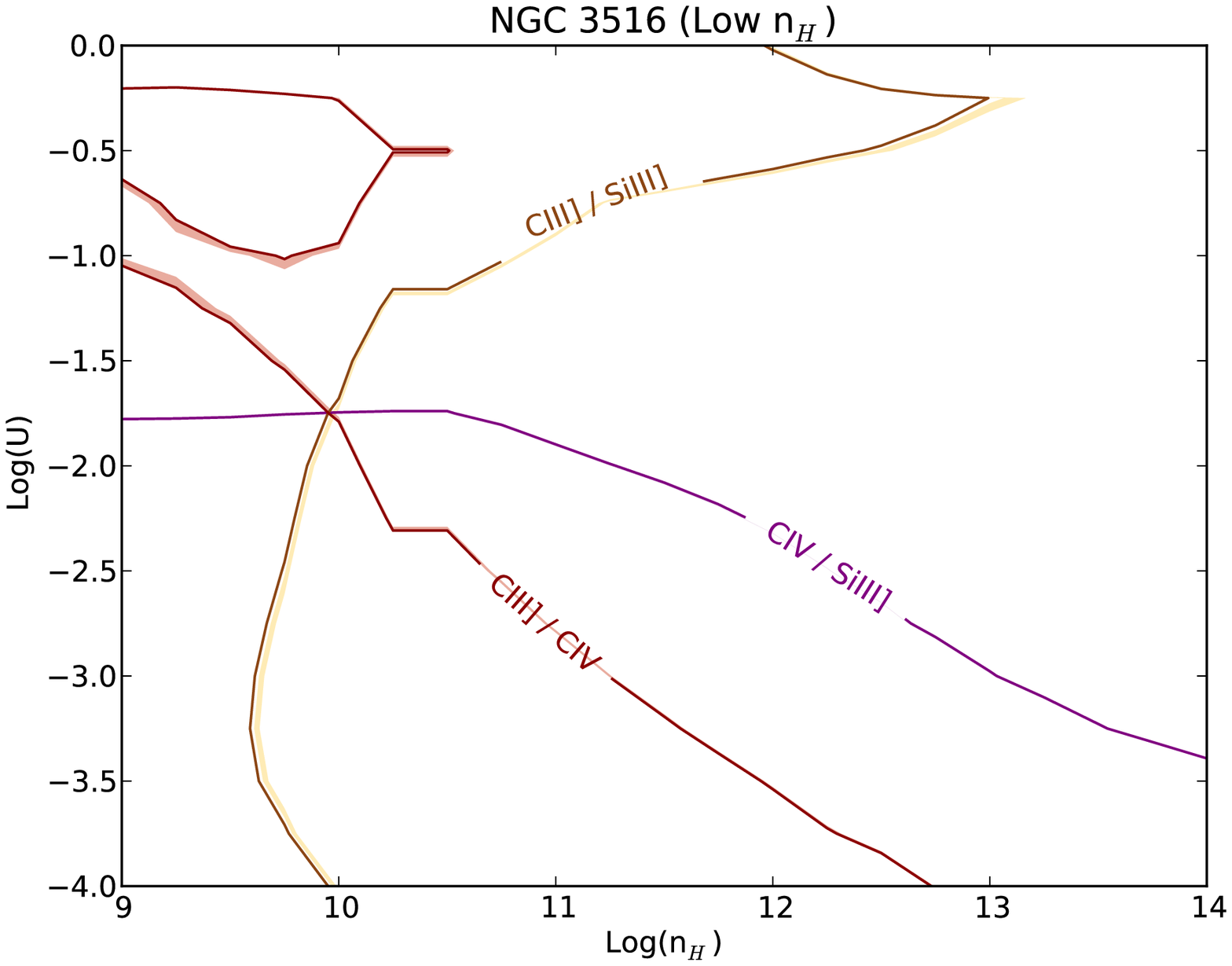}\\
\includegraphics[scale=0.35]{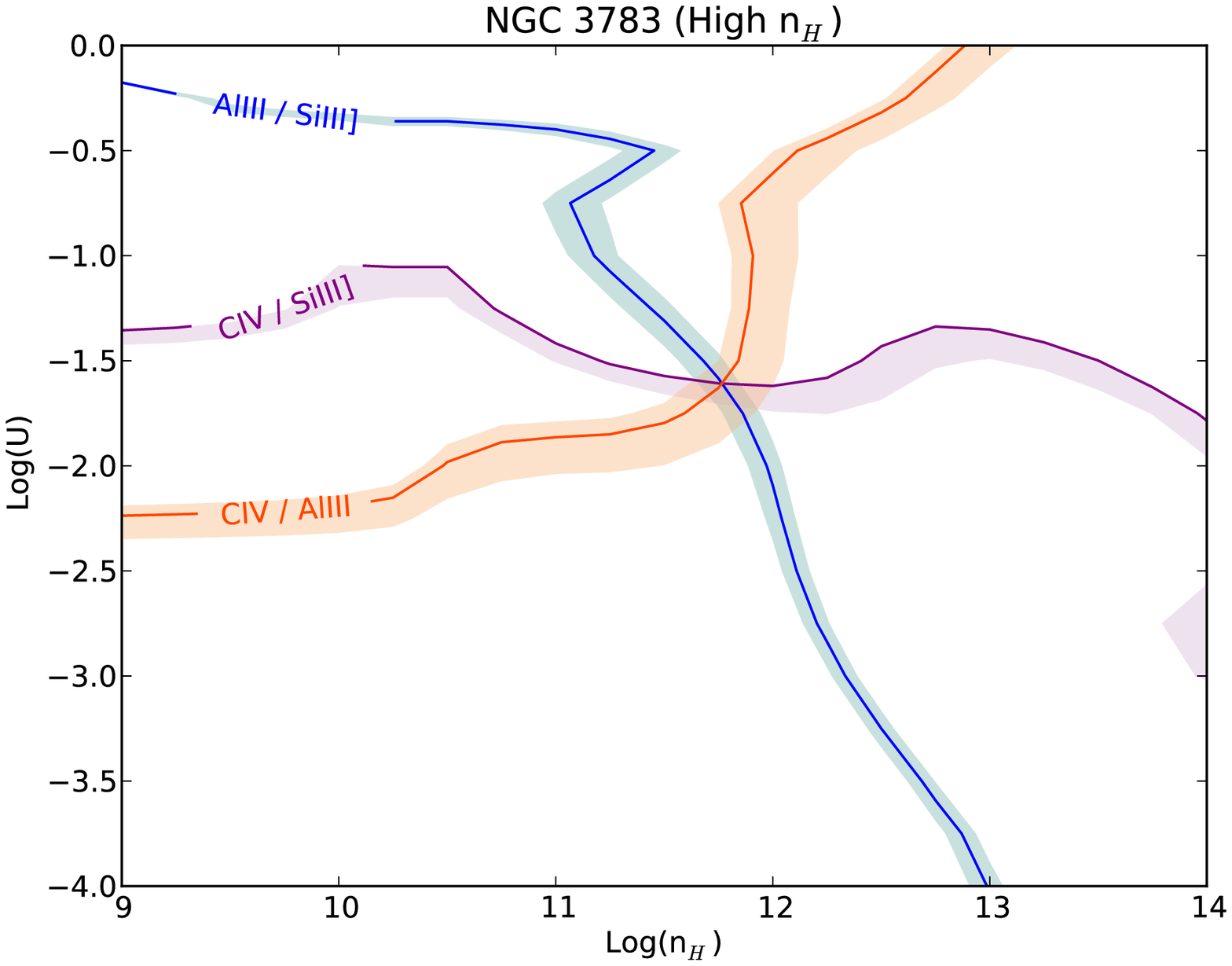}\includegraphics[scale=0.35]{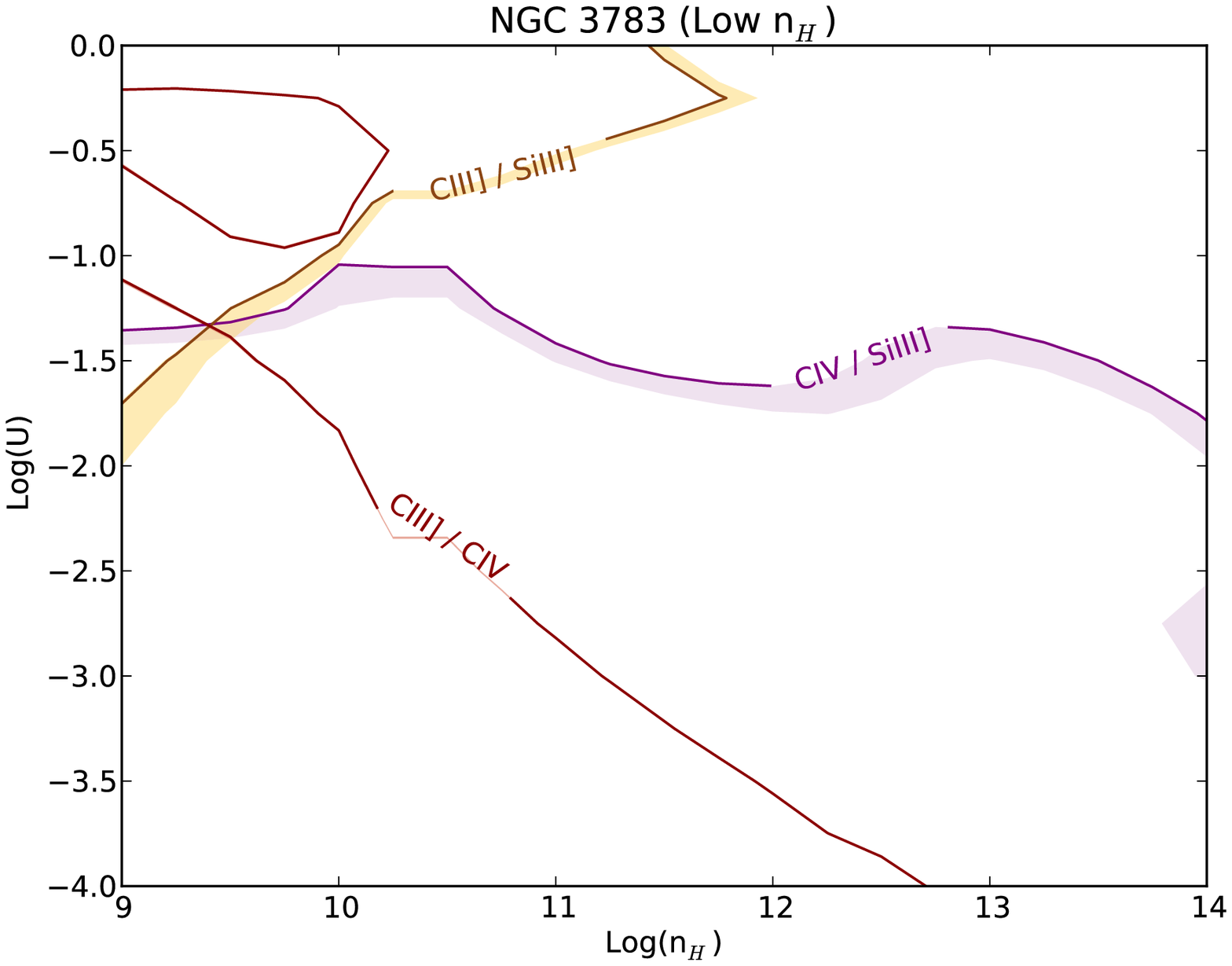}\\
\centerline{Fig. 2. --- Continued.}
\clearpage
\includegraphics[scale=0.35]{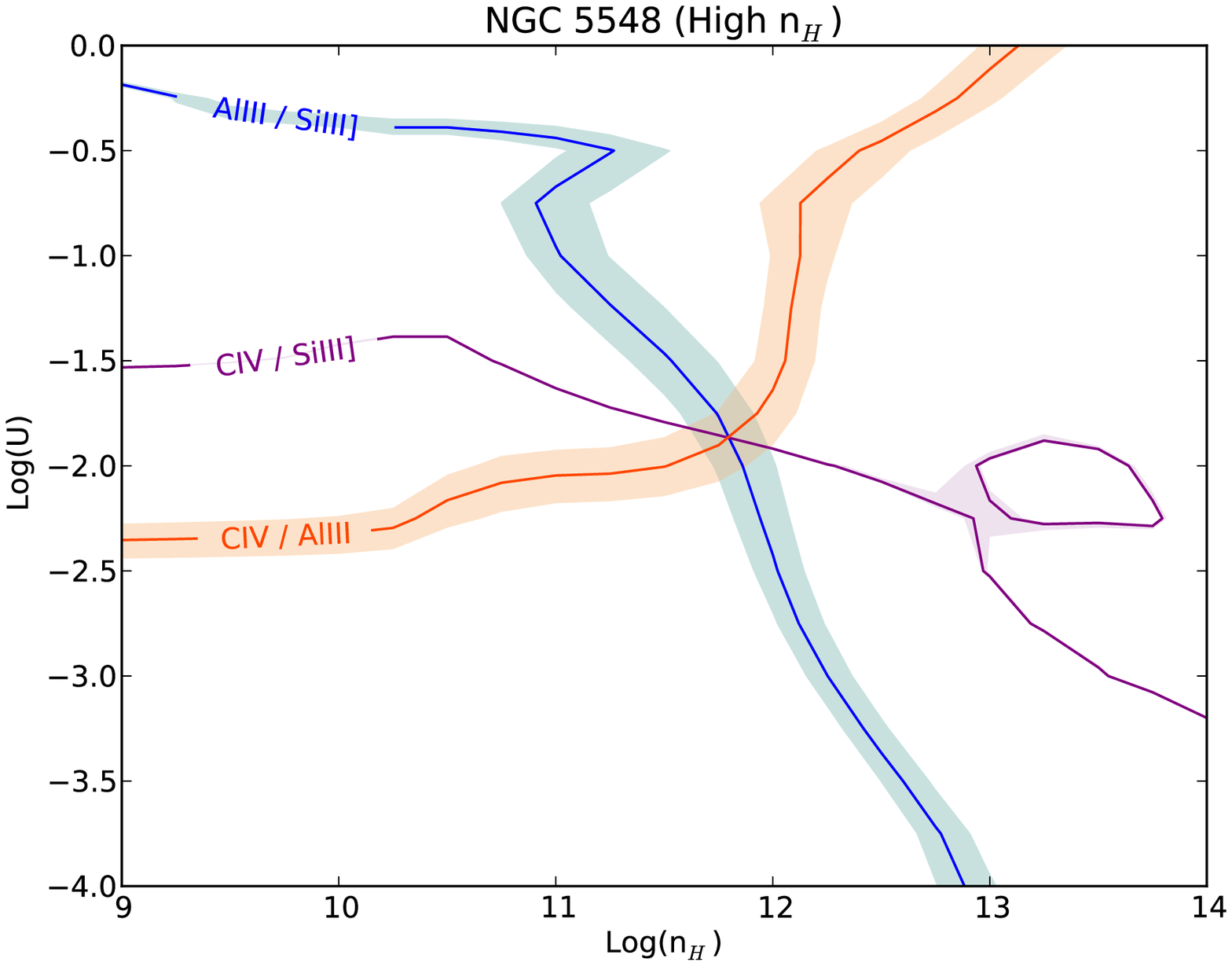}\includegraphics[scale=0.35]{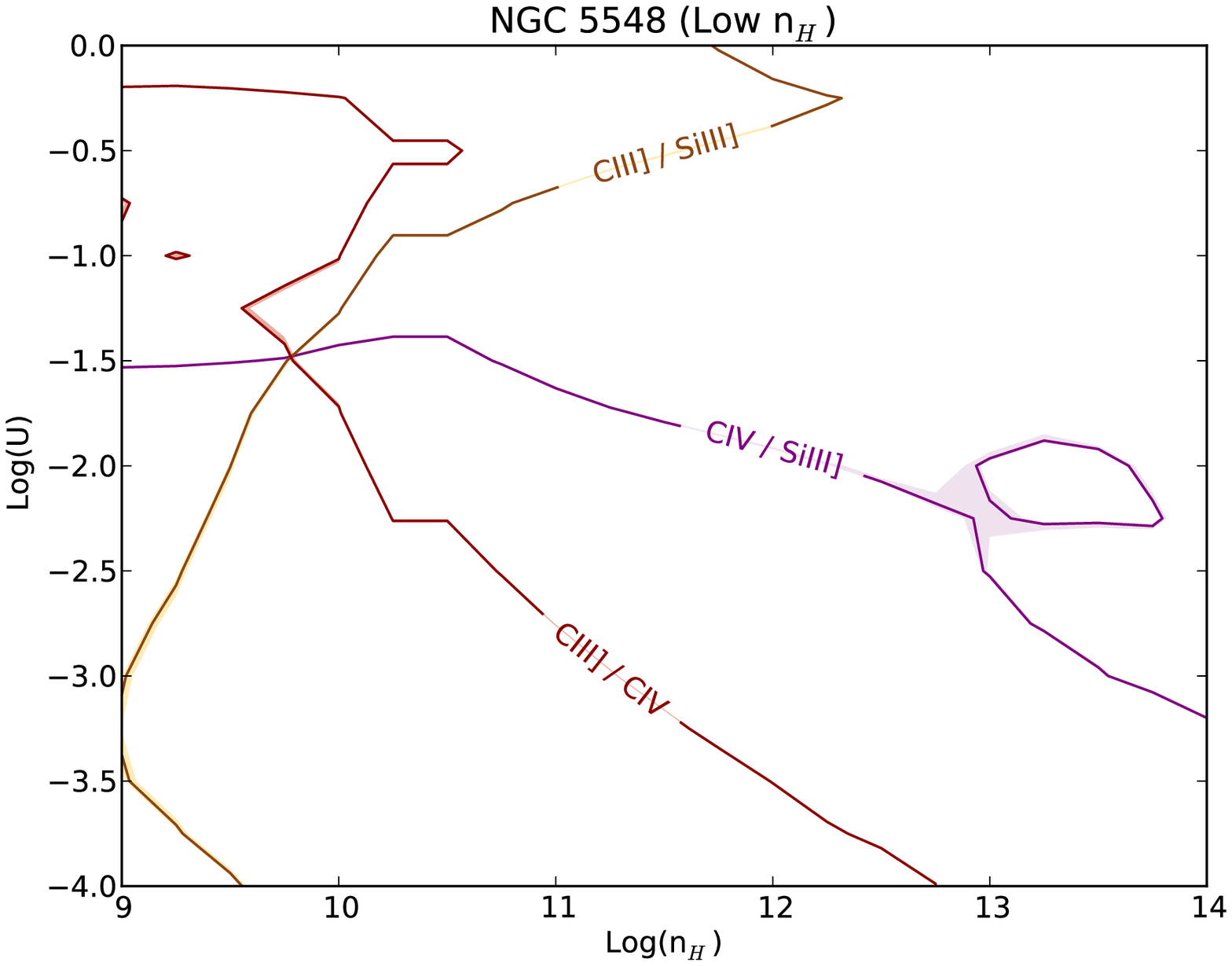}\\
\includegraphics[scale=0.35]{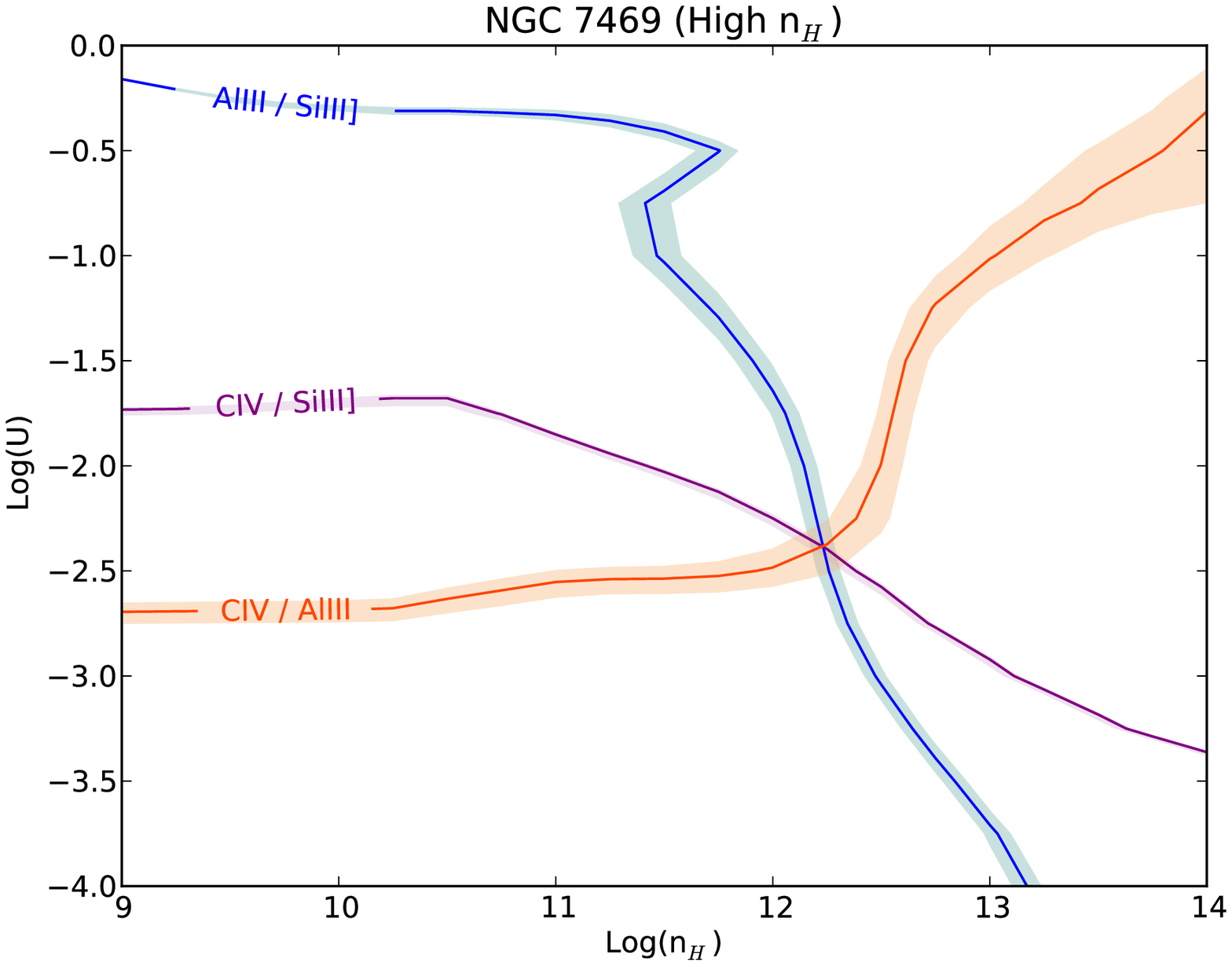}\includegraphics[scale=0.35]{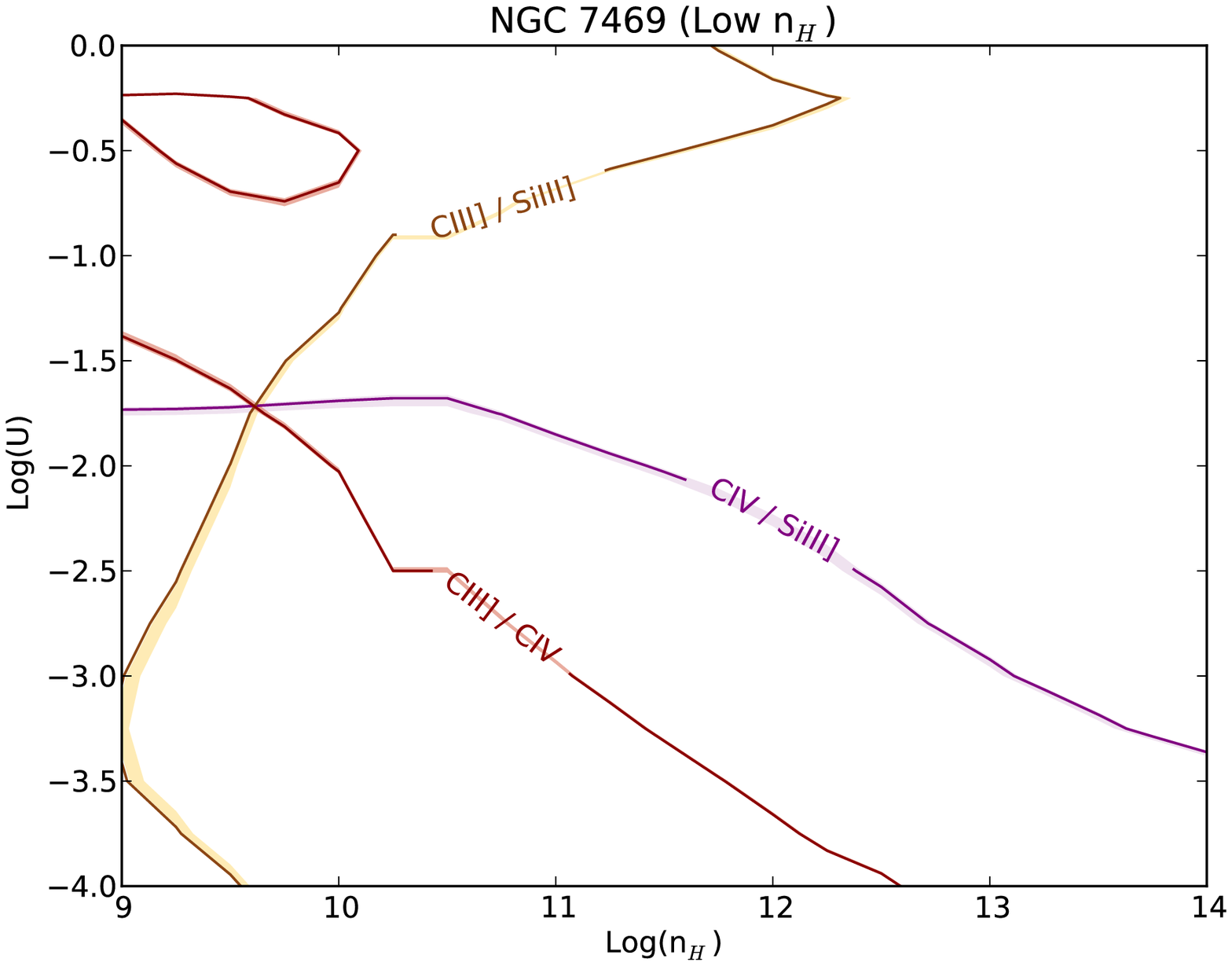}\\
\includegraphics[scale=0.35]{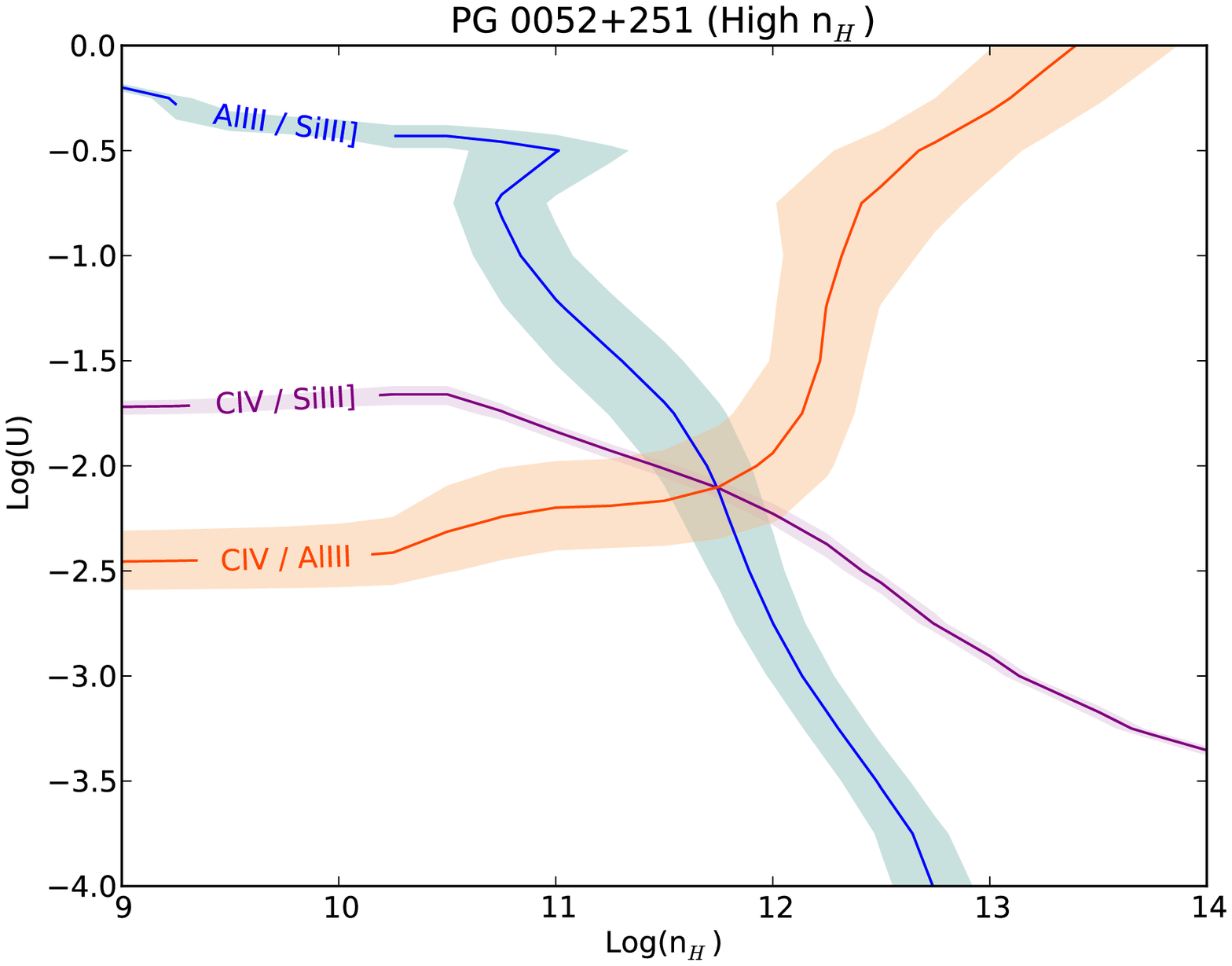}\includegraphics[scale=0.35]{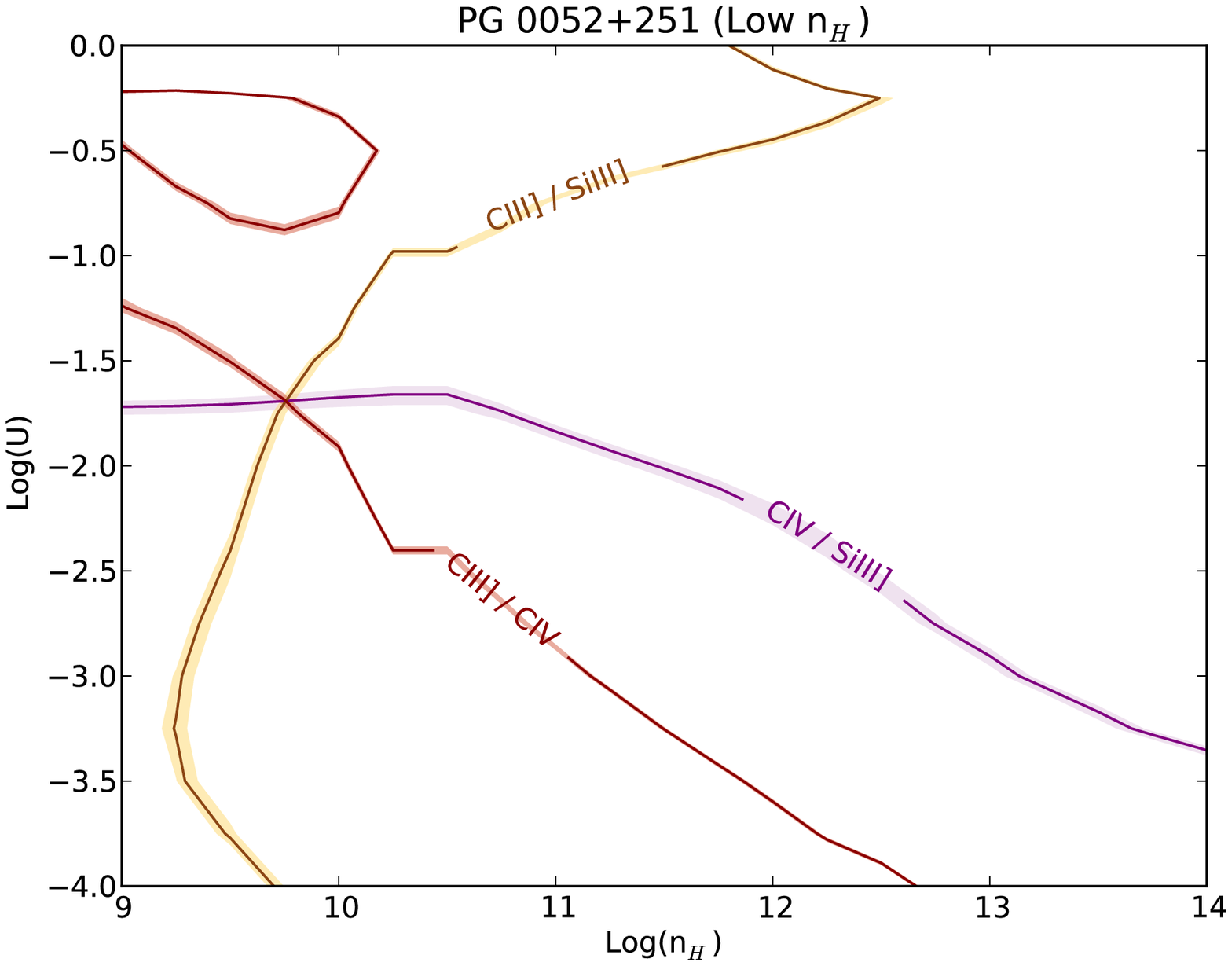}\\
\centerline{Fig. 2. --- Continued.}
\clearpage
\includegraphics[scale=0.35]{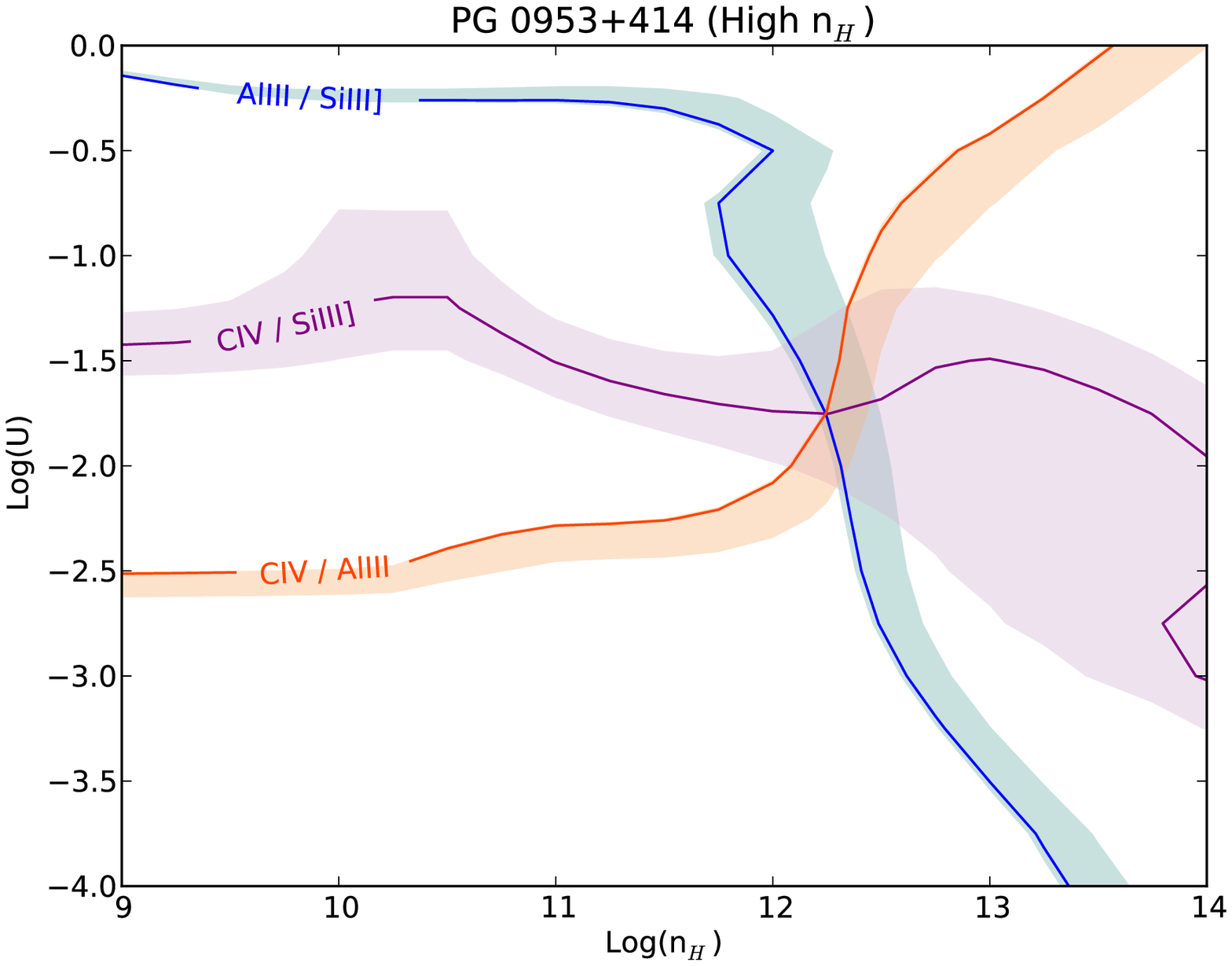}\includegraphics[scale=0.35]{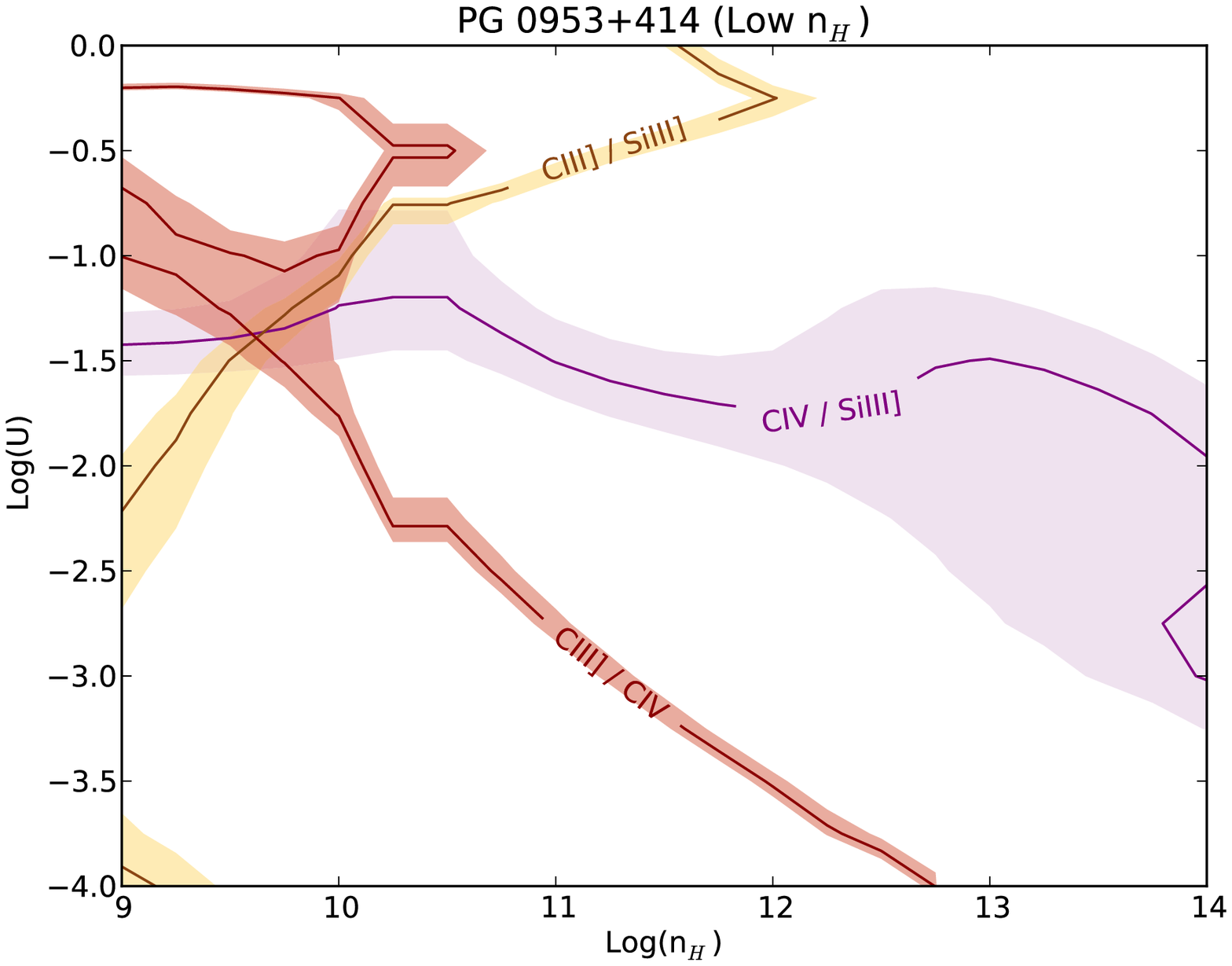}\\
\includegraphics[scale=0.35]{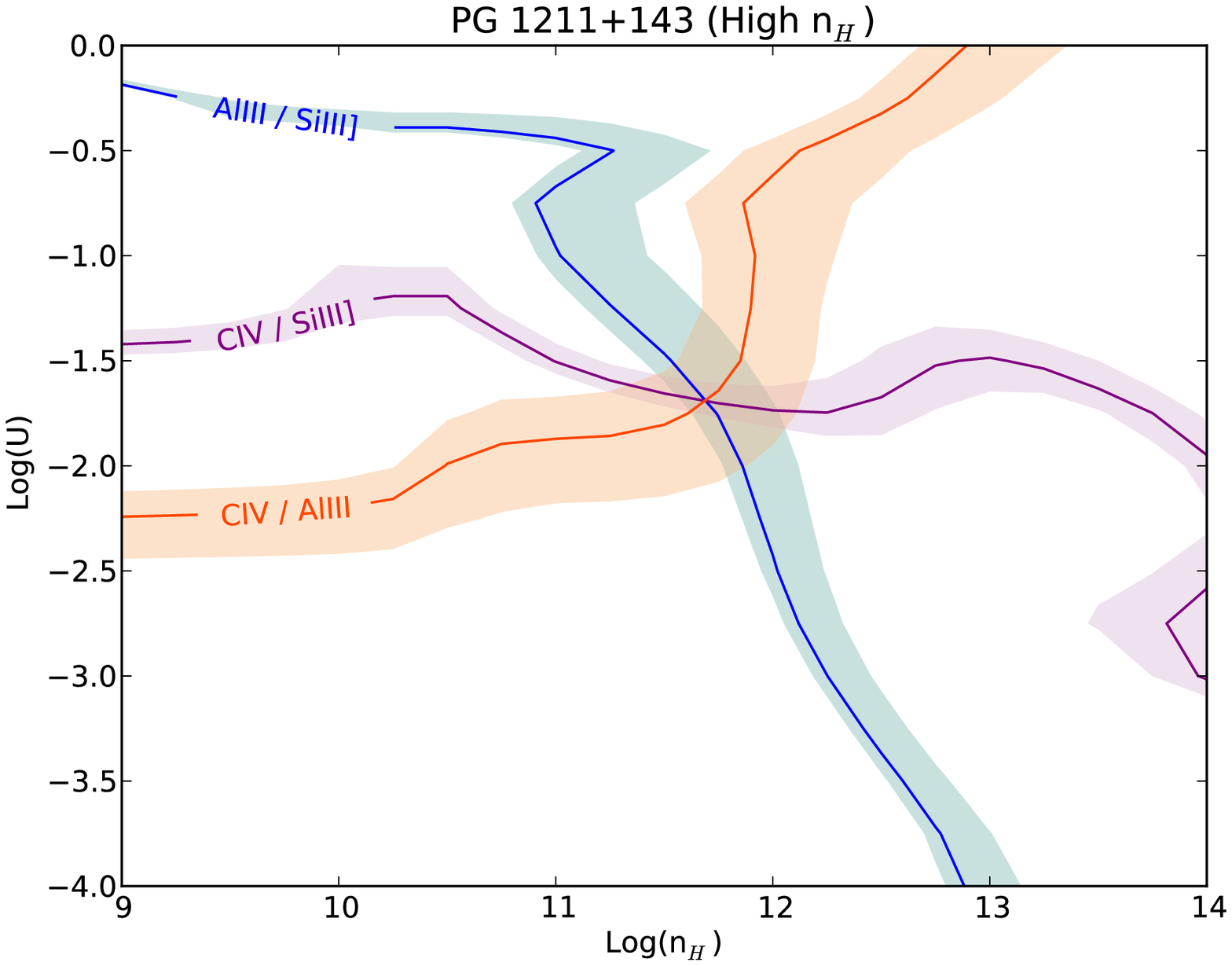}\includegraphics[scale=0.35]{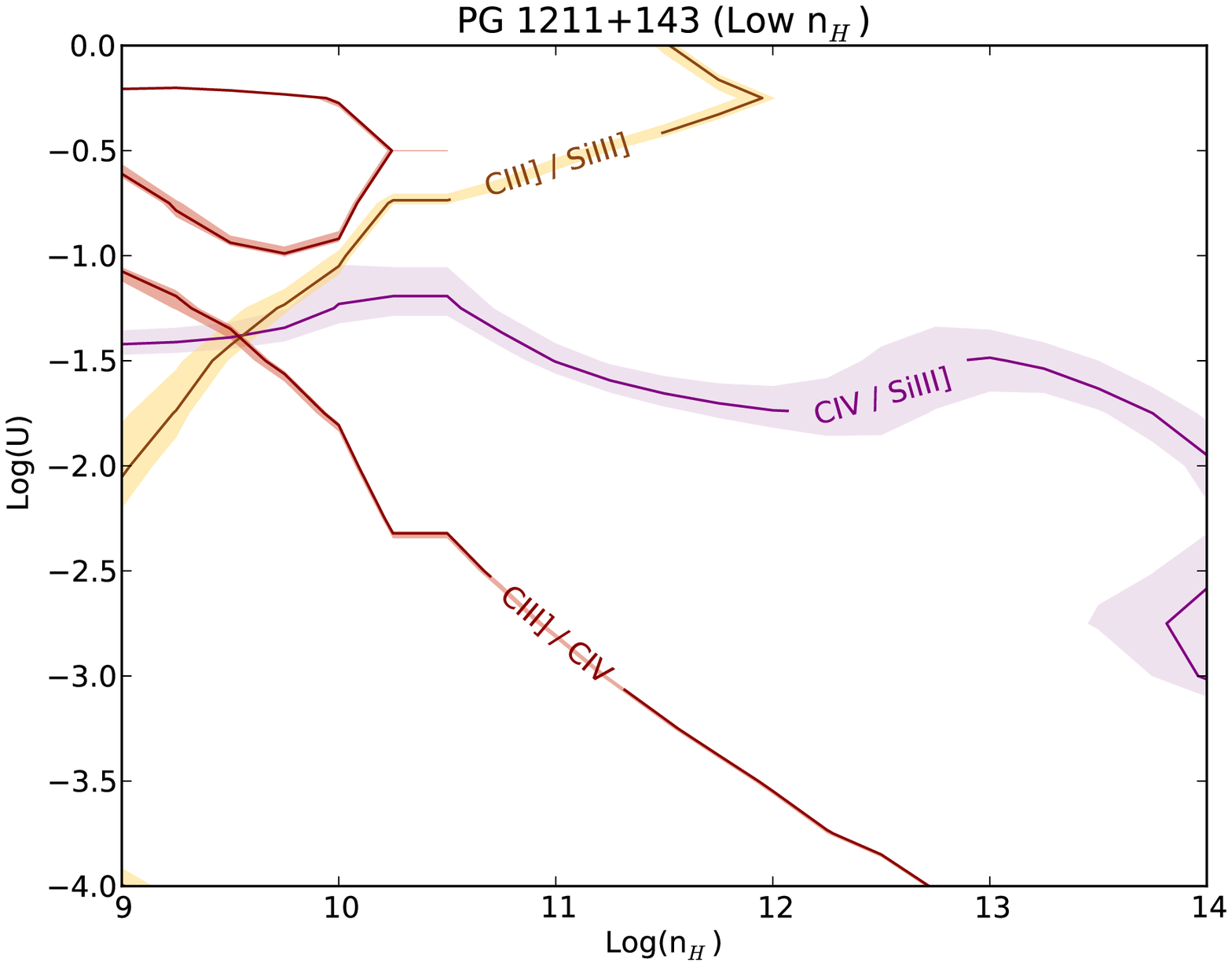}\\
\includegraphics[scale=0.35]{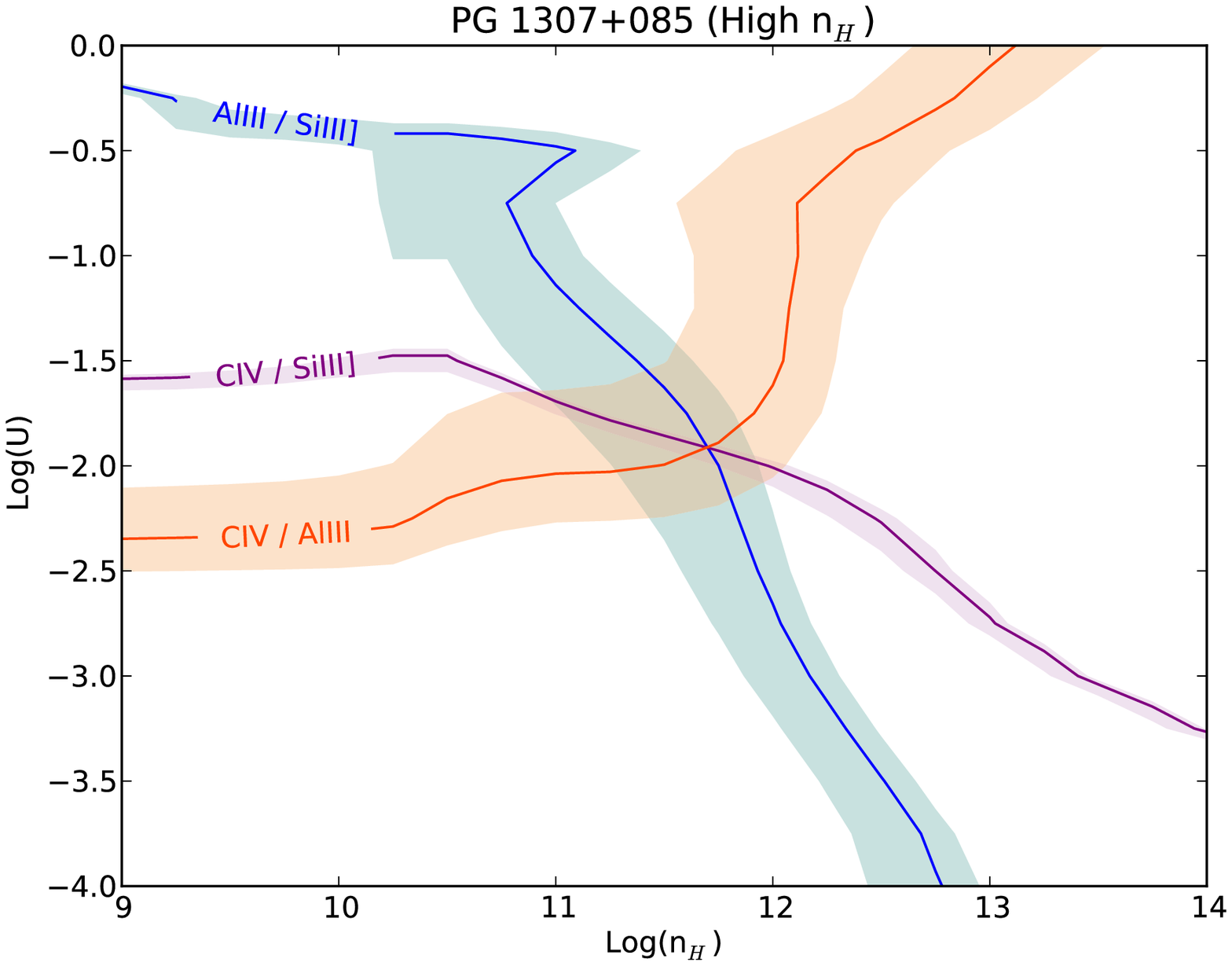}\includegraphics[scale=0.35]{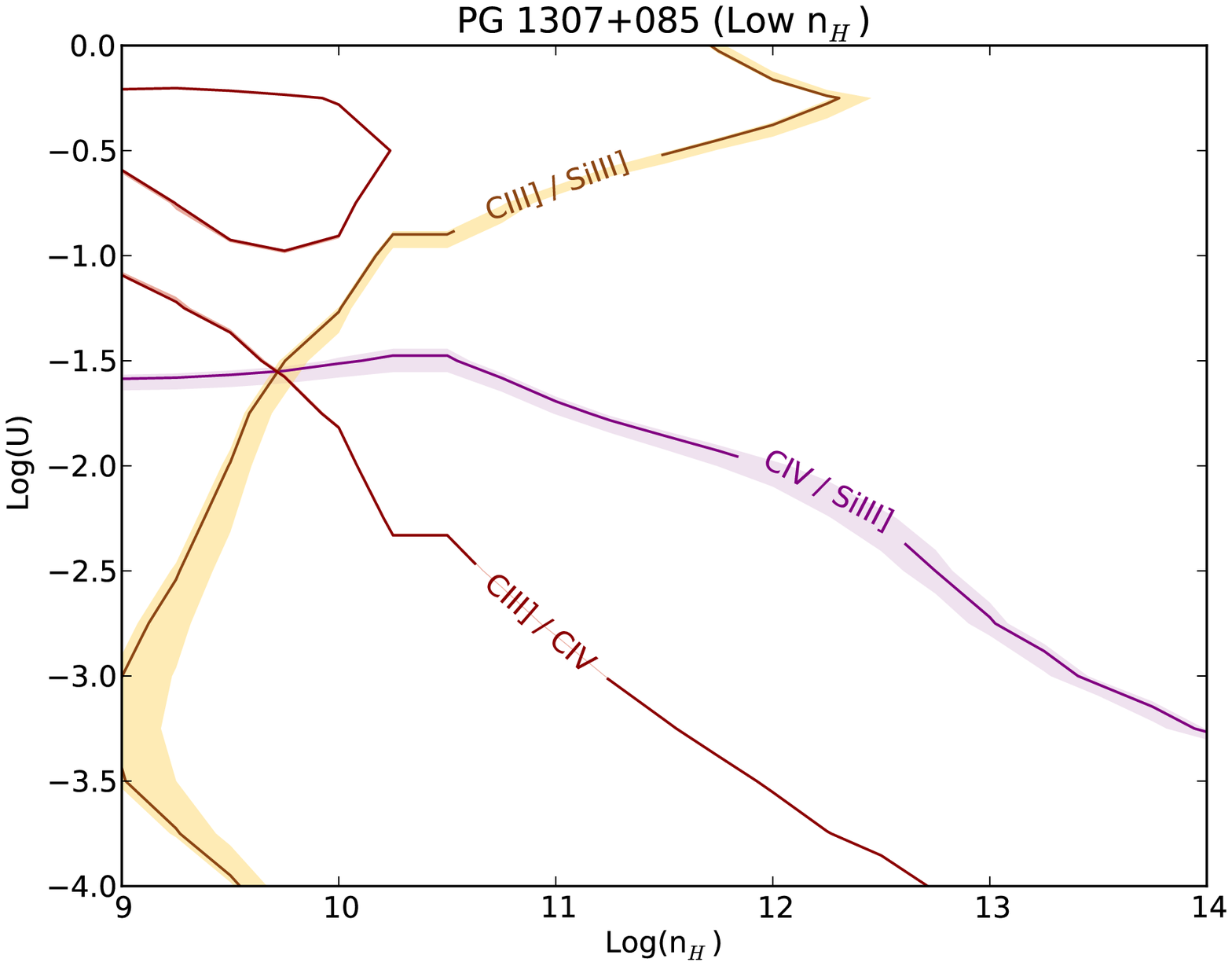}\\
\centerline{Fig. 2. --- Continued.}
\clearpage
\includegraphics[scale=0.35]{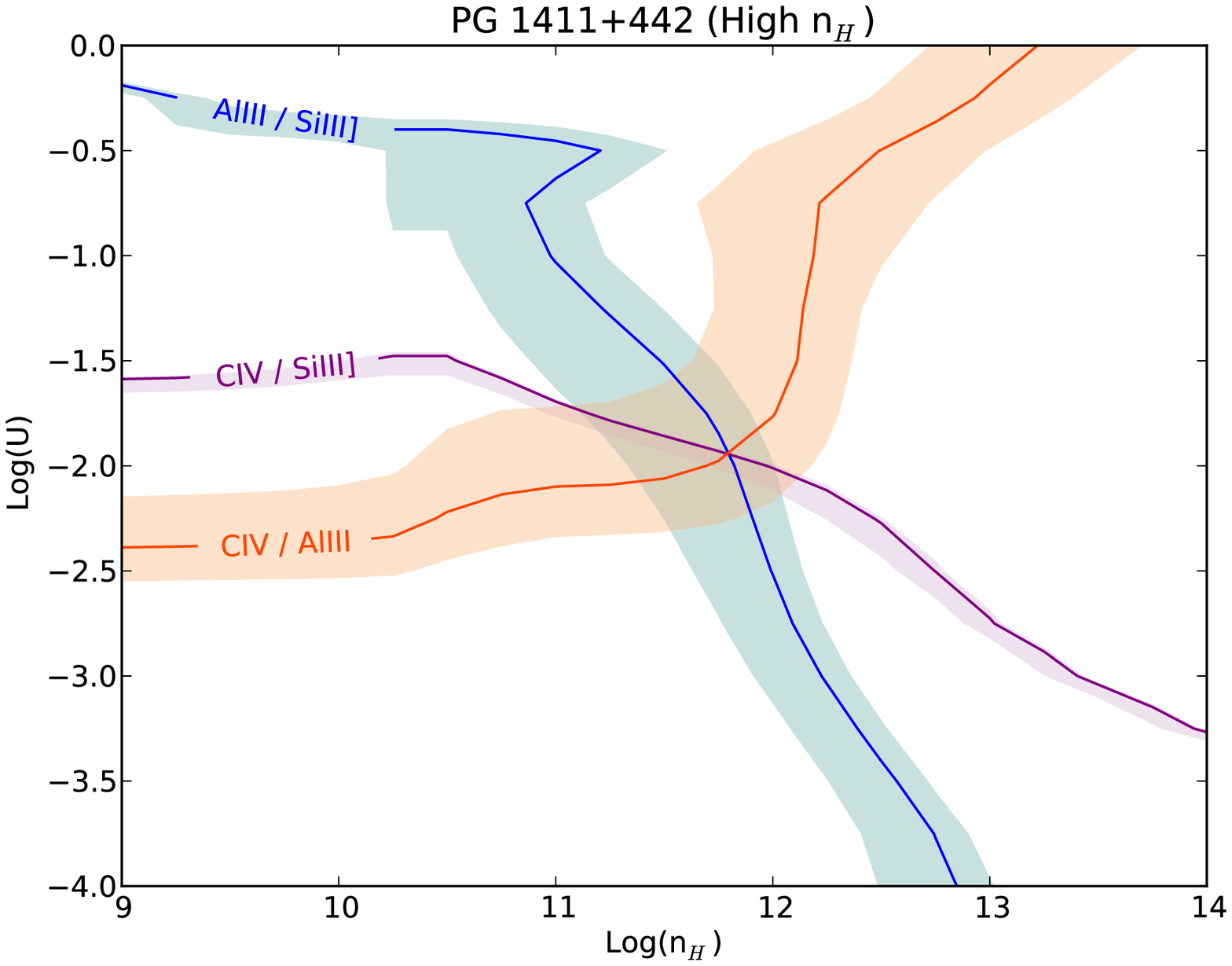}\includegraphics[scale=0.35]{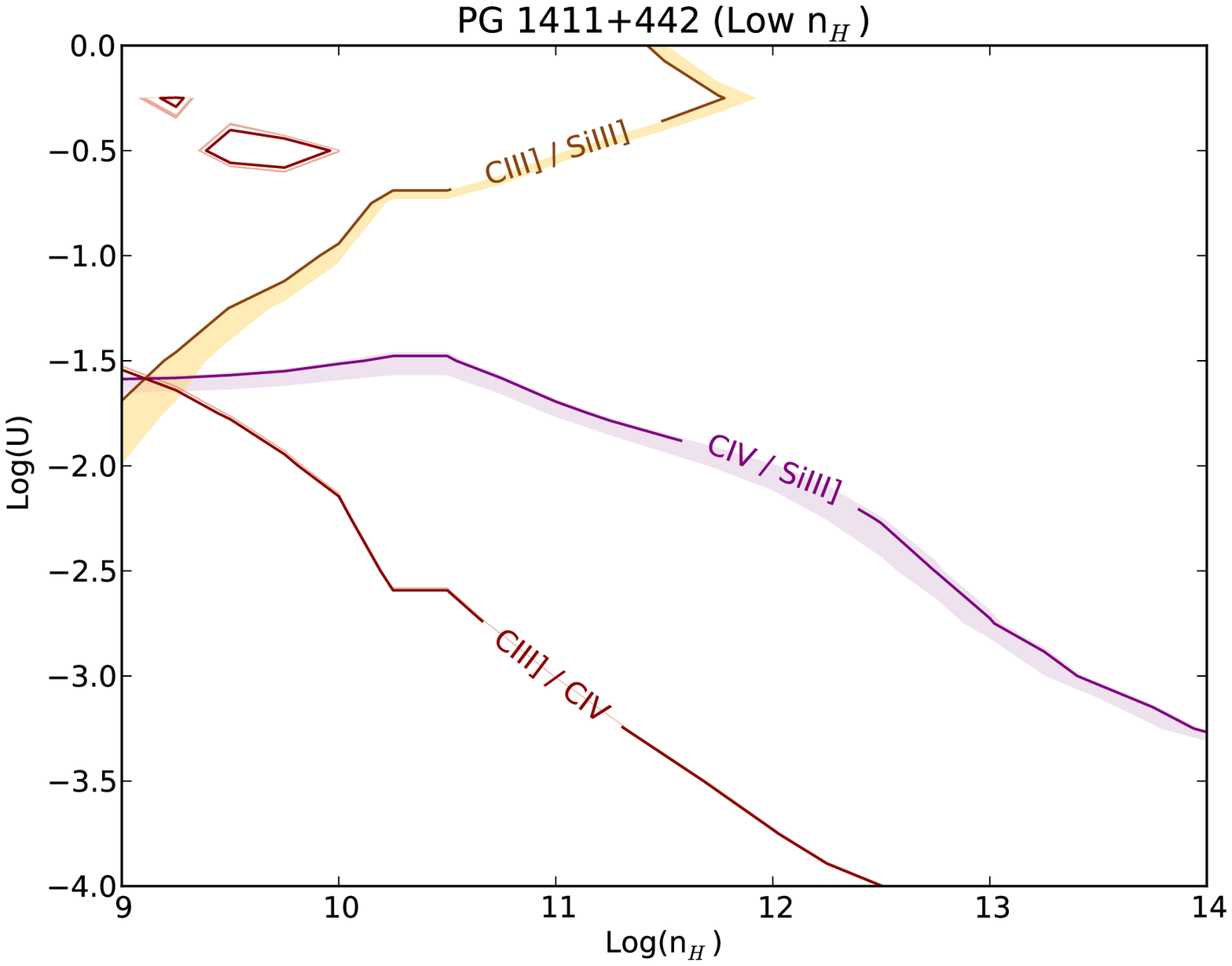}\\
\centerline{Fig. 2. --- Continued.}

\vfill

\vfill

\end{document}